\documentclass[conference]{IEEEtran}
\IEEEoverridecommandlockouts
\usepackage{cite}
\usepackage{amsmath,amssymb,amsfonts}
\usepackage{algorithmic}
\usepackage{graphicx}
\usepackage{textcomp}
\usepackage{xcolor}
\usepackage[none]{hyphenat}
\usepackage[ruled,vlined, linesnumbered]{algorithm2e}
\usepackage{subfigure}
\usepackage{balance}  
\usepackage{url}
\usepackage{amssymb}
\usepackage{amsmath}
\usepackage{epstopdf}
\usepackage{amsmath,epsfig,amsfonts,amssymb,graphicx}
\usepackage{subfigure}
\usepackage{array}
\usepackage{color}
\usepackage{enumerate}
\usepackage{verbatim}
\usepackage{colortbl}
\usepackage{multirow}
\usepackage{ntheorem}

\newcommand{\algorithmicbreak}{\textbf{break}}
\newcommand{\Break}{\algorithmicbreak}
\newtheorem{mydef}{Definition}
\newtheorem{lemma}{Lemma}

\newtheorem{corollary}{Corollary}
\def\BibTeX{{\rm B\kern-.05em{\sc i\kern-.025em b}\kern-.08em
    T\kern-.1667em\lower.7ex\hbox{E}\kern-.125emX}}

\expandafter\def\expandafter\UrlBreaks\expandafter{\UrlBreaks
  \do\a\do\b\do\c\do\d\do\e\do\f\do\g\do\h\do\i\do\j%
  \do\k\do\l\do\m\do\n\do\o\do\p\do\q\do\r\do\s\do\t%
  \do\u\do\v\do\w\do\x\do\y\do\z\do\A\do\B\do\C\do\D%
  \do\E\do\F\do\G\do\H\do\I\do\J\do\K\do\L\do\M\do\N%
  \do\O\do\P\do\Q\do\R\do\S\do\T\do\U\do\V\do\W\do\X%
  \do\Y\do\Z}

\begin{document}

\SetKwInput{KwIn}{Input}
\SetKwInput{KwOut}{Output}
\title{Influence-aware Task Assignment in Spatial Crowdsourcing (Technical Report)}

\author{Xuanhao~Chen\textsuperscript{1}, Yan~Zhao\textsuperscript{2}, Kai~Zheng\textsuperscript{1*}, Bin~Yang\textsuperscript{2},
Christian S.~Jensen\textsuperscript{2}
\\
\textsuperscript{1}University of Electronic Science and Technology of China, China\\
\textsuperscript{2}Department of Computer Science, Aalborg University, Denmark\\
xhc@std.uestc.edu.cn, yanz@cs.aau.dk, zhengkai@uestc.edu.cn, \{byang, csj\}@cs.aau.dk\\
\thanks{\textsuperscript{*}Corresponding author: Kai~Zheng.}
}

\maketitle

\begin{abstract}
With the widespread diffusion of smartphones, Spatial Crowdsourcing (SC), which aims to assign spatial tasks to mobile workers, has drawn increasing attention in both academia and industry. One of the major issues is how to best assign tasks to workers. Given a worker and a task, the worker will choose to accept the task based on her affinity towards the task, and the worker can propagate the information of the task to attract more workers to perform it. These factors can be measured as worker-task influence. Since workers' affinities towards tasks are different and task issuers may ask workers who performed tasks to propagate the information of tasks to attract more workers to perform them, it is important to analyze worker-task influence when making assignments. We propose and solve a novel influence-aware task assignment problem in SC, where tasks are assigned to workers in a manner that achieves high worker-task influence. In particular, we aim to maximize the number of assigned tasks and worker-task influence. To solve the problem, we first determine workers' affinities towards tasks by identifying workers' historical task-performing patterns. Next, a Historical Acceptance approach is developed to measure workers' willingness of performing a task, i.e., the probability of workers visiting the location of the task when they are informed. Next, we propose a Random reverse reachable-based Propagation Optimization algorithm that exploits reverse reachable sets to calculate the probability of workers being informed about tasks in a social network. Based on worker-task influence derived from the above three factors, we propose three influence-aware task assignment algorithms that aim to maximize the number of assigned tasks and worker-task influence. Extensive experiments on two real-world datasets offer detailed insight into the effectiveness of our solutions.


\end{abstract}
\section{Introduction}
With the near-ubiquitous diffusion of smartphones and similar devices, a new kind of crowdsourcing has emerged, namely Spatial Crowdsourcing (SC), where smartphone users serve as workers that perform tasks at specific physical locations. In SC, examples of spatial tasks include reporting local hot spots, taking photos or videos of a POI, and monitoring traffic conditions~\cite{zhao2020preference}.

SC has received substantial attention in the last years~\cite{cheng2016task,cheng2017prediction,song2017trichromatic,tong2018dynamic,tong2017flexible,tong2018unified,xia2019profit,zhao2017destination,ye2021task}. Studies exist that aim to maximize the total number of completed tasks~\cite{kazemi2012geocrowd}, the diversity score of assignments~\cite{cheng2014reliable}, the number of completed tasks for a worker with an optimal schedule~\cite{deng2013maximizing}, etc. These studies generally focus on the spatio-temporal information of workers and tasks during task assignment, while they do not consider worker-task influence, i.e., how to ensure that assigned tasks satisfy workers' affinities towards tasks and are well-known among workers who are likely to visit the locations of tasks. Visiting the location of a task is equivalent to accepting the task. 
In real-world scenarios, different workers prefer different kinds of tasks. Moreover, when completing tasks, workers can propagate information on available tasks to their friends through social networks. Workers who are informed can choose to perform tasks based on their historical task-performing patterns. 
It is important to analyze such phenomena when assigning tasks. For example, the owner of a new restaurant may want to publish a leaflet distribution task to promote the restaurant as widely as possible. Some free meal coupons and VIP cards are offered to workers who accept the task and help to propagate the news about the restaurant. If we only consider spatio-temporal information, an available worker who close to the restaurant at the current time will be assigned the task, but the worker may not be able to promote the restaurant widely. Thus, the case will not be a successful promotion. In addition, the real-time locations of workers are temporary, which ignores the worker's historical task-performing patters. Moreover, by analyzing the social networks that workers are in, we can obtain valuable insights about interactions among workers, which can be further utilized to improve the quality of spatial task assignments.

Recent studies have explored the effects of social impact in task assignment, where social network features are used to extract preference of worker groups~\cite{li2020consensus,li2020group}. However, these studies do not consider the interactions among workers, which include information propagation patterns and social network structures.
Different workers have different abilities to propagate information~\cite{tang2018online} and different probabilities to visit the locations of tasks~\cite{chen2020efficient}. This indicates that different workers contribute differently to worker-task influence. Moreover, it is important to infer task execution behaviors based on historical task-performing records of workers. Several approaches use past task-performing patterns to deduce worker preferences for tasks~\cite{yuen2012task,zhao2020preference}, but they do not analyze the willingness of workers to perform tasks, i.e., the probabilities that workers will visit the task locations. If a worker previously performed tasks near a new task, the worker is more likely to visit the location of the new task~\cite{to2014framework}. Lastly, we are not aware of any existing task assignment techniques that combine social networks and historical task-performing patterns to determine worker-task influence, which is a key factor for improving the quality of task assignment in SC.

\begin{figure}
\centerline{\includegraphics[scale=0.3]{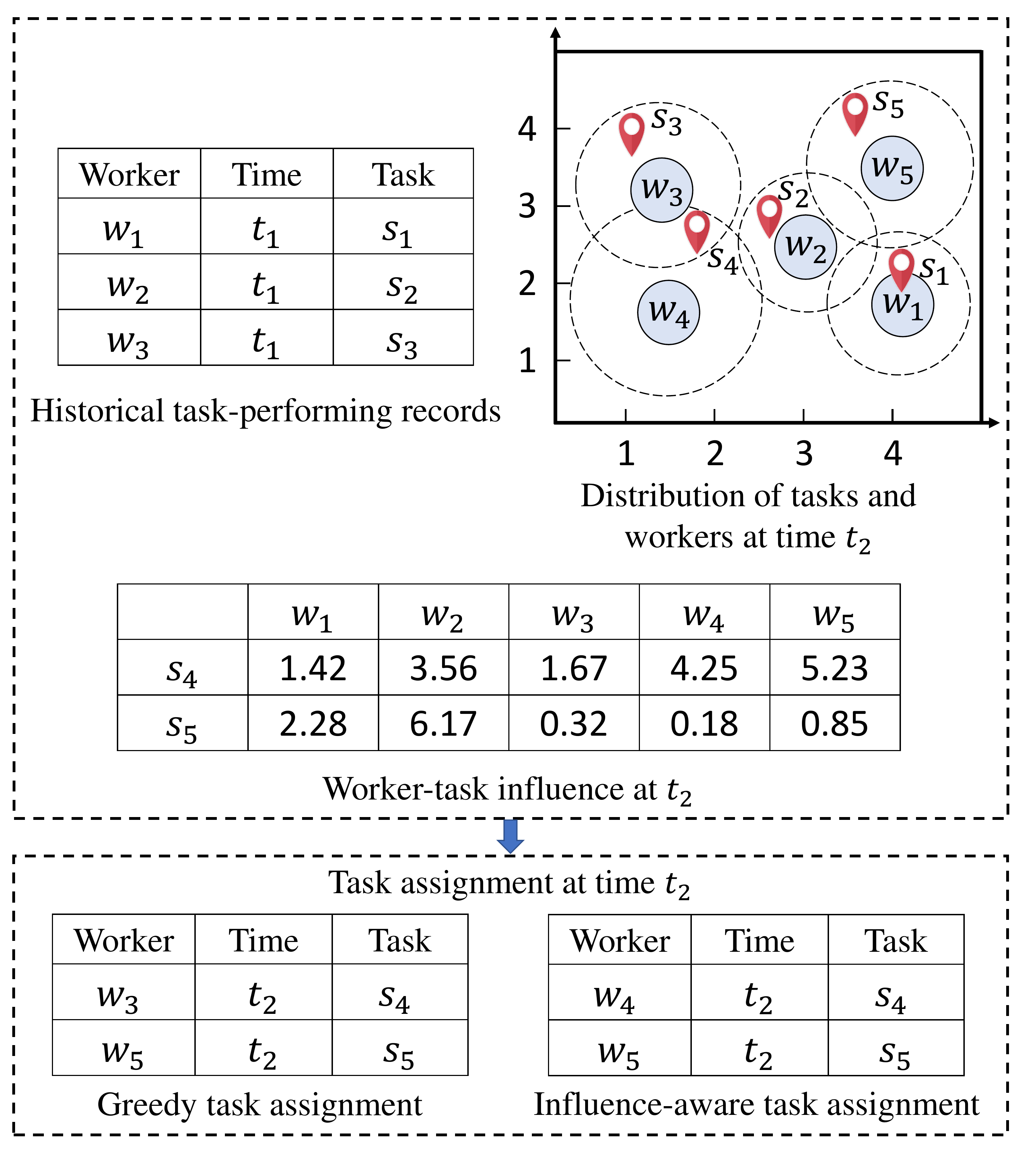}}
\vspace{-0.5cm}
\caption{Running Example}
\label{fig:example}
\vspace{-0.7cm}
\end{figure}

To address these challenges, we propose the Influence-aware Task Assignment (ITA) problem, where the objective is to assign tasks to suitable workers so as to maximize both the total number of assigned tasks and worker-task influence, which consists of workers' affinities (namely worker-task affinity) towards tasks, the probability (namely worker willingness) of workers visiting the locations of tasks and the probability (namely worker propagation) of workers being informed about tasks in social networks. Larger worker-task influence means that workers' affinities towards tasks are larger, and the number of people who are willing to visit the locations of tasks after informed is larger. An example of the ITA problem is illustrated in Figure~\ref{fig:example}. Workers, $w_1$, $w_2$, and $w_3$ performed tasks, $s_1$, $s_2$, and $s_3$, at time $t_1$, respectively. At time $t_2$, workers $w_4$ and $w_5$ are online, and tasks $s_4$ and $s_5$ become available. These are tasks published by new restaurants that ask workers to take photos and then advertise the restaurants on social media. The requirement of tasks is to increase the number of people who are willing to visit the location of the restaurant after knowing about it, i.e., enlarging worker-task influence. The circle around each worker denotes the reachable region of the worker at the current time. 
Because of the budgets of the restaurant, only one worker is required to perform each task, while the worker who are assigned the task should enlarge worker-task influence. A simple greedy approach is to assign tasks to the nearest worker, which gives the task assignment $\{(s_4,w_3),(s_5,w_5)\}$, where the value of worker-task influence is $1.67+0.85=2.52$. However, adopting an influence-aware task assignment approach, we can achieve a higher worker-task influence with assignment $\{(s_4,w_4),(s_5,w_5)\}$, where the value of worker-task influence is $4.25+0.85=5.1$.

Worker-task influence can be computed by worker-task affinity (i.e., workers' affinities towards tasks), worker willingness (i.e., the probability of workers visiting the locations of tasks) and worker propagation (i.e, the probability of workers being informed about tasks in social networks.). However, there exists a challenge of how to combine worker-task influence with existing objectives such as maximizing the number of assigned tasks. In other words, influence-aware assignment should optimize for worker-task influence without sacrificing other objectives. To achieve this, we propose a Data-driven Influence-aware Task Assignment (DITA) framework, consisting of two primary components. First, worker-task influence is calculated, 
which not only considers online interactions among workers, but also captures workers' historical task-performing patterns and real-time task assignments mode. Second, we design three algorithms to maximize the overall task assignments by giving higher priorities to workers who generate higher worker-task influence at every time instance.


The paper's contributions can be summarized as follows:

i) We formalize and study an Influence-aware Task Assignment (ITA) problem in the context of SC.
  To the best of our knowledge, this is the first study in SC that considers worker-task influence 
  and assigns tasks based on the influence.

ii) We calculate worker-task influence by taking into account worker-task affinity, worker willingness and worker propagation.

iii) We design three alternative algorithms to solve the ITA problem, including basic Influence-aware Assignment, Entropy-based Influence-aware Assignment, and Distance-based Influence-aware Assignment.

iv) We conduct extensive experiments on two real-world datasets to offer insight into the effectiveness of the proposed methods.


\section{PROBLEM STATEMENT}
\label{sec:problem}
We present necessary preliminaries, define the problem addressed, and give an overview of our solution framework. Table \ref{notation} lists notation used throughout the paper.

\subsection{Preliminary Concepts}
\label{sec:notation}

\begin{mydef}[Spatial Task]\label{def:st}
A spatial task, denoted by $s=(l, p, \varphi, C)$, has a location $s.l$, a publication time $s.p$, a valid time $\varphi$ (meaning that it will expire at $s.p+s.\varphi$), and multiple category labels $s.C$.
\end{mydef}


\begin{mydef}[Worker]
\label{def:worker}
A worker, denoted by $w=(l,r)$, consists of a location $w.l$ and a reachable distance $w.r$. The reachable range of worker $w$ is a circle with center $w.l$ and radius $w.r$, within which $w$ can accept assignments.
\end{mydef}

A spatial task $s$ can be completed only if a worker arrives at its location before the expiration deadline $s.p+s.\varphi$. With single-task assignment mode, the SC server assigns each task to one worker
at a time. 

\begin{mydef}[Worker-Task Influence]
\label{def:ws.if}
Given a worker $w$ and a task $s$. Worker-task influence (calculated in Section \ref{sec:methodology}), denoted as $\mathit{if}(w,s)$, consists of $w$'s affinity towards $s$, the probability of other workers visiting the location of $s$ after informed by $w$, and the probability of other workers who are informed by $w$ through social networks.
\end{mydef}

\begin{table}
\centering
\caption{Summary of Notation}
\label{notation}
\vspace{-0.2cm}
\begin{tabular}{|l|l|}
\hline
Symbol & Definition\\\hline
\hline
$s$ & Spatial task\\\hline
$s.l$ & Location of spatial task $s$\\\hline
$s.p$ & Publication time of spatial  task $s$\\\hline
$s.\varphi$ & Valid time of spatial task $s$\\\hline
$s.C$ & Categories of spatial task $s$\\\hline
$S$ & A spatial task set\\\hline
$w$ & Worker\\\hline
$w.l$ & Location of worker $w$\\\hline
$w.r$ & Reachable distance of worker $w$\\\hline
$W$  & A worker set \\\hline
$\mathit{if}(w,s)$ & Worker-task influence of worker $w$ and spatial task $s$\\\hline
$A$ & A task assignment \\\hline
$|A|$ & The total number of assigned tasks in task assignment $A$\\\hline
$A_{opt}$ & The optimal task assignment\\\hline
$\mathbb{A}$ & A task assignment set \\\hline
\end{tabular}
\vspace{-0.6cm}
\end{table}

\begin{mydef}[Spatial Task Assignment]
Given a set of tasks $S$ and a set of workers $W$, a spatial task assignment, denoted by $A$,
consists of a set of worker-task pairs of the form $(s,w)$, where task $s$ is assigned to worker $w$ satisfying the spatio-temporal constraints, and where each worker or task can be assigned at most once.
\end{mydef}

We use $|A|$ to denote the total number of assigned tasks in task assignment $A$.
The problem investigated is stated as follows:

\textbf{ITA Problem Statement}.
Given a set of workers and a set of tasks at the current time in an SC platform, our problem is to find a task assignment $\mathit{A_{\mathit{opt}}}$
that achieves the following goals:

1) primary optimization goal: maximize the total number of assigned tasks (i.e., $\forall$ $\mathit{A_i \in \mathbb{A}}$ ($|\mathit{A_i}|$$\le$$|\mathit{\mathit{A_{opt}}}|$)), where $\mathbb{A}$ denotes all possible assignments; and

2) secondary optimization goal: maximize worker-task influence of assignments.


\begin{lemma}
The ITA problem is NP-hard.
\end{lemma}

\begin{proof*}
We can prove the lemma through a reduction from the 0-1 knapsack problem, which is described as follows: Given a set $U$ with $n$ items, in which each item $u_i$ is labelled with a weight $l_i$ and a value $h_i$, the 0-1 knapsack problem is to find a subset $U^*$ of $U$ that maximizes $\sum_{u_i\in U^*}h_i$ subjected to $\sum_{u_i\in U^*}l_i\leq L$, where $L$ is the maximum weight capacity.

Consider the following instance of the ITA problem. Given a task set $S$ with $n$ tasks, each task $s_i\in S$ is associated with a worker (corresponding to the weight $l_i=1$ of the 0-1 knapsack problem). Here, the number of workers is sufficiently large. 
The value $h_i$ of each task $s_i$ that is a function related to task completion and worker-task influence, is at least as hard as the $h_i$ (that is a constant) in the $0$-$1$ knapsack problem, so that this difference does not make our problem easier. In addition, we have $L$ workers. Therefore, the ITA problem is to identify a task subset $S^*$ of $S$ that maximizes $\sum_{s_i\in S}h_i$ subjected to $\sum_{s_i\in S}l_i\leq L$. 

If the ITA problem instance can be solved in polynominal time, a $0$-$1$ knapsack problem can be solved by being transformed to the corresponding ITA problem instance and then it can be solved in polynominal time. This contradicts the fact that the $0$-$1$ knapsack problem is NP-hard~\cite{vazirani2013approximation}, and so there cannot be an efficient solution (i.e., in polynominal time) to the ITA problem instance that is then NP-hard. Since the ITA problem instance is NP-hard, the ITA problem is also NP-hard.
\end{proof*}
\subsection{Framework Overview}
\begin{figure}
\vspace{-0.3cm}
\centerline{\includegraphics[scale=0.33]{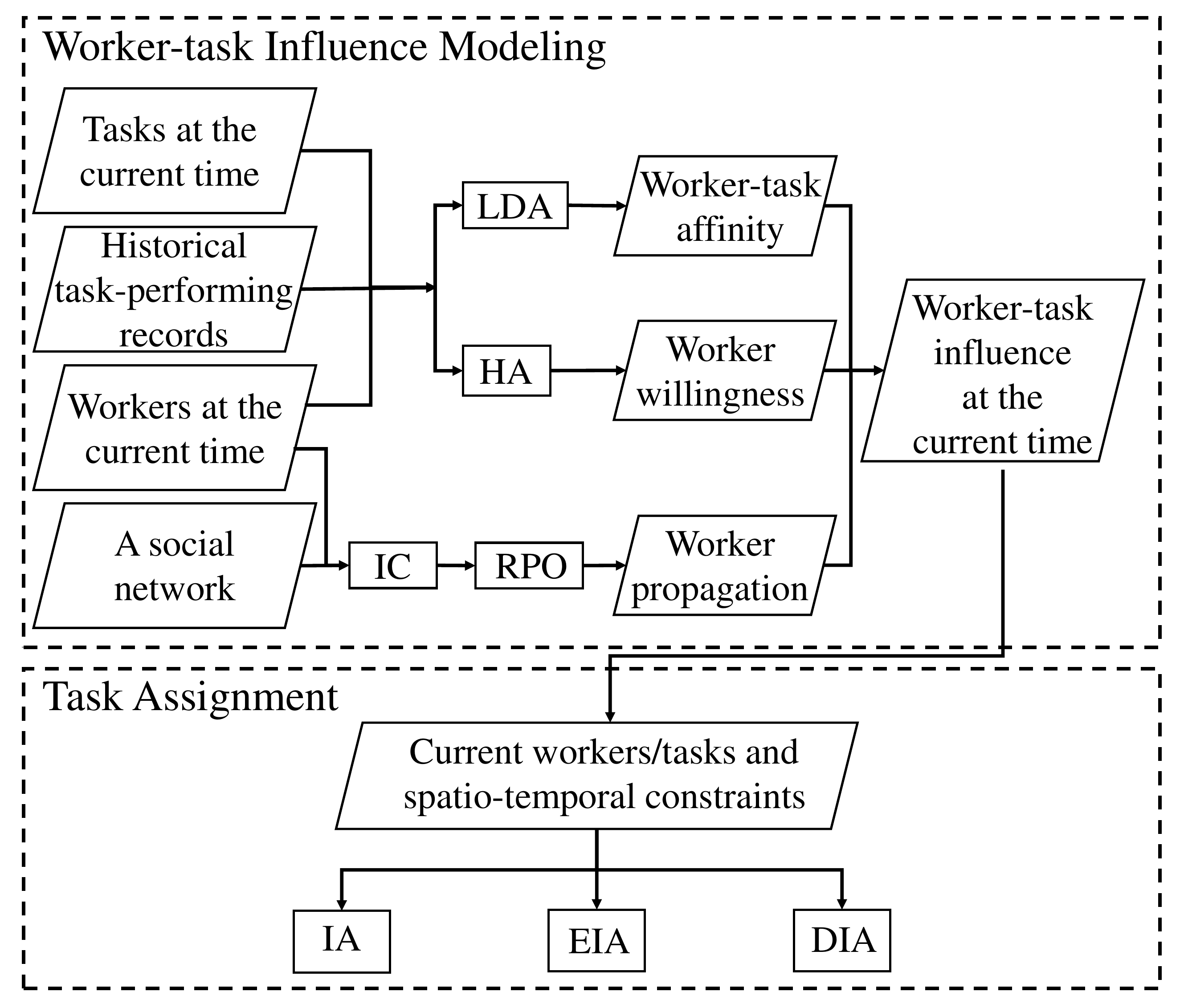}}
\vspace{-0.4cm}
\caption{DITA Framework}
\label{fig:framework}
\vspace{-0.7cm}
\end{figure}

We propose a framework, Data-driven Influence-aware Task Assignment (DITA), to solve the ITA problem. The framework has two components: worker-task influence modeling and task assignment, as shown in Figure~\ref{fig:framework}.


The first component aims to calculate worker-task influence. 
Specifically, we employ Latent Dirichlet Allocation (LDA) to measure workers' affinities (i.e., worker-task affinity) towards tasks, where we treat the categories of tasks that workers have already completed as documents to train the LDA model, and then the categories of tasks and workers at the current time are input into the trained LDA model to compute the worker-task affinity. For worker willingness calculation, we propose a Historical Acceptance (HA) algorithm to measure the probability of a worker visiting the location of a task based on the task-performing history of the worker and the real-time locations of the worker and task. 
For worker propagation, we first exploit an Independent Cascade (IC) model to simulate information propagation process of tasks in a given social network, and then we propose a Random reverse reachable-based Propagation
Optimization (RPO) algorithm to calculate worker propagation based on IC and social networks.


In the task assignment component, considering the spatio-temporal constraints (i.e., the reachable regions of workers and expiration times of tasks) of workers and tasks, 
we optimize the task assignment based on worker-task influence at each time instance and propose
a basic Influence-aware Assignment (IA) method. Taking worker-task influence and location entropy into account, we propose an Entropy-based Influence-aware Assignment (EIA) method. Moreover, a Distance-based Influence-aware Assignment (DIA) method which considers worker-task influence and workers' travel costs is developed.

\section{Worker-task Influence Calculation}
\label{sec:methodology}
We proceed to detail how to calculate worker-task influence for a worker and a task. We cover worker-task affinity, worker willingness, worker propagation and worker-task influence.

\subsection{Worker-Task Affinity Calculation}
\label{sec:category affinity calculation}
In SC, different workers exhibit different affinities (i.e., preferences) for the same categories of tasks, leading to different task-performing behaviors. For example,  a worker may like to report a hot spot, while another worker may prefer to monitor traffic conditions. Since task categories contain semantic information (e.g., restaurant) and the Latent Dirichlet Allocation (LDA) model~\cite{blei2003latent} performs well at modeling semantic affinity (i.e., semantic matching) between text documents by learning topics, we employ it to quantify worker-task affinity.

In LDA, a document is regarded as a set of words generated by several topics, 
where each topic is described by terms following a probability distribution. 
The modeling process can be formalized as follows:
\begin{equation*}
\footnotesize
    P(v_i|d)=\sum_{j=1}^{|\mathit{Top}|}P(v_i|t_j)P(t_j|d)
\end{equation*}

Here $P(v_i|d)$ is the probability of term $v_i$ for a document $d$ and $|\mathit{Top}|$ is the number of topics. Next, $P(v_i|t_j)$ is the probability of $v_i$ within topic $t_j$, and $P(t_j|d)$ is the probability of picking a term from $t_j$ in document $d$. LDA estimates the topic-term distribution, $P(v_i|t_j)$, and the document-topic distribution, $P(t_j|d)$, using Dirichlet priors. It iterates multiple times over each term in $d$ until the parameters in LDA converge. This way, we get the topic distribution of each document. Each topic is a probability distribution over a set of words. In the LDA model, words that are related semantically have high probability of belonging to the same topic.

\begin{figure}
\vspace{-0.3cm}
\centerline{\includegraphics[scale=0.33]{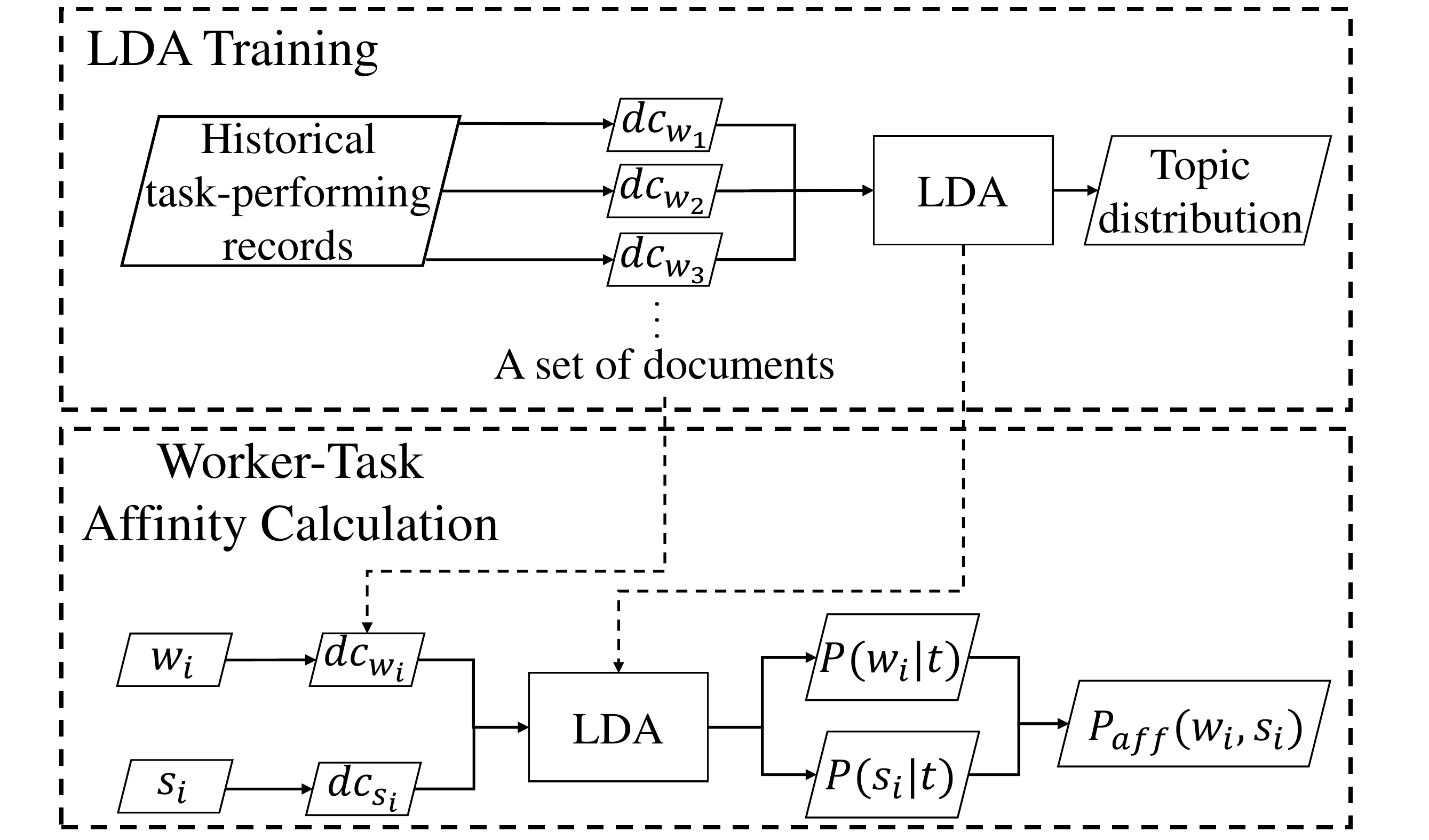}}
\vspace{-0.3cm}
\caption{Worker-task Affinity Calculation}
\label{fig:lda}
\vspace{-0.6cm}
\end{figure}

In order to adapt LDA to model worker-task affinity, we treat each task category as a word, and we treat the categories of tasks in the historical task-performing records of worker $w_i$ as a document, denoted by $\mathit{dc}_{w_i}$. The documents across all workers on an SC platform form a set of documents that is used to train the LDA model, cf. Figure~\ref{fig:lda}. 
Based on the documents, LDA can learn topics. Each topic is represented by a probability distribution over categories. For worker $w_i$ and task $s_i$ at the current time,
we can use the trained LDA model to calculate the topic distribution, where the topic distribution of $w_i$ is calculated from the historical task-performing records that reflect the preferred category distribution of $w_i$, and the topic distribution of $s_i$ is calculated based on its categories. Then the learned topics and the document $\mathit{dc}_{s_i}$ formed by the categories of the location of $s_i$ are used to estimate the worker-task affinity, $P_\mathit{aff}$, as follows:
\begin{equation*}
\label{equa:category affinity}
\footnotesize
 P_\mathit{aff}(w_i,s_i)=\sum\nolimits_{t\in \mathit{Top}}P(w_i|t)\cdot P(s_i|t),
\end{equation*}where $t$ denotes a topic and $\mathit{Top}$ is a set of learned topics. Further, $P(w_i|t)$ and $P(s_i|t)$ quantify how well topic $t$ matches the topic distribution of $w_i$'s historical task-performing records and the topic distribution of task $s_i$, respectively. A larger $P_\mathit{aff}(w_i,s_i)$ value indicates that $w_i$ is more likely to perform $s_i$, since the preferred category distribution of $w_i$ and that of the task $s_i$ are correlated better.

\subsection{Worker Willingness Calculation}
\label{sec:worker willingness calculation}
In general, different workers exhibit different willingness to visit the location of a task. Previous studies only consider real-time locations of workers and tasks~\cite{gong2018location,tao2020differentially} when assigning tasks. However, measuring a worker's willingness to visit the location of a task according to the distance between the worker's real-time location and the task location represents an incomplete picture. The real-time location is temporary, and this ignores the worker's historical task-performing patters.

To tackle this issue, we propose a Historical Acceptance (HA) approach to measure the willingness of a worker $w$ to visit the locations of particular tasks based on the worker's historical task-performing records (denoted as $S_w$) and the real-time locations of workers and tasks, where $S_w=\{(s_1,t^a_{s_1},t^l_{s_1}),(s_2,t^a_{s_2},t^l_{s_2}),\ldots,(s_n,t^a_{s_n},t^l_{s_n})\}$ and each triplet $(s_i,t^a_{s_i},t^l_{s_i})$ consists of task $s_i$, 
a task arrival time $t^a_{s_i}$, and a task completion time $t^l_{s_i}$. The worker willingness is measured as the probability that a worker moves from the locations of the tasks they have performed to the location of the current task.

In particular, HA computes worker willingness in terms of stationary distribution modeling of workers' historical mobility and movement probability density calculation.

\subsubsection{Stationary Distribution Modeling of Workers' Historical Mobility}
\label{sec:rwr}
The stationary distribution of a worker's historical mobility
captures the probability that a worker $w$ stays at the location of a performed task $s_i$ denoted as $P_w(w,s_i)$. This probability can be computed using the \emph{Random Walk with Restart} (RWR) method, which is an efficient approach to simulating the movement of objects~\cite{han2011data}. In order to adapt the RWR method to compute the stationary distribution of a worker's historical mobility, we exploit workers' historical task-performing records $S_w$ (ordered by check-in time) to construct an $n\times n$ weight matrix for worker $w$ ($w\in W$), where $n$ is the number of tasks performed by $w$. The weight of item $w'_\mathit{ij}$ in the $i$-th row and $j$-th column is set to $1/\sum_j^{n}m_\mathit{ij}$, where $m_\mathit{ij}=1$ if $w$ performed tasks at the $j$-th location; otherwise, $m_\mathit{ij}=0$.

\subsubsection{Movement Probability Density Calculation}
\label{sec:fw}
The movement probability density of worker $w$ is the probability density of moving from the location of task $s_i$ to the location of the next task, $s_{i+1}$, denoted as $f_w(d(s_i,s_{i+1}))$, where $f_w$ is a probability density function and $d(s_i,s_{i+1})$ is the distance between the location of $s_i$ and the location of $s_{i+1}$. Previous studies~\cite{zhu2014exploiting,zhu2015modeling} show that the movements of workers are self-similar. Since the random variable described by the Pareto distribution obeys the self-similarity property~\cite{singhai2007novel}, we choose the Pareto distribution to measure the movement probability density of worker $w$, denoted as $f_w(x;\pi;\omega)=\frac{\pi\omega^\pi}{x^{\pi+1}}$, where $x$ is the distance between locations of tasks, $\pi$ is a shape parameter that can be calculated using maximum likelihood estimation, and $\omega$ is the minimum value of $x$. Since real-time tasks are not known in advance, we use $S_w$ (ordered by check-in time) to compute $f_w$. Note that different task-performing orders will lead to different $f_w$. As a worker may perform several tasks at the same location, the minimum value of $d(s_i,s_{i+1})$ is 0, where $s_i$ and $s_{i+1}$ are tasks in $S_w$. We set $x_i$ of $f_w$ to $d(s_i,s_{i+1})+1$ to avoid $x_i$ being 0. In this case, $\omega=1$. Based on $x_i$, we employ maximum likelihood estimation to estimate $\pi$ of $f_w$. 
Further, $\pi$ can be calculated using the following equation:
\begin{equation*}
\footnotesize
\label{alpha}
\frac{d}{d \pi} \prod\limits_{i}^{|S_w|-1} \frac{\pi}{x_{i}^{\pi+1}}=0
\end{equation*}where $|S_w|$ is the number of performed tasks of $w$ and $x_{i}\geq1$.

Accordingly, $\pi$ is given by Equation~\ref{pi}.
\begin{equation}
\footnotesize
\begin{aligned}
\label{pi}
\begin{array}{c}
{\pi=\frac{|S_w|-1}{\sum_{i}^{|S_w|-1} \ln x_{i}}, \text { where } \sum\limits_{i}^{|S_w|-1} \ln x_{i} \neq 0}
\end{array}
\end{aligned}
\end{equation}

Since a worker may stay at different locations of performed tasks, we first need to compute the probability of worker $w$ staying at the location of a performed task $s_i$, and then we combine the probability with the probability that $w$ moves from the location of $s_i$ to the location of the current task $s$ to compute worker willingness. Based on Sections~\ref{sec:rwr} and \ref{sec:fw}, the willingness of $w$ to visit the location of $s$ can be calculated as follows:
\begin{equation}
\footnotesize
\begin{aligned}
\label{equ:offline}
  P_\mathit{wil}(w,s)&=\sum\limits_{s_i}^{S_w}P_w(w,s_i)\cdot\int_{d(s_i,s)}^\infty f_w(x)dx \\
  &=\sum\limits_{s_i}^{S_w}P_w(w,s_i)\cdot (d(s_i,s)+1)^{-\pi}
\end{aligned}
\end{equation}

\subsection{Worker Propagation Calculation}
\label{sec:task awareness calculation}
When knowing information of a task, a worker has the potential to propagate the task's information to other workers independently through social media. We propose \emph{worker propagation} to measure the probability of workers being known tasks.

There are two main challenges when computing worker propagation. First, since the information propagation in a social network is complex, it is important to simulate the propagation process reasonably. Second, based on the propagation process, the computation of worker propagation should be completed in limited time since we aim to assign tasks to workers online. To address these challenges, we propose an approximation method, called Random reverse reachable-based Propagation Optimization (RPO), for calculating the worker propagation. 


\subsubsection{Random Reverse Reachable Set Generation}\label{sec:rr}
We first detail how to generate Random Reverse Reachable (RRR) sets for workers, which will be used in the RPO method. The definition of an RRR set follows.
\begin{mydef}[Random Reverse Reachable Set]
\label{def:rr}
Given a social network $G=(W,E)$, constructing the reverse graph $G'$ of $G$ and selecting a worker $w_i$ uniformly at random from $G'$, a subgraph $g_i$ is a directed graph sampled from $G'$ under a given propagation model. A random reverse reachable (RRR) set for $w_i$ is a set of workers in $g_i$ that can reach $w_i$.

\end{mydef}

To generate an RRR set for each worker, it is important to select a suitable propagation model.
Independent Cascade (IC)~\cite{kempe2003maximizing,chen2010scalable,borgs2014maximizing,tang2015influence,stoica2019fairness,chen2020efficient,chen2021community} is a commonly-used propagation model, where users inform their neighbors independently.
In our ITA problem, a worker knowing a task has the potential to propagate information about the task to the neighbors independently, which can be well modeled by IC. Therefore, we use the IC model to
simulate the information propagation process of tasks and sample subgraphs from $G'$ to generate RRR sets.

The IC model is an iterative model. At the beginning of IC model, a worker $w_s$ who knows task $s$ is selected to inform the neighbors independently. In each iteration, 
if a worker $w_i$ has more than one neighbor knowing the task information in the current iteration, the worker will be informed by these neighbors independently. The probability, $P_{\mathit{k}}(w_i)$, of worker $w_i$ being informed by the neighbors in the $\mathit{k}$-th iteration is calculated as follows:
\begin{equation*}
\footnotesize
P_\mathit{k}(w_i)=1-\prod_{w_j\in \mathit{NE}_{k-1}(w_i)}(1-P_j(w_j,w_i)),
\end{equation*}where $\mathit{NE}_{k-1}(w_i)$ is the neighbors of $w_i$ who know a given task in the $(k-1)$-th iteration, and $P_j(w_j,w_i)$ is an in-degree-based probability that neighbor $w_j$ of $w_i$ informs $w_i$, which is a ratio between $1$ and $w_i$'s in-degree. 

Workers who are informed have only one chance to inform their neighbors. In the $k$-th iteration of the IC model, a worker who is not informed is added to a subgraph $g_s$ with the probability $P_{k}$, and the edges connecting the worker and the neighbors who are informed in the $(k-1)$-th iteration are added to $g_s$ with the probability $P_j$.
When no new workers are informed, the propagation process terminates, and $g_s$ is constructed.
Accordingly, the RRR set of worker $w_s$ is a set of workers that can reach $w_s$ along a finite number of edges in the directed graph $g_s$.


\subsubsection{Random Reverse Reachable-based Propagation Optimization Method} Next, we present the Random reverse reachable-based Propagation Optimization (RPO) method.

Based on Definition~\ref{def:rr}, if a worker $w_s$ appears in an RRR set of another worker $w_i$, the propagation process from $w_s$ should have a certain probability to inform $w_i$. Specifically, we can get following lemma.
\begin{lemma}
\label{lemma:rr}
Given two workers $w_s$, and $w_i$, the probability that $w_i$ is informed by $w_s$ under a propagation process equals the probability that $w_s$ belongs to an RRR set of $w_i$~\cite{borgs2014maximizing}.
\end{lemma}

Given a set of $N$ RRR sets, $\mathbb{R}=\{R_1,R_2,\ldots,R_N\}$, and a worker $w_s$ in $G$, let $\mathbb{R}_i$ ($\mathbb{R}_i\subseteq\mathbb{R}$) be a set of RRR sets that are generated by worker $w_i$ in $G$. 
Based on Lemma~\ref{lemma:rr} and the linearity of expectation, we can calculate the informed probability $P_\mathit{pro}(w_s,w_i)$ of $w_i$ being informed by $w_s$, as follows:
\begin{equation}
\footnotesize
\label{equa:imm}
P_\mathit{pro}(w_s,w_i)=\frac{|W|}{N}\cdot\mathbb{E}\left[\sum\nolimits_{j=1}^{|\mathbb{R}_i|} v_j\right],
\end{equation}where $v_j=0$ if $\{w_s\}\cap R_j=\emptyset$; otherwise, $v_j=1$. $R_j$ is the $j$-th set of $\mathbb{R}_i$, $|W|$ is the number of workers in $G$, $|\mathbb{R}_i|$ is the size of $\mathbb{R}_i$, and $\mathbb{E}\left[\sum_j^{|\mathbb{R}_i|}v_j\right]$ is the expected number of RRR sets generated by $w_i$ that cover $w_s$. To ensure that the estimation of $P_\mathit{pro}(w_s,w_i)$ is accurate, it is essential that $N$ is sufficiently  large to ensure that $\sum_j^{|\mathbb{R}_i|}v_j$ not deviate significantly from its expectation. The analysis of how to choose a setting for $N$ is covered in Section~\ref{sec:fa}.

The whole process of the RPO method is covered in Algorithm~\ref{alg:rrpo}. 
The main computational challenge is the huge search space when enumerating all possible RRR sets of each worker, which increases exponentially with respect to the number
of workers. Therefore, it is important to obtain a limited number of RRR sets that make it possible to guarantee an approximation ratio of the probability that workers are informed. To achieve this, we propose two lower bounds on the number of RRR sets (an iteration-based lower bound $N_\mathit{R}(k)$ and a threshold-based lower bound $N_R'(\gamma)$).
Specifically, given a worker $w_s$ who knows task $s$ and a social network $G=(W,E)$ as input, 
Algorithm~\ref{alg:rrpo} iteratively generates $N_\mathit{R}(k)$ RRR sets (stored in $\mathbb{R}$) based on $\mathit{G}$ (lines~\ref{alg:l4}--\ref{alg:l6}), where $N_\mathit{R}(k)$ is the iteration-based lower bound on the number of RRR sets.
Then for each worker $w_i\in W$, the algorithm computes the number $N_p(w_i)$ of workers which $w_i$ can propagate the task information to based on the current $\mathbb{R}$ and then finds the maximal $N_p(w_i)$, denoted as $N_p^{opt} = \max_{w_i\in W}{N_p(w_i)}$ (lines~\ref{alg:l7}--\ref{alg:l8}).
If $N_p^{opt}$ is larger than a threshold $\gamma$, a threshold-based lower bound $N_R'(\gamma)$ on the number of RRR sets is computed based on the threshold $\gamma$ and $N_p^{opt}$ (lines~\ref{alg:l9}--\ref{alg:l11}); otherwise,  $\mathbb{R}$ is set to $\emptyset$ (lines~\ref{alg:l12}--\ref{alg:l13}). 
Next, the algorithm continues to generate $(N_R'(\gamma)-|\mathbb{R}|)$ RRR sets when the size of the current $\mathbb{R}$ is too small, i.e., $|\mathbb{R}|<N_R'(\gamma)$ that is computed in the iteration (lines~\ref{alg:l15}--\ref{alg:l16}). Getting a set $\mathbb{R}$ of suitable RRR sets, we can compute $P_\mathit{pro}(w_s,w_i)$ ($w_i\in W$) (according to Equation~\ref{equa:imm}) and output the worker propagation for a worker, i.e., $\mathit{WP}_{w_s}\leftarrow(P_\mathit{pro}(w_s,w_1),\ldots,P_\mathit{pro}(w_s,w_{|W|}))$ (lines~\ref{alg:l17}--\ref{alg:l20}). The computation of $\mathit{k}$, $N_\mathit{R}(k)$, $N_p(w_i)$, $N_p^{opt}$, $N_R'(\gamma)$, $\gamma$ and the approximation ratio are discussed in Section~\ref{sec:fa}.


\begin{algorithm}
\footnotesize
\label{alg:rrpo}
\caption{RPO}
\KwIn{a worker $w_s$ who knows task $s$; a social network $G=(W,E)$ 
}
\KwOut{worker propagation $\mathit{WP}_{w_s}$ of $w_s$}

$\mathbb{R}\leftarrow \emptyset$;\\ \label{alg:l1}
$k\leftarrow |W|/2$;\\
\Repeat{$k = 2$}{

   Compute  $N_\mathit{R}(k)$;\\ \label{alg:l4}
   \tcp{\scriptsize{$N_\mathit{R}(k)$ is the iteration-based lower bound of the number of RRR sets.}}
    Generate $N_\mathit{R}(k)$ RRR sets based on $\mathit{G}$ (according to Section~\ref{sec:rr});\\
    Insert these RRR sets into $\mathbb{R}$;\\ \label{alg:l6}
        Compute $N_p(w_i)$ ($w_i\in W$) based on $\mathbb{R}$;\\ \label{alg:l7}
        \tcp{\scriptsize{$N_p(w_i)$ denotes the number of workers which $w_i$ can propagate the task information to.}}
    Find the maximal $N_p(w_i)$, denoted as $N_p^{opt}$;\\ \label{alg:l8}
    \If{$N_p^{opt}\geq \gamma$}
    {   \label{alg:l9}
        Compute $N_R'(\gamma)$ based on $N_p^{opt}$;\\
        \tcp{\scriptsize{$N_R'(\gamma)$ is the threshold-based lower bound of the number of RRR sets.}}
        \Break;\\ \label{alg:l11}
    }
    \Else
    {   \label{alg:l12}
       $\mathbb{R}\leftarrow \emptyset$\; \label{alg:l13}
       $k\leftarrow k/2$;\\
    }
}
\If{$|\mathbb{R}|< N_R'(\gamma)$}
{   \label{alg:l15}
    Generate $(N_R'(\gamma)-|\mathbb{R}|)$ RRR sets and insert them into $\mathbb{R}$\; \label{alg:l16}
}
\For{each $w_i \in W\setminus \{w_s\}$}
{   \label{alg:l17}
    Compute $P_\mathit{pro}(w_s,w_i)$ based on $\mathbb{R}$ (according to Equation~\ref{equa:imm})\;
}
$\mathit{WP}_{w_s}\leftarrow(P_\mathit{pro}(w_s,w_1),\ldots,P_\mathit{pro}(w_s,w_{|W|}))$\;
Return $\mathit{WP}_s$ \label{alg:l20}
\end{algorithm}

The complexity of RPO is dominated by the generation of RRR sets, which takes $O(|E|+|\mathbb{R}|\cdot |M|\cdot |E|/|W|)$ time, where $|E|$ is the number of edges in $G$, $|\mathbb{R}|$ is the size of $\mathbb{R}$, $|M|$ is the number of workers who can be informed by the greedy informed worker (see Definition~\ref{def:gis}), and $|W|$ is the number of workers in $G$.

\subsection{Worker-Task Influence Calculation}
We combine worker-task affinity $P_\mathit{aff}$, worker willingness $P_\mathit{wil}$, and worker propagation $WP_{w_s}$ to calculate the worker-task influence, $\mathit{if}(w_{s},s)$, of worker $w_{s}$ and task $s$, as follows:
\begin{equation*}
\label{tp}
\footnotesize
\mathit{if}(w_s,s) = P_\mathit{aff}(w_s,s)\cdot \sum_{w_i\in W\setminus \{w_s\}} P_\mathit{wil}(w_i,s)\cdot P_\mathit{pro}(w_s,w_i),
\end{equation*}where $w_s$ is a worker who knows the information of $s$, $W$ denotes the worker set in the social network, and $P_\mathit{pro}(w_s,w_i)$ is the $i$-th value in $WP_{w_s}$.

\subsection{Feasibility Analysis}
\label{sec:fa}
In order to guarantee an approximation ratio of computing worker propagation based on Random Reverse Reachable (RRR) sets, we present a feasibility analysis of computing a suitable number $N$ of RRR sets. First, we introduce the notions of informed range and martingale and then use these to propose lemmas that facilitate the computation of the number of RRR sets. Then corresponding proofs are provided to guarantee a high approximation ratio between the estimated worker propagation and the worker propagation computed based on RRR sets. The notions of informed range and martingale~\cite{chung2006concentration} are defined as follows:

\begin{mydef}[Informed Range]
Given a worker $w_s$ and an RRR set $\mathbb{R}=\{R_1,R_2,\ldots,R_N\}$, the informed range $\sigma(w_s)$ of $w_s$ is the estimated fraction of workers that are informed by $w_s$.
\begin{equation*}
\footnotesize
\label{equa:sigma}
\sigma(w_s)=\sum\nolimits_{i=1}^{|W|}P_\mathit{pro}(w_s,w_i)=\frac{|W|}{N}\cdot\mathbb{E}\left[\sum\nolimits_{j=1}^N v_j\right],
\end{equation*}
\end{mydef} where $v_j=0$ if $\{w_s\}\cap R_j=\emptyset$; otherwise, $v_j=1$.

\begin{mydef}[Martingale]
A sequence of random variables $x_1,x_2,\ldots$ is a martingale if and only if $\mathbb{E}[|x_i|]<+\infty$ and $\mathbb{E}[x_i|x_1,x_2,\ldots,x_{i-1}]=x_{i-1}$ for any i.
\end{mydef}

An important property of martingales~\cite{chung2006concentration} is shown as follows:
\begin{lemma}
\label{lemma:mar}
Given a martingale $x_1,x_2,\ldots$ such that $|x_1|\leq l_1$ and $|x_j-x_{j-1}|\leq l_1$ for $j\in[2,i]$, and $\mathit{Var}[x_1]+\sum_{j=2}^i\mathit{Var}[x_j|x_1,x_2,\ldots,x_{j-1}]\leq l_2$, for any $\epsilon>0$,
\begin{equation*}
\footnotesize
Pr\left[x_i-\mathbb{E}[x_i]\geq\epsilon\right] \leq \exp\left(-\frac{\epsilon^2}{\frac{2}{3}l_1\epsilon+2l_2}\right),
\end{equation*}
\end{lemma}where $\mathit{Var}[\cdot]$ is the variance of a random variable.

Given a worker $w_s$, since each worker $w_i$ is selected uniformly at random to generate $R_i$ and the generation of $R_i$ is independent of $R_1,R_2,\ldots,R_{i-1}$, we have $\mathbb{E}[v_i|v_1,v_2,\ldots,v_{i-1}]=\mathbb{E}[v_i]=\sigma(w_s)/|W|$. Let $\alpha=\sigma(w_s)/|W|$ and $x_i=\sum_{j=1}^i(v_j-\alpha)$. It is clear that $\mathbb{E}[x_i]=0$ and $\mathbb{E}[x_i|x_1,x_2,\ldots,x_{i-1}]=x_{i-1}$, which indicates that $x_1,x_2,\ldots,x_N$ is a martingale. Moreover, based on $x_i=\sum_{j=1}^i(v_j-\alpha)$, it is clear that $|x_1|\leq 1$ and $|x_i-x_{i-1}|\leq1$ for any $i\in[2,N]$. Combining this with the independence of $R_i$, we have:
\begin{equation}
\label{equa:var}
\footnotesize
\begin{aligned}
\mathit{Var}[x_1] + &\sum\nolimits_{i=2}^N \mathit{Var}[x_i|x_1,x_2,\ldots,x_{i-1}]=N\alpha(1-\alpha)\leq N\alpha
\end{aligned}
\end{equation}

Based on Lemma~\ref{lemma:mar} and Equation~\ref{equa:var}, the following corollary is derived.
\begin{corollary}
\label{c:1}
For any $\epsilon>0$,
\begin{equation*}
\footnotesize
Pr\left[\sum\nolimits_j^N v_j-N\cdot\alpha\geq \epsilon\cdot N\cdot\alpha\right]\leq
    \exp\left(-\frac{\epsilon^2}{2+\frac{2}{3}\epsilon} \cdot N\cdot\alpha\right)
\end{equation*}
\end{corollary}

We obtain the corollary by applying Lemma~\ref{lemma:mar} to $-x_1,-x_2,\ldots,-x_N$.
\begin{corollary}
\label{c:2}
For any $\epsilon>0$,
\begin{equation*}
\footnotesize
Pr\left[\sum\nolimits_j^N v_j-N\cdot\alpha\leq -\epsilon\cdot N\cdot\alpha\right]\leq
    \exp\left(-\frac{\epsilon^2}{2}\cdot N\cdot\alpha\right)
\end{equation*}
\end{corollary}

Given a set $\mathbb{R}=\{R_1,R_2,\ldots,R_N\}$ of RRR sets, let $f_R(w_s)$ be the fraction of RRR sets in $\mathbb{R}$ that cover $w_s$. It is clear that the number $N_p(w_s)$ of workers who can be informed by $w_s$ is $|W|\cdot f_R(w_s)$. Based on Corollary~\ref{c:2}, we have following lemma.
\begin{lemma}
\label{lemma:lambda11}
Let $\lambda\in(0,1)$, $\epsilon>0$, and
\begin{equation}
\footnotesize
\label{equ:lambda11}
N^{'}=\frac{2|W|\cdot\ln(1/\lambda)}{\sigma(w_s)\cdot\epsilon^2}
\end{equation}
If $N^{'}\leq N$, $N_p(w_s)\geq(1-\epsilon)\cdot\sigma(w_s)$ holds with at least probability $1-\lambda$.
\end{lemma}

Lemma~\ref{lemma:lambda11} shows that when $N'$ is sizable, the calculation of worker propagation based on Equation~\ref{equ:lambda11} guarantees a $(1-\epsilon)$-approximate solution. However, the value of $\sigma$ is different for different workers, which makes it hard to derive a suitable value of $N'$. Moreover, $N'$ should be as smaller as possible to reduce the computation time. To address this issue, we derive a lower bound on $N^{'}$. Let $w_s^\tau$ be a worker with maximum informed range, i.e., $\sigma(w_s^\tau)\geq\sigma(w_s)$, for any $w_s$ in $G$. We can rewrite Lemma~\ref{lemma:lambda11} as follows:

\begin{lemma}
\label{lemma:lambda1}
Let $\lambda\in(0,1)$, $\epsilon>0$, and
\begin{equation*}
\label{equa:lambda1}
\footnotesize
N_R'(\gamma)=\frac{2|W|\cdot\ln(1/\lambda)}{\sigma(w_s^\tau)\cdot\epsilon^2}
\end{equation*}
If $N_R'(\gamma)\leq N$, $N_p(w_s^\tau)\geq(1-\epsilon)\cdot\sigma(w_s^\tau)$ holds with at least probability $1-\lambda$.
\end{lemma}

The $\gamma$ in Lemma~\ref{lemma:lambda1} is a threshold, to be computed in Lemma~\ref{lemma:lambda3}. Since we aim to guarantee an approximation that is as high as possible, the setting of $\lambda$ should be as low as possible (e.g., $\lambda=1/|W|^o$, where $o\geq 1$).

However, in real cases, $\sigma(w_s^\tau)$ is unknown in advance. 
To address this problem, we derive a lower bound on $\sigma(w_s^\tau)$ with the help of a so-called greedy informed worker that can be calculated in advance. A greedy informed worker is defined as follows:
\begin{mydef}[Greedy Informed Worker]
\label{def:gis}
A greedy informed worker $w_s^\theta$ is a worker generated by a greedy approach that maximizes $f_R$: $w_s^\theta=\arg\max\limits_{w_i\in W}f_R(w_i)$. 
\end{mydef}

In order to use $w_s^\theta$ to derive the lower bound on $\sigma(w_s^\tau)$, we construct a test $T(\cdot)$ related to $w_s^\theta$ on a set of values (denoted as $K=\{k_1,k_2,\ldots\}$) and run the test on $K$. If $k_i>\sigma(w_s^\tau)$ then $T(k_i)=\mathit{false}$ with a high probability, and $k_i$ can be considered as a lower bound of $\sigma(w_i^\tau)$. Let $N_p^\mathit{opt}=|W|\cdot f_R(w_s^\theta)$. Based on Corollary~\ref{c:1}, we can construct $T(\cdot)$ based on following lemma:
\begin{lemma}
\label{lemma:lambda3}
Given $\epsilon^*>0$, let $k_i\in K$, $\gamma=(1+\epsilon^*)\cdot k_i$, $\lambda^*\in(0,1)$, and
\begin{equation*}
\footnotesize
\label{equa:delta2}
N\geq N_{R}(k_i)=\frac{\left(2+\frac{2}{3}\epsilon^*\right)\cdot\left(\ln(|W|)+\ln(1/\lambda^*)\right)\cdot |W|}
{{\epsilon^{*}}^2\cdot k_i}
\end{equation*}
If $\sigma(w_s^\tau)<k_i$, we get $N_p^\mathit{opt}<\gamma$ with at least probability $1-\lambda^*$.
\end{lemma}

Based on Lemma~\ref{lemma:lambda3}, it is easy to see that if $N_p^\mathit{opt}\geq\gamma$, then $\sigma(w_s^\tau)\geq k_i$ holds at least with probability $1-\lambda^*$, and $\sigma(w_s^\tau)$ can be set to $N_p^\mathit{opt}\cdot k_i/\gamma$. We can set $K=\{|W|/2,|W|/4,|W|/8,\ldots,2\}$ and then run the test $T(k_i): N_p^\mathit{opt}\geq\gamma$ on $O(\log_2 |W|)$ values of $K$ to compute the lower of $\sigma(w_s^\tau)$. Moreover, since we need to guarantee the approximation with high probability (e.g., $1-1/|W|^o$), the setting of $\lambda^*$ can be $\frac{1}{|W|^o\cdot\log_2|W|}$.

Based on Lemmas~\ref{lemma:lambda1} and~\ref{lemma:lambda3}, $\mathit{N}$ should be set to $\max\{N_R'(\gamma),N_R(k_i)\}$ to guarantee the approximation. However, it is difficult to choose suitable settings for $\epsilon$ and $\epsilon^*$ that minimize $\max\{N_R'(\gamma),N_R(k_i)\}$. To address that problem, $\max\{N_R'(\gamma),N_R(k_i)\}$ can be approximated with a simple function of $\epsilon$ and $\epsilon^*$, and then $\epsilon^*=\sqrt{2}\epsilon$ is derived as the minimizer of the function.

The proofs of above lemmas are shown as follows:

\textbf{Proof of Lemma~\ref{lemma:lambda11}.} Given any worker $w_s$, let $\alpha=\mathbb{E}[f_R(w_s)]$, by Lemma~\ref{lemma:rr}, $\alpha=\mathbb{E}[f_R(w_s)]=\sigma(w_s)/|W|$.

Based on Corollary~\ref{c:2}, we have
\begin{equation*}
\footnotesize
\begin{aligned}
&Pr[N_p(w_s)\leq (1-\epsilon)\cdot\sigma(w_s)]\\
&=Pr[f_R(w_s)-\alpha\leq -\epsilon\cdot\alpha]\\
&=Pr[N\cdot f_R(w_s)-N\cdot\alpha\leq-\epsilon\cdot N\cdot\alpha]\\
&=Pr\left[\sum\nolimits_j^N v_j-N\cdot\alpha\leq -\epsilon\cdot N\cdot\alpha\right]\\
&\leq \exp(-\epsilon^2\cdot N\cdot\alpha/2) \\
&\leq \exp(-\epsilon^2\cdot N'\cdot\alpha/2)=\lambda
\end{aligned}
\end{equation*}

The proof of Lemma~\ref{lemma:lambda1} is similar to that of Lemma~\ref{lemma:lambda11} when setting $\alpha=\mathbb{E}[f_R(w_s^\tau)]$.

\textbf{Proof of Lemma~\ref{lemma:lambda3}.} Given any worker $w_s$, let $\alpha=\mathbb{E}[f_R(w_s)]$. The following equation can be derived.
\begin{equation*}
\footnotesize
\label{equa:lambda3}
\alpha=\mathbb{E}[f_R(w_s)]\leq\mathbb{E}[f_R(w_s^\tau)]=\sigma(w_s^\tau)/|W|<k_i/|W|
\end{equation*}
Let $\beta=\frac{(1+\epsilon^*)\cdot k_i}{|W|\cdot\alpha}-1$. By Equation~\ref{equa:lambda3}, we can get $\beta>\frac{\epsilon^*\cdot k_i}{|W|\cdot\alpha}>\epsilon^*$. Based on Corollary~\ref{c:1}, we have
\begin{equation*}
\footnotesize
\begin{aligned}
&Pr\left[N_p(w_s)\geq \gamma\right]\\
&=Pr\left[f_R(w_s)-\alpha\geq \left(\frac{\gamma}{|W|\cdot\alpha}-1\right)\cdot \alpha\right]\\
&=Pr\left[N\cdot f_R(w_s)-N\cdot\alpha\geq \left(\frac{\gamma}{|W|\cdot\alpha}-1\right)\cdot N\cdot\alpha\right]\\
&\leq \exp\left(-\frac{\beta^2\cdot N\cdot\alpha}{2+\frac{2}{3}\beta}\right)\leq \exp\left(-\frac{\beta^2\cdot N_R(k_i)\cdot\alpha}{2+\frac{2}{3}\beta}\right)\\
&<\exp\left(-\frac{1}{2/\epsilon^*+2/3}\cdot
\frac{(2+2\epsilon^*/3)\cdot\ln\left(|W|/\lambda^*\right)}
{\epsilon^*}\right)\\
&=\lambda^*/|W|
\end{aligned}
\end{equation*}
According to the union bound and $N_p(w_s)\leq N_p^\mathit{opt}$, we have $N_p^\mathit{opt}<\gamma$ with at least $1-\lambda^*$ probability.

\section{Influence-aware Task Assignment}
\label{sec:algorithm}
We propose three algorithms, including basic, Entropy-based, and Distance-based Influence-aware Assignment, abbreviated IA, EIA, and DIA, respectively, that solve the ITA problem.
\subsection{Influence-aware Assignment}
Taking worker-task influence as the priority of task assignment, we propose a basic Influence-aware Assignment (IA) algorithm to solve the ITA problem by transforming it to a Minimum Cost Maximum Flow (MCMF)~\cite{kazemi2012geocrowd} problem.

To adapt MCMF to the ITA problem, we first construct a task assignment graph based on the available workers and tasks. Specifically, given a set of workers $W=\{w_1,w_2,\ldots\}$, and a set of tasks $S=\{s_1,s_2,\ldots\}$ at time $t$, we construct a graph $G=(N,E)$, where $N$ and $E$ denote sets of nodes and edges, respectively. Let $|N|=|W|+|S|+2$ and $|E|=|W|+|S|+m$, where $m$ is the number of available assignments for all workers. Since tasks expire at their deadlines and workers only accept tasks in their reachable range, the available assignments for worker $w$, denoted as $w.A$, should satisfy the following conditions:

i) task $s$ is located in the reachable circular range of
worker $w$, i.e., $d(w.l,s.l)\leq w.r$.

ii) worker $w$ has enough time to reach the location
of $s$ before it expires, i.e., $t+t(w.l,s.l)\leq s.p+s.\varphi$. 

We use $d(w.l,s.l)$ to denote the Euclidean distance between $w.l$ and $s.l$, and use $t(w.l,s.l)$ to denote the travel time from $w.l$ to $s.l$. For the sake of simplicity, we assume all the workers share the same travel speed, meaning that the travel time and distance are equivalent. However, the proposed algorithms can also address the cases where workers are moving at different speeds. Let $|w.A|$ be the number of available assignments for worker $w$, and thus $m$ can be derived by summing $|w.A|$ for all workers: $m=\sum_{w\in W}|w.A|$.

\begin{figure}
\centerline{\includegraphics[scale=0.33]{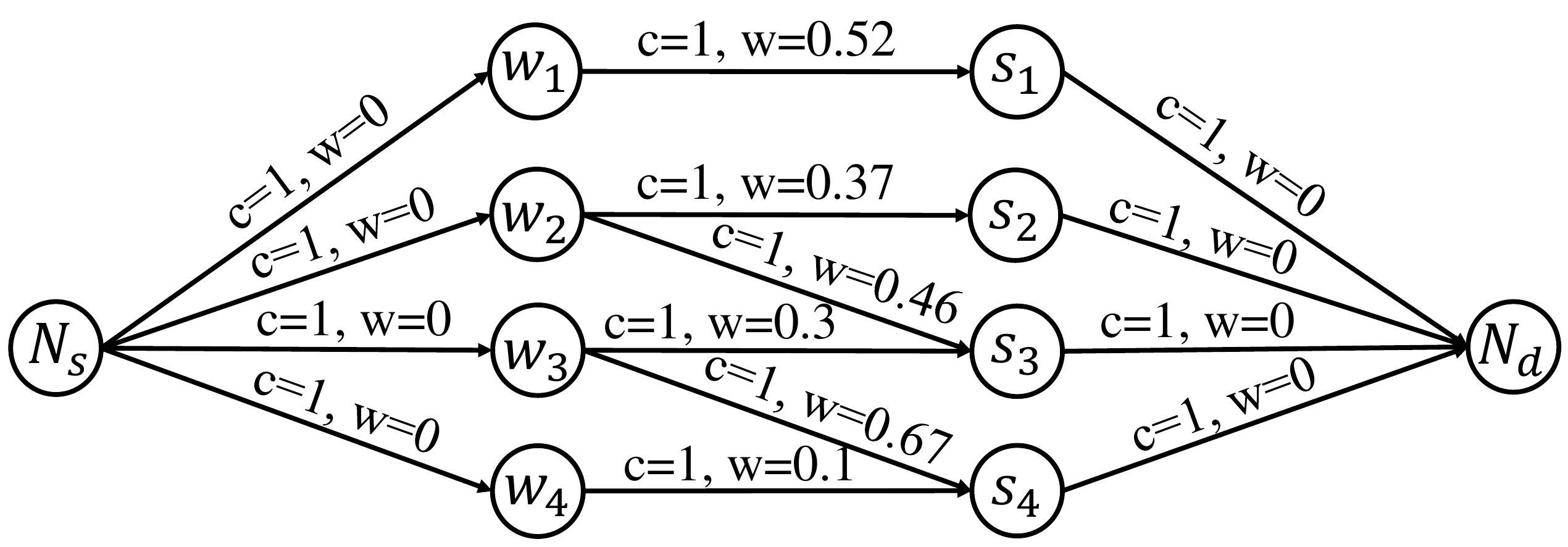}}
\vspace{-0.3cm}
\caption{Task Assignment Graph}
\label{fig:mcmf}
\vspace{-0.5cm}
\end{figure}

In graph $G$, nodes $n_i$ and $n_{|W|+j}$ correspond to a worker $w_i$ and a task $s_j$, respectively. Moreover, we add two new nodes (denoted as $n_0$ and $n_{|W|+|S|+1}$) as the source ($N_s$) and destination ($N_d$), respectively. An example graph $G$ for four workers and four tasks at the same time is illustrated in Figure~\ref{fig:mcmf}. The graph is generated by following steps:

i) $N_s$ connects all worker nodes, and the capacities of the corresponding edges are set to 1, i.e., $c=1$, since each worker can perform only one task at a time. The costs of these edges are set to 0.

ii) Each task node connects with $N_d$, and the capacities of the corresponding edges are set to 1, indicating that each task can be assigned to at most 1 worker. The costs of these edges are set to 0.

iii) If the assignment $(s_j,w_i)$ is available, i.e., $(s_j,w_i)\in w.A$, we add an edge from worker node $n_i$ to task node $n_{|W|+j}$. The capacities of the corresponding edges are set to 1, and the cost (denoted as $w(n_i,n_{|W|+j})$) is the ratio between 1 and the worker-task influence, $\mathit{if}(w_i,s_j)$, of $w_i$ and  $s_j$, i.e., $w(n_i,n_{|W|+j})=\frac{1}{\mathit{if}(w_i,s_j)+1}$.

Then the task assignment problem is converted into an MCMF problem in the directed graph $G$ from $N_s$ to $N_d$, which is to achieve the maximum flow (i.e., maximizing the task assignments) while minimize the cost (i.e., maximizing worker-task influence). The Ford-Fulkerson algorithm~\cite{ford1956maximal} is employed to compute the maximum flow of the graph, and then linear programming is used to minimize the cost of the flow~\cite{kazemi2012geocrowd}. 
\subsection{Entropy-based Influence-aware Assignment (EIA)}
In SC, each task has a location. If many workers are close to a task, i.e., the relative proportion of workers close to the task is high, the task is more likely to be completed. Considering that location entropy~\cite{cranshaw2010bridging,kazemi2012geocrowd} is an efficient metric to measure the total number of workers in the location of a task as well as the relative proportion of their visits to that location, we use it to measure the relative proportion of workers in the location of a specific task. Lower location entropy indicates that the distribution of the visits to that task is restricted to only a few workers. To maximize the total number of task assignments, a task located in a region with smaller location entropy should be given higher priority when making assignments. Let $\mathit{Num}_w$ denote the historical number of visits of worker $w$ to the location of task $s$, and let $\mathit{Num}_s$ denote the total number of visits of all workers to the location of task $s$. Then the location entropy $s.e$ of task $s$, is computed as follows:
\begin{equation*}
\footnotesize
s.e=-\sum\nolimits_{w\in W_s}P_s(w)\cdot\ln P_s(w),
\end{equation*}where $W_s$ is a set of workers that have performed task $s$ historically, and $P_s(w)=\mathit{Num}_w/\mathit{Num}_s$.

Considering worker-task influence and location entropy, 
EIA adapts IA by setting the cost $w(n_i,n_{|W|+j})$ of each edge that connects $w_i$ and $s_j$ to $(s.e+1)/(\mathit{if}(w_i,s_j)+1)$, where $\mathit{if}(w_i,s_j)$ is the worker-task influence of worker $w_i$ and task $s_j$.

\subsection{Distance-based Influence-aware Assignment (DIA)}
The IA algorithm fails to consider travel costs between the locations of workers and tasks. Workers are more likely to perform nearby tasks~\cite{to2014framework,zhao2020preference}, and travel cost is a critical factor when workers choose which tasks to perform. We compute the travel cost between a worker $w_i$ and a task $s_j$, denoted as $d(w_i.l,s_j.l)$, using Euclidean distance. Workers who are closer to tasks will be given higher priority to perform them. To achieve this, we propose a Distance-based Influence-aware Assignment (DIA) algorithm that uses travel costs to discount worker-task influence. Specifically, DIA modifies IA by setting the cost $w(n_i,n_{|W|+j})$ of each edge that connects $w_i$ and $s_j$ to $1/(F(w_i.l,s_j.l)\cdot \mathit{if}(w_i,s_j)+1)$, where $F(w_i.l,s_j.l)=1-\min (1,d(w_i.l,s_j.l)/w_i.r)$.

\section{Experimental Evaluation}
\label{sec:experiment}
\subsection{Experimental Setup}
Due to the lack of benchmarks for spatial crowsourcing task assignment algorithms, two check-in datasets consisting of social networks of workers, and workers' check-ins from Brightkite (BK)~\cite{cho2011friendship} and FourSquare (FS)~\cite{likhyani2017locate} are used to simulate a spatial crowdsourcing scenario. This is common practice when evaluating SC platforms~\cite{zhao2020predictive,cheng2017prediction,wang2021task,deng2013maximizing}. Since BK does not contain category information of venues, we exact categories of the venues with the aid of the FourSquare API\footnote{https://developer.foursquare.com/docs/}. BK has 58,228 users, 214,078 social connections, and 4,491,143 check-ins from April 2008 to October 2010. FS has 11,326 users, 47,164 social connections, and 1,385,223 check-ins from January to December 2011.

We assume that all users are workers since users who check in at different spots are good candidates to perform nearby spatial tasks, and we assume that their locations are those of the most recent check-ins. Moreover, we set the time granularity to one day, during which the available tasks and workers are entered into our framework. We also assume that users who check in at a time instance are available workers for that time instance, and we assume that a worker is online until the worker is assigned a task. For each check-in venue, we use its
location and the earliest check-in time of the day as the location
and publication time of a task. Further,
the categories of check-in locations are regarded as task categories. We set the number of topics used to extract worker-task affinity to 50, i.e., $\mathit{|Top|}=50$. The informed probability of each social network edge, $e$, is set to $1/\mathit{id}_e$~\cite{chen2010scalable,jung2012irie,tang2015influence}, i.e., $P_j=1/\mathit{id}_e$, where $\mathit{id}_e$ denotes the number of edges with the same end point with $e$. 
The parameters $\epsilon$ and $o$ in the Random reverse reachable-based Propagation Optimization approach are set to 0.1 and 1, respectively. 
Travel costs are calculated using Euclidean distance, and the speeds of workers is set to 5 km/h. The default values of all parameters used in the experiments are summarized in Table~\ref{parameters}. In task assignment experiments, we run the algorithms over 4 days of a month on BK and FS, and we report average results. All experiments are run on a Linux (Ubuntu 16.04) machine with Intel(R) Xeon(R) E5-2650 v4 2.20GHz processor and 256G memory.

\begin{table}[htbp]
\vspace{-0.4cm}
\centering
\caption{Parameter Settings}
\label{parameters}
\vspace{-0.3cm}
\begin{tabular}{|l|l|}
\hline
Parameter & Default value\\\hline
\hline
Number of tasks $|S|$ & 1500\\\hline
Number of workers $|W|$ & 1200\\\hline
Valid time of tasks $\varphi$ & 5 h\\\hline
Workers' reachable radius $r$ & 25 km\\
\hline
\end{tabular}
\vspace{-0.4cm}
\end{table}
\subsection{Experimental Results}
\subsubsection{Influence Modeling Performance}
We first evaluate the performance of worker-task affinity,
worker willingness, and
worker propagation and their impact on worker-task influence. We consider the IA algorithm and three variants of it to study the contribution to worker-task influence  of the three aspects. The methods are as follows:

i) IA: Our basic Influence-aware Assignment algorithm, which considers worker-task influence and aims to maximize total task assignment and worker-task influence.

ii) IA-WP: A variant of IA that considers worker willingness and worker propagation.

iii) IA-AP: A variant of IA that considers worker-task affinity and worker propagation.

iv) IA-AW: A variant of IA that considers worker-task affinity and worker willingness.

Since we aim to maximize the influence of tasks, we propose Average Influence, $\mathit{AI}$, to evaluate the performance of each algorithm, which is calculated as follows:
\begin{equation}
\footnotesize
\label{equ:atp}
\mathit{AI}=\frac{\sum_{(s,w)\in A}\mathit{if}(w,s)}{\mathit{|A|}},
\end{equation}where $\mathit{if}(w,s)$ is the worker-task influence of worker $w$ and task $s$, and $\mathit{|A|}$ is the number of assignments.

\begin{figure}[htbp]
\centering
\subfigure[Average Influence on BK] {\includegraphics[width=0.23\textwidth]{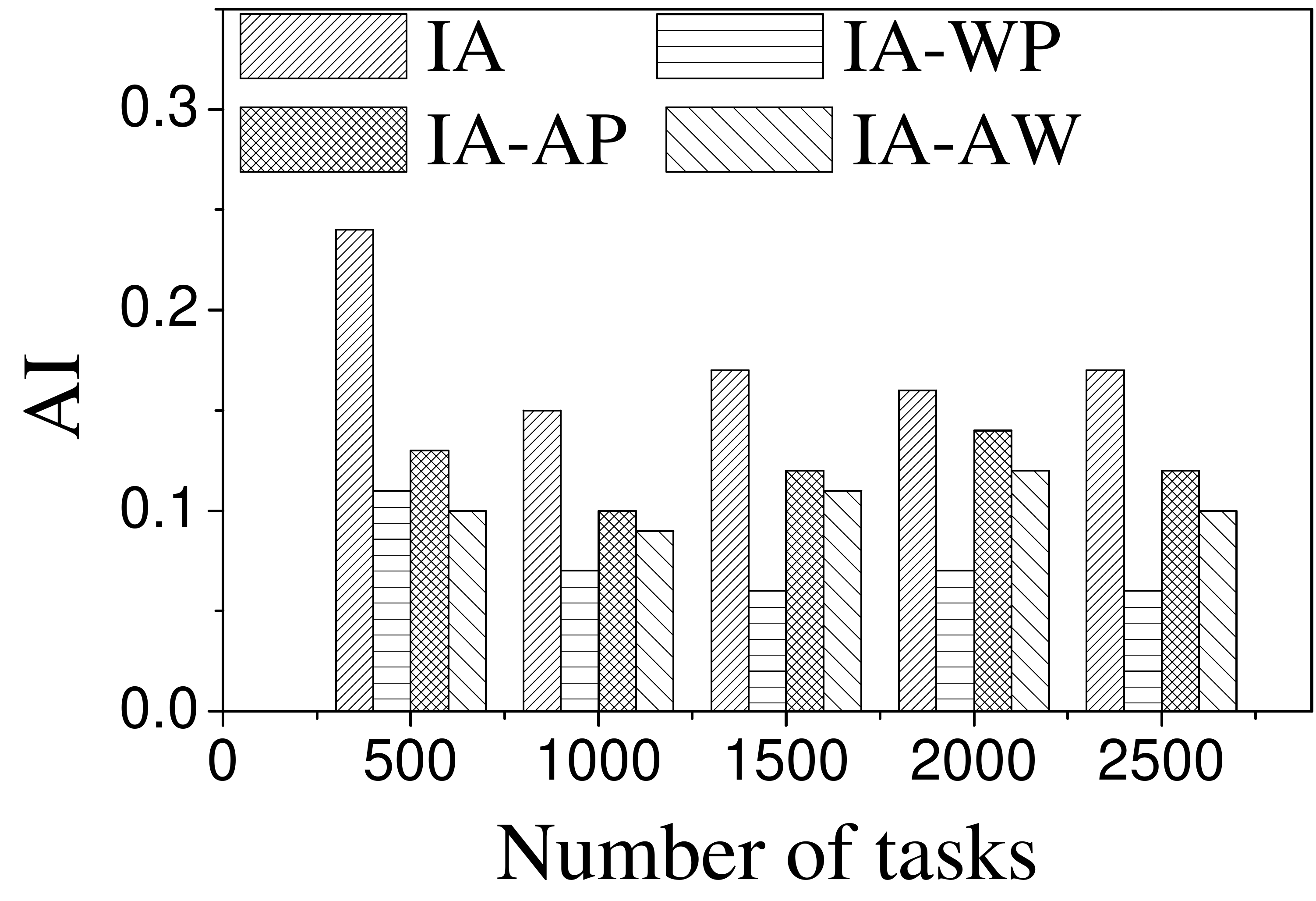}\label{fig:var-s-bk-cpu}}
\subfigure[Average Influence on FS] {\includegraphics[width=0.23\textwidth]{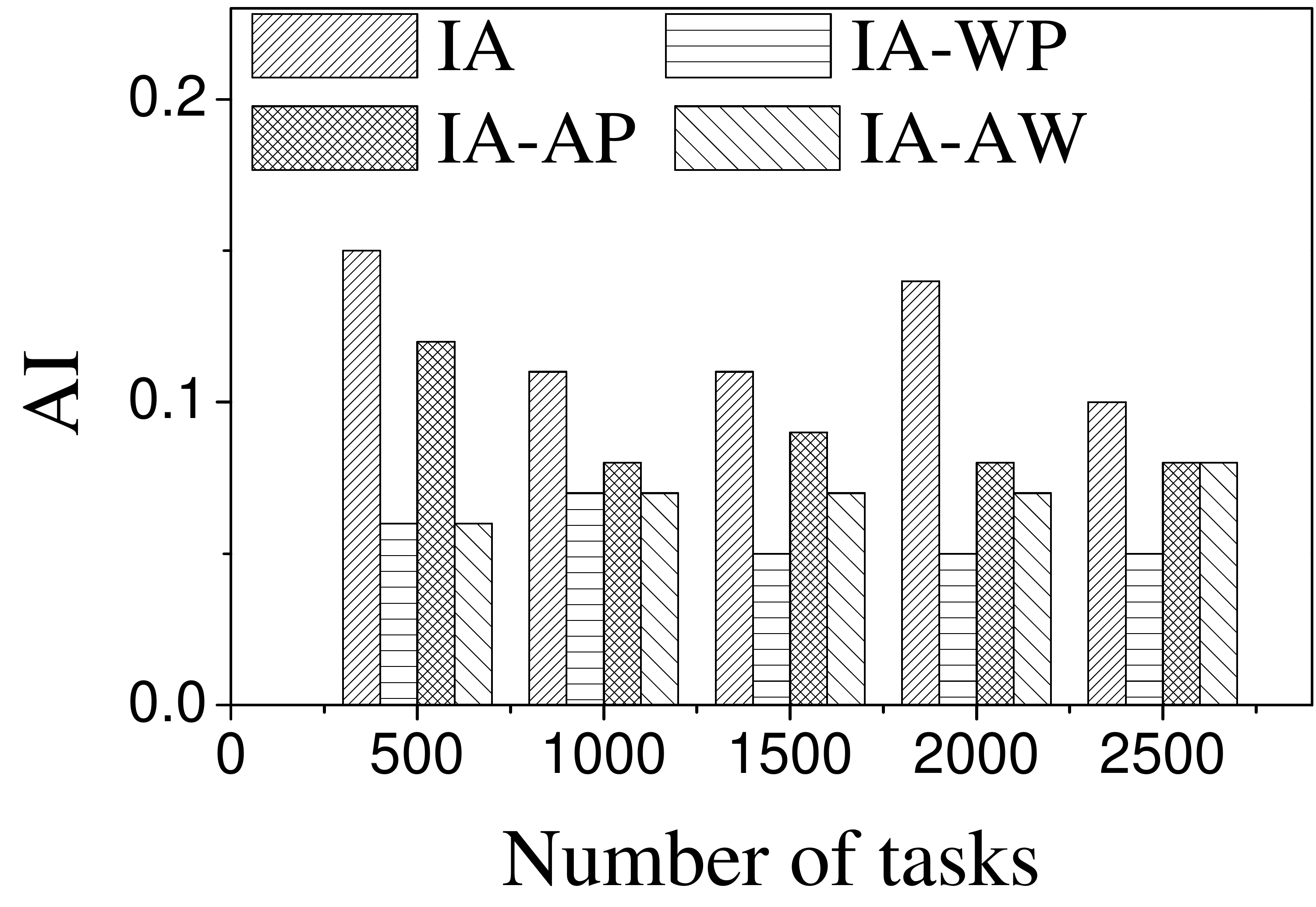}\label{fig:var-s-fs-cpu}}
\vskip -9pt
\caption{Effect of $|S|$}
\label{fig:vs}
\end{figure}

\emph{Effect of $|S|$:} We first study the effect of $|S|$ on AI. As illustrated in Figure~\ref{fig:vs}, IA achieves the largest AI, followed by IA-AP, for any $|S|$. The reason is that IA considers worker-task affinity, worker willingness and worker propagation, while none of the variants consider all aspects. IA-AP performs better than IA-WP and IA-AW. This may be due to the fact that the average probability of workers visiting locations of tasks is small, which means that the weight of worker willingness on computing task influence is smaller than that of worker-task affinity and worker propagation. Another observation is that AI of IA is highest when $|S|=500$. The reason is that the number of workers who can generate larger worker-task influence is small and that most of them are selected when $|S|=500$.

\begin{figure}[htbp]
\centering
\subfigure[Average Influence on BK] {\includegraphics[width=0.23\textwidth]{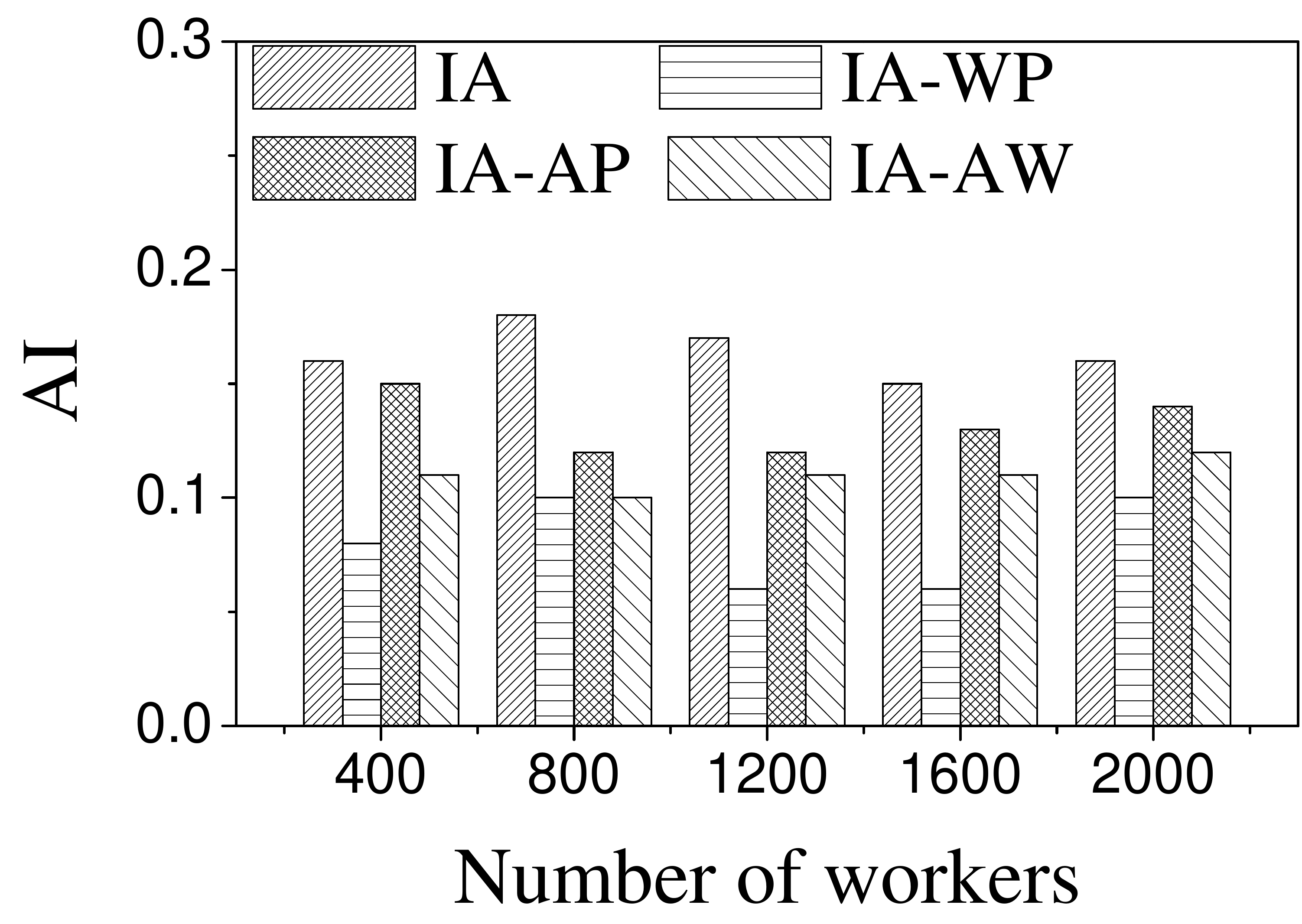}\label{fig:var-w-bk-cpu}}
\subfigure[Average Influence on FS] {\includegraphics[width=0.23\textwidth]{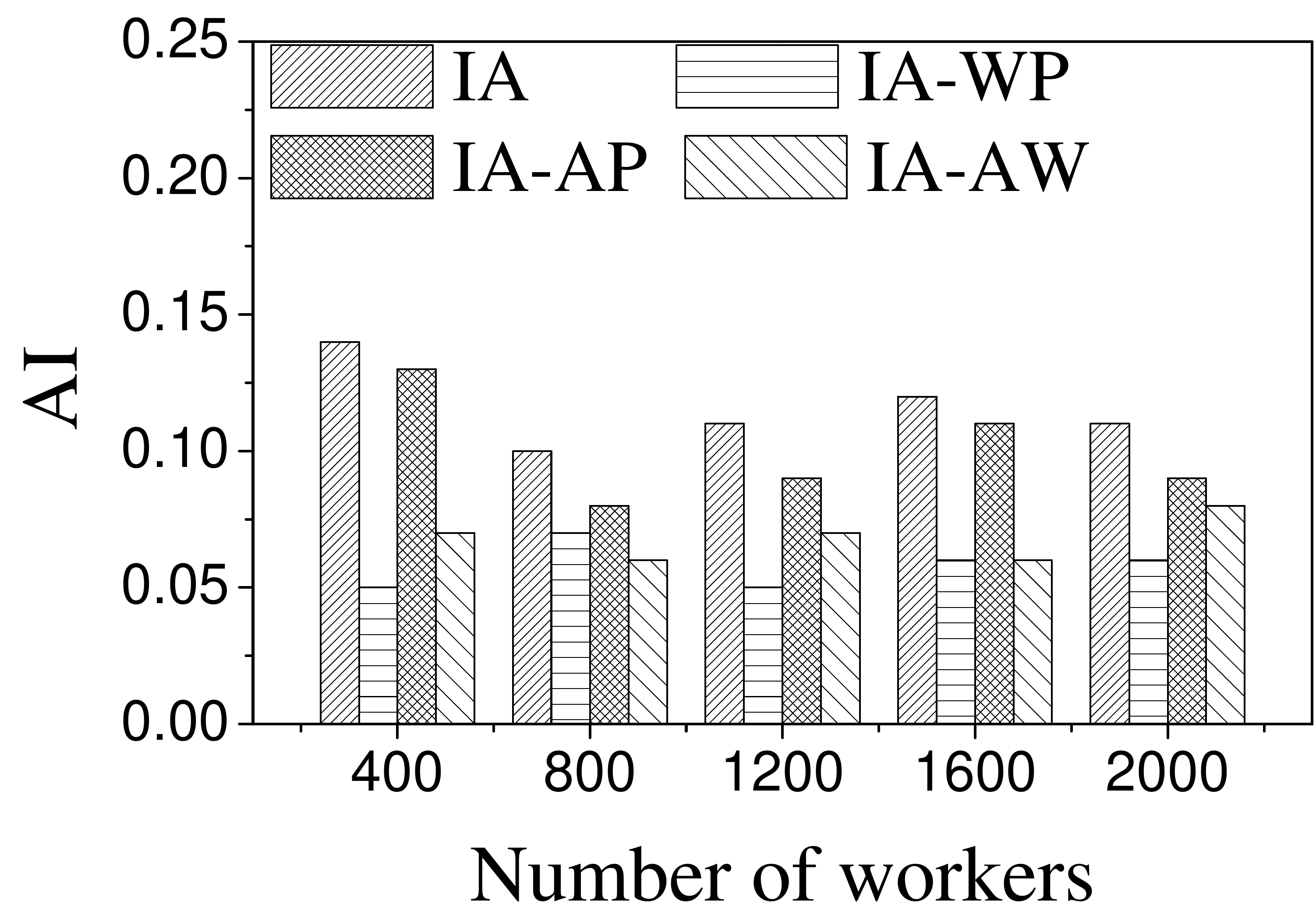}\label{fig:var-w-fs-cpu}}
\vskip -9pt
\caption{Effect of $|W|$}
\label{fig:vw}
\end{figure}

\emph{Effect of $|W|$:} Next we study the effect of $|W|$. From Figure \ref{fig:vw}, we can see that the AI of IA-WP is lowest among these algorithms in most cases, which means the weight of worker-task affinity plays a more important role on modeling worker-task influence than worker willingness and worker propagation.

\begin{figure}[htbp]
\centering
\subfigure[Average Influence on BK] {\includegraphics[width=0.235\textwidth]{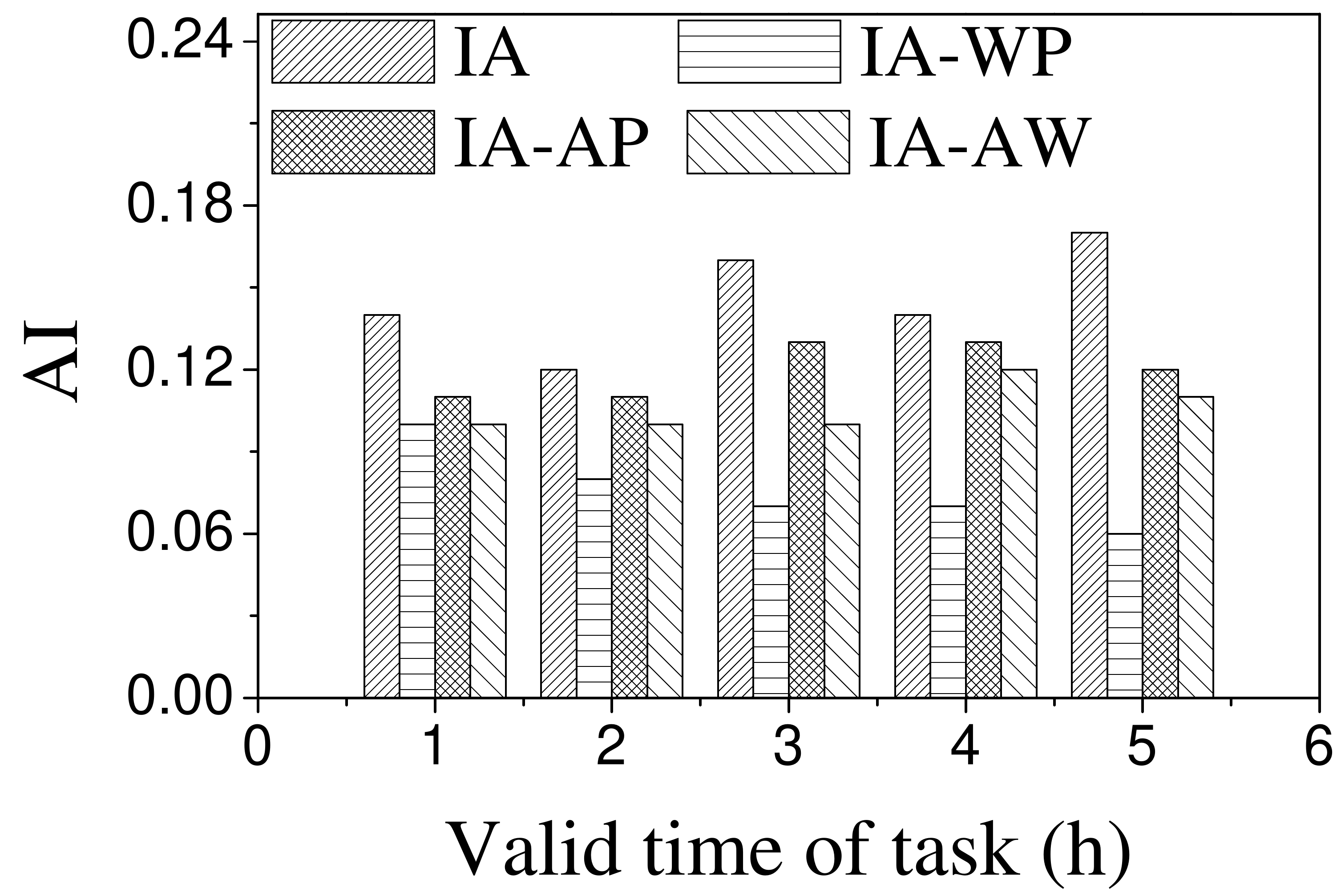}\label{fig:var-s-bk-cpu}}
\subfigure[Average Influence on FS] {\includegraphics[width=0.235\textwidth]{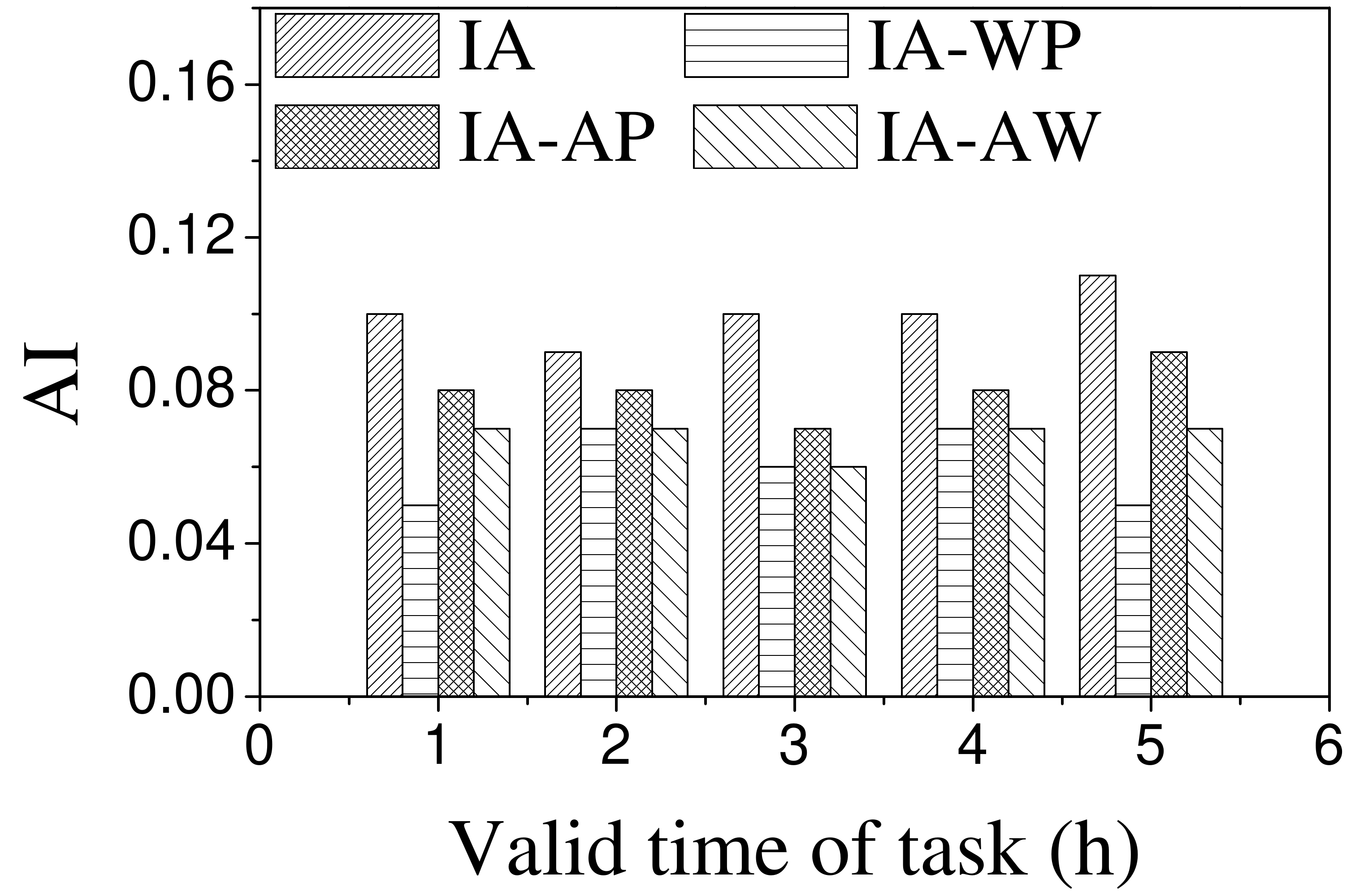}\label{fig:var-s-fs-cpu}}
\vskip -9pt
\caption{Effect of $\varphi$}
\label{fig:vphi}
\end{figure}

\emph{Effect of $\varphi$:} Next, we study the effect of $\varphi$. As can be seen in Figure \ref{fig:vphi}, the AI of all methods changes randomly. This may be due to the fact that the number of the available tasks increases when $\varphi$ grows, which means workers would be assigned tasks with higher AI on any $\varphi$.

\begin{figure}[htbp]
\centering
\subfigure[Average Influence on BK] {\includegraphics[width=0.23\textwidth]{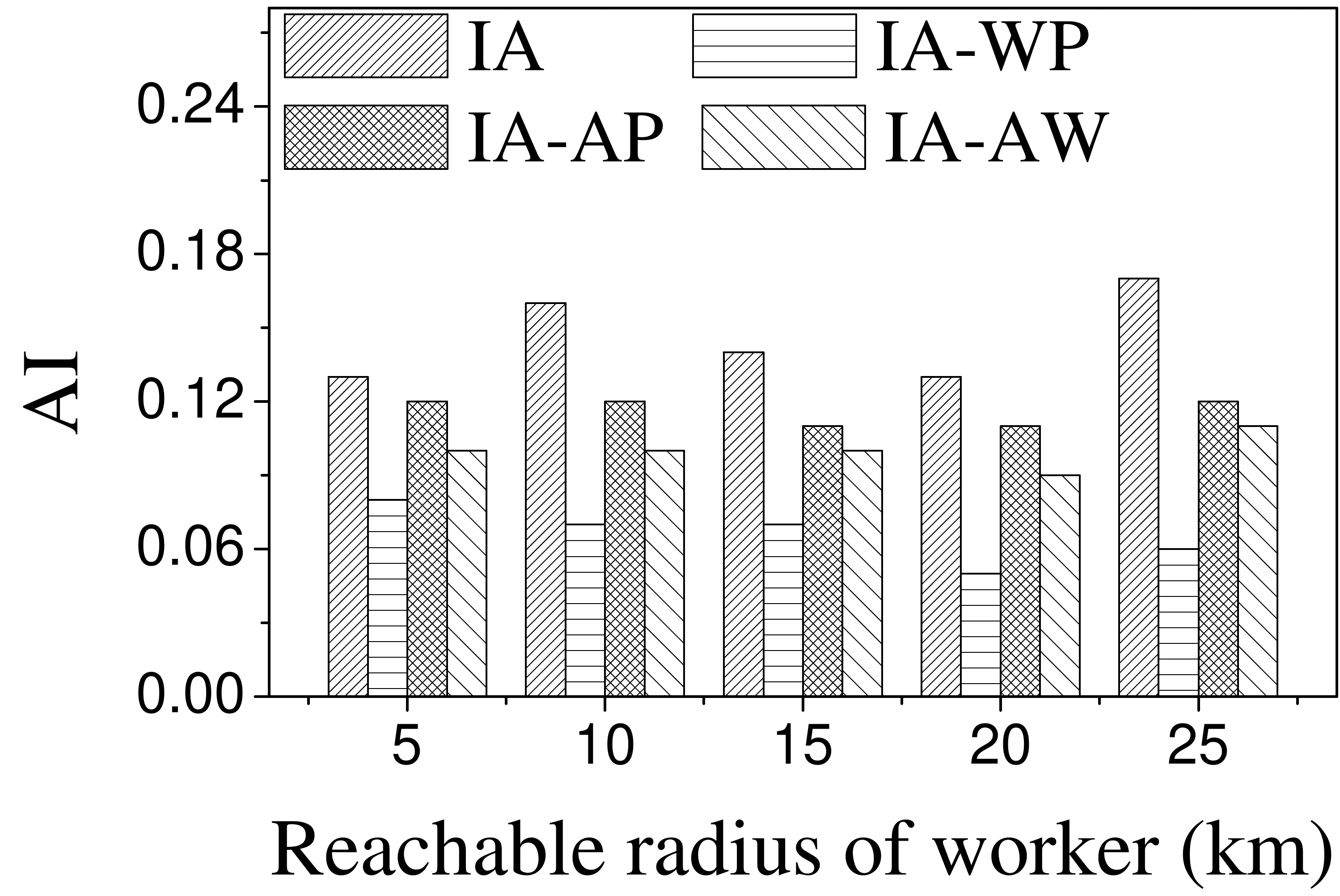}\label{fig:var-s-bk-cpu}}
\subfigure[Average Influence on FS] {\includegraphics[width=0.23\textwidth]{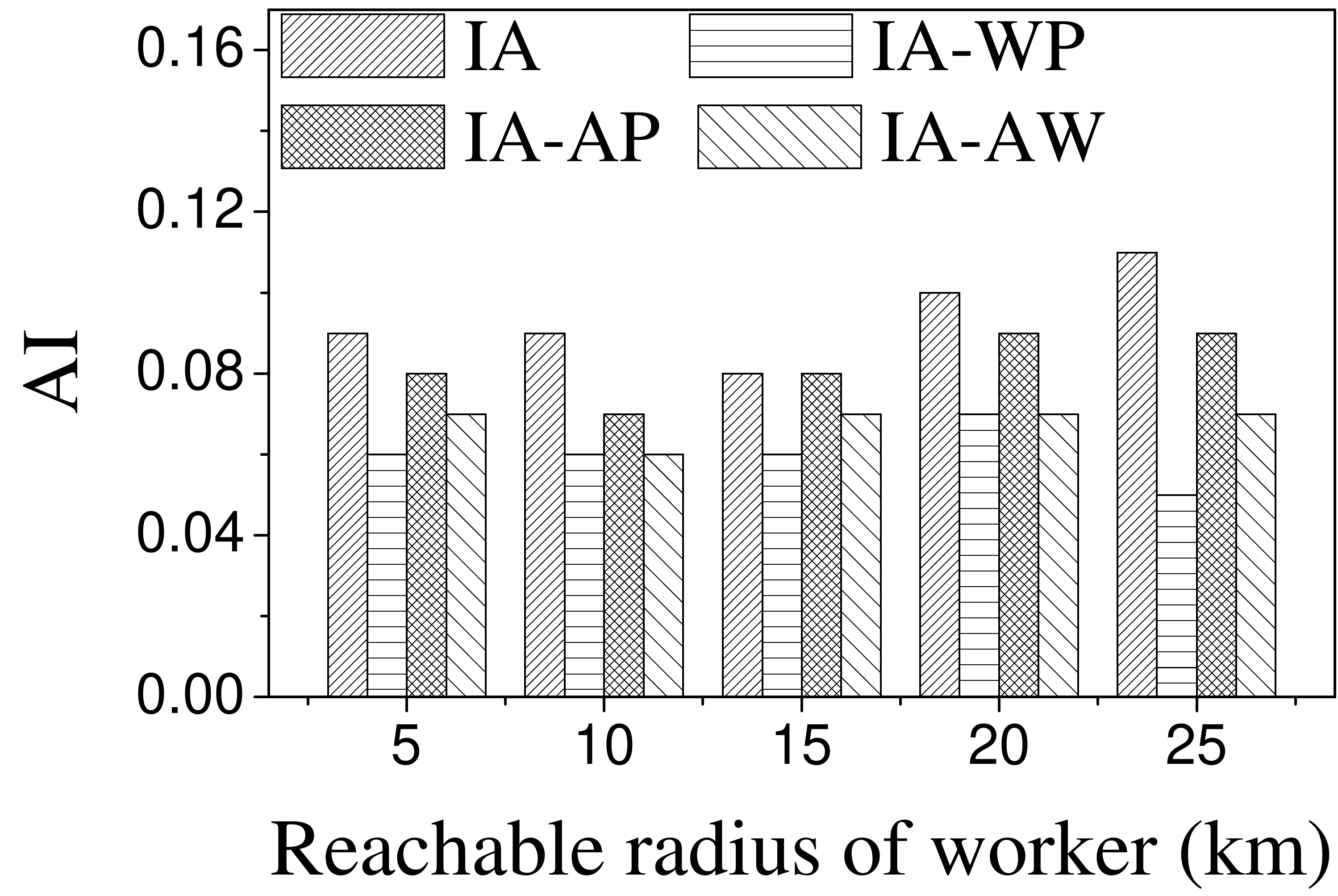}\label{fig:var-s-fs-cpu}}
\vskip -9pt
\caption{Effect of $r$}
\label{fig:vr}
\end{figure}

\emph{Effect of $r$:} As expected, the AI of all approaches changes randomly with the change of $r$ (see Figure \ref{fig:vr}), since the number of available tasks grows when $r$ increases. AI of IA is highest when $r=25$. The reason is that a larger $r$ means that workers have more chances to perform different tasks, which means the assignment is with higher probability to have larger AI.

\subsubsection{Performance of Influence-aware Task Assignment}
Next, we evaluate the different task assignment algorithms. 

\begin{figure*}
\centering
\subfigure[CPU Time] {\includegraphics[width=0.19\textwidth]{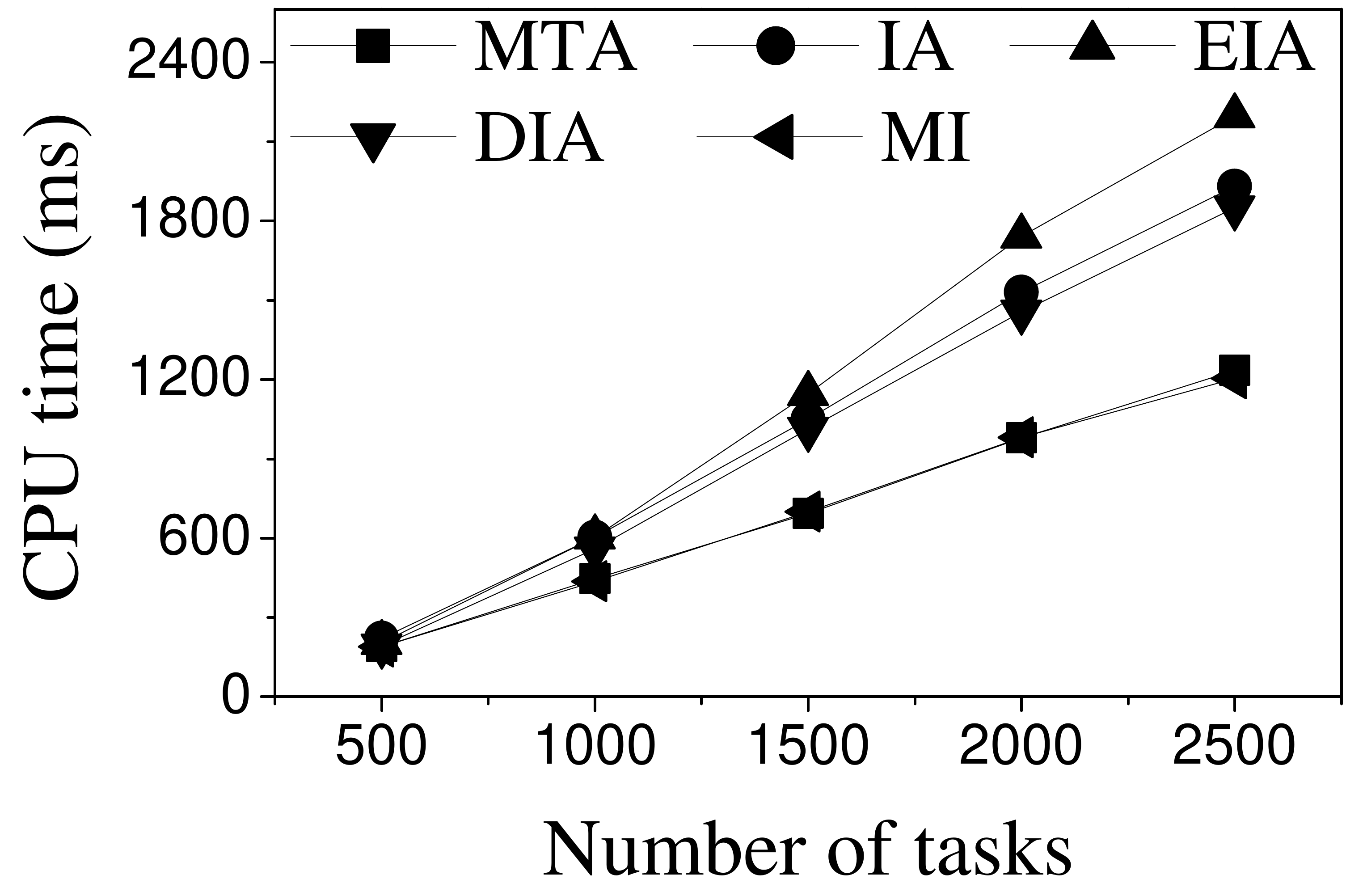}\label{fig:s-bk-cpu}}
\subfigure[Number of Assigned Tasks] {\includegraphics[width=0.19\textwidth]{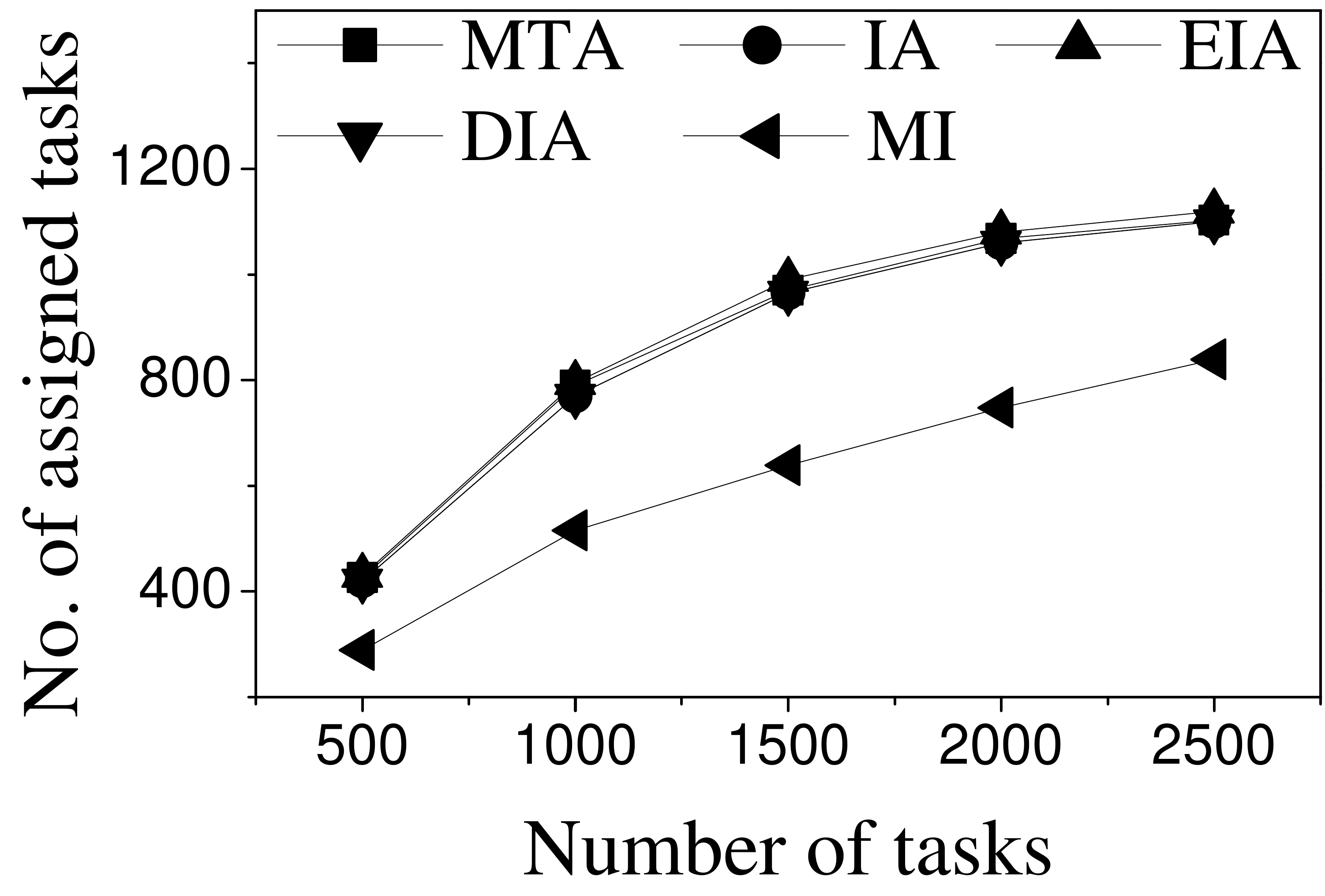}\label{fig:s-bk-s}}
\subfigure[Average Influence] {\includegraphics[width=0.185\textwidth]{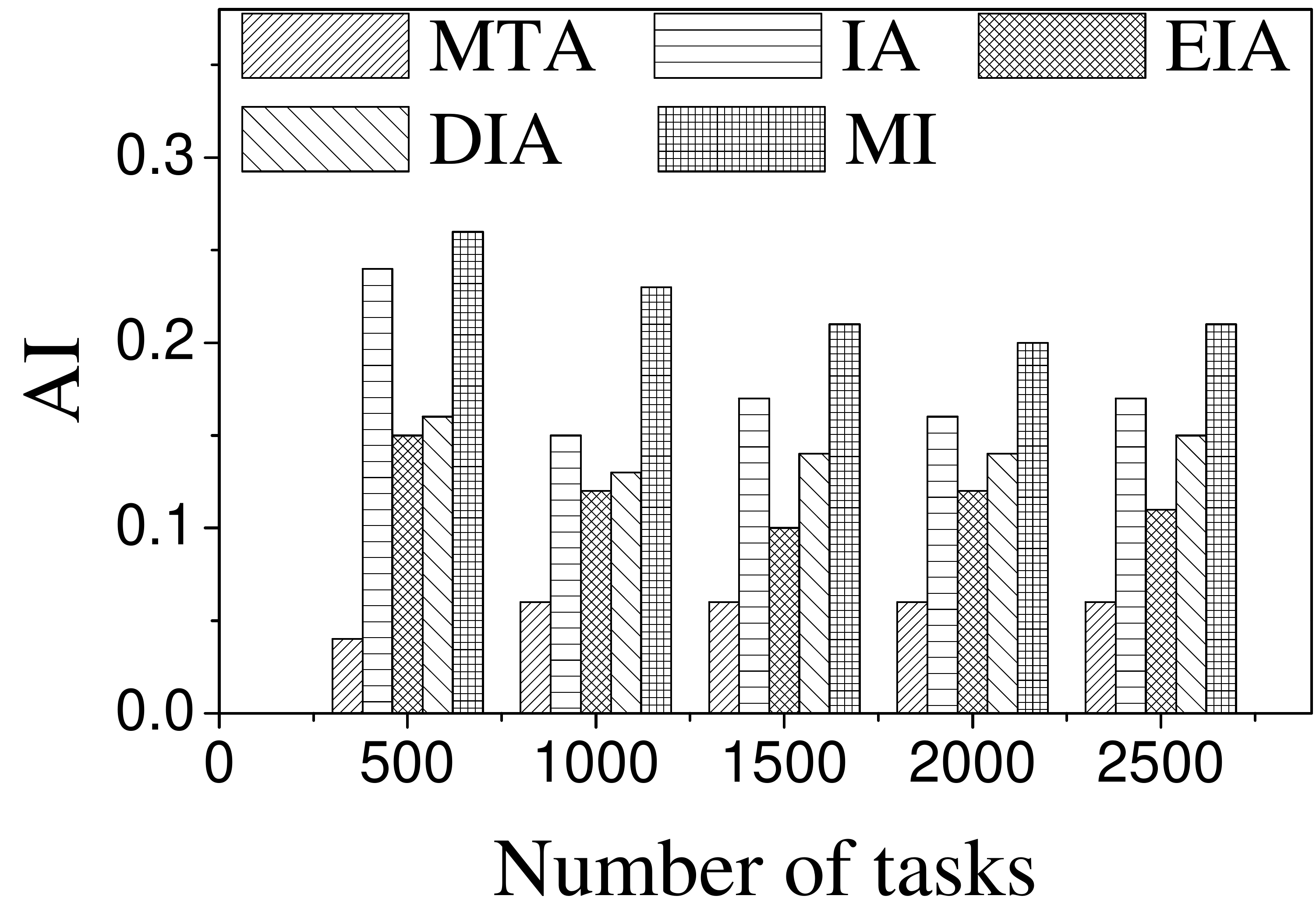}\label{fig:s-bk-ai}}
\subfigure[Average Propagation] {\includegraphics[width=0.184\textwidth]{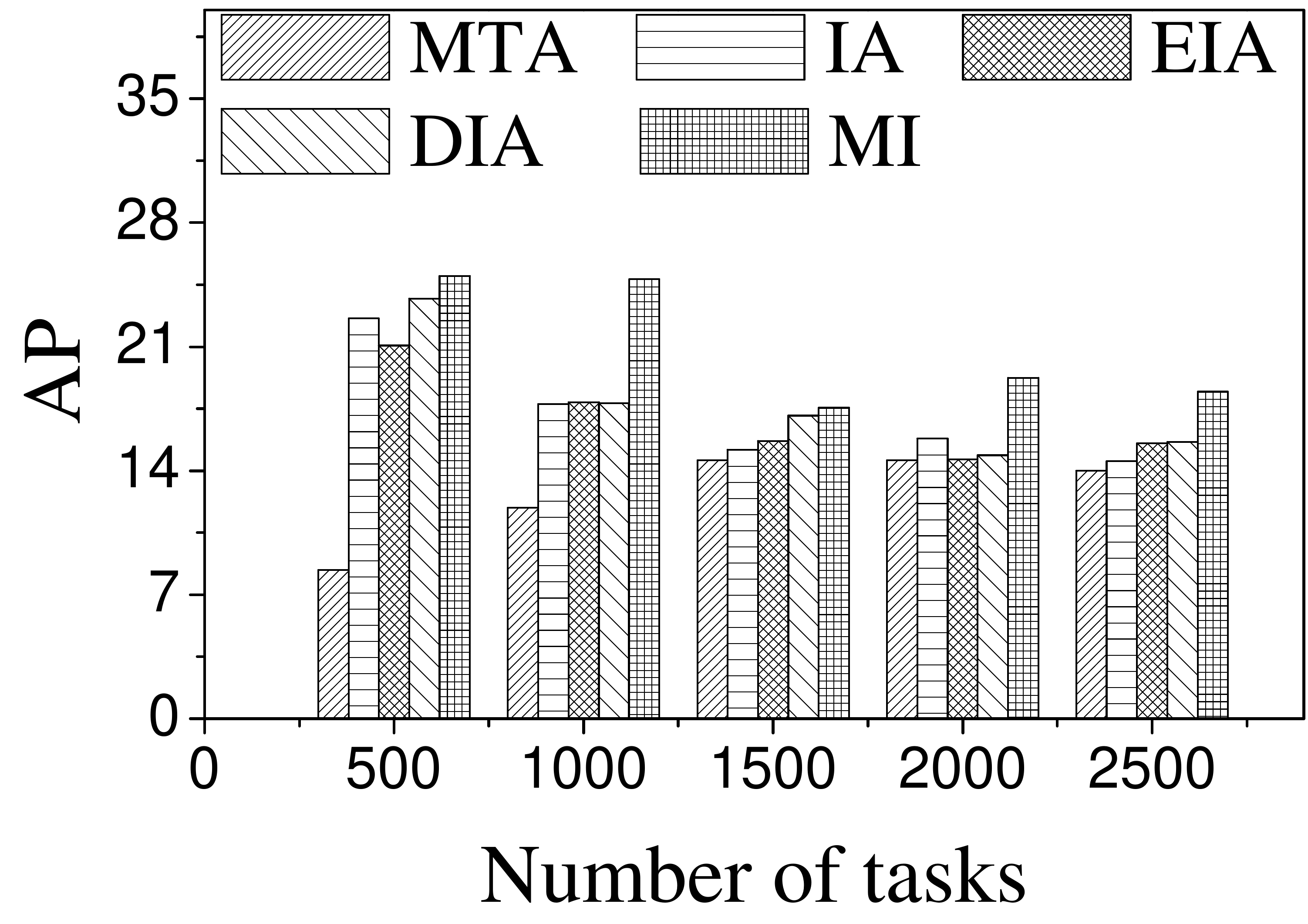}\label{fig:s-bk-ap}}
\subfigure[Travel Cost] {\includegraphics[width=0.185\textwidth]{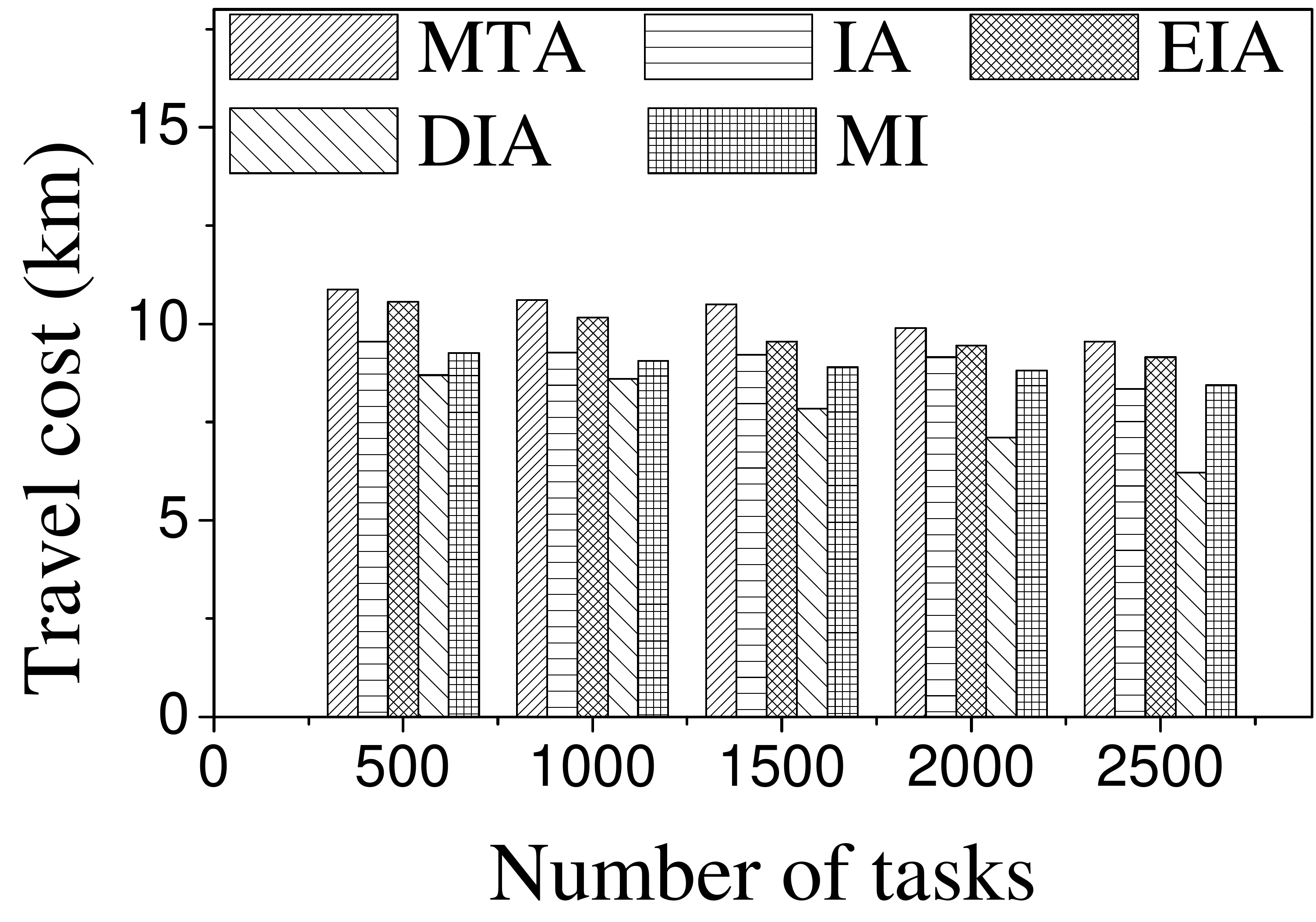}\label{fig:s-bk-d}}
\vskip -9pt
\caption{Effect of $|S|$ on BK}
\label{fig:s-bk}
\end{figure*}

\begin{figure*}
\centering
\subfigure[CPU Time] {\includegraphics[width=0.19\textwidth]{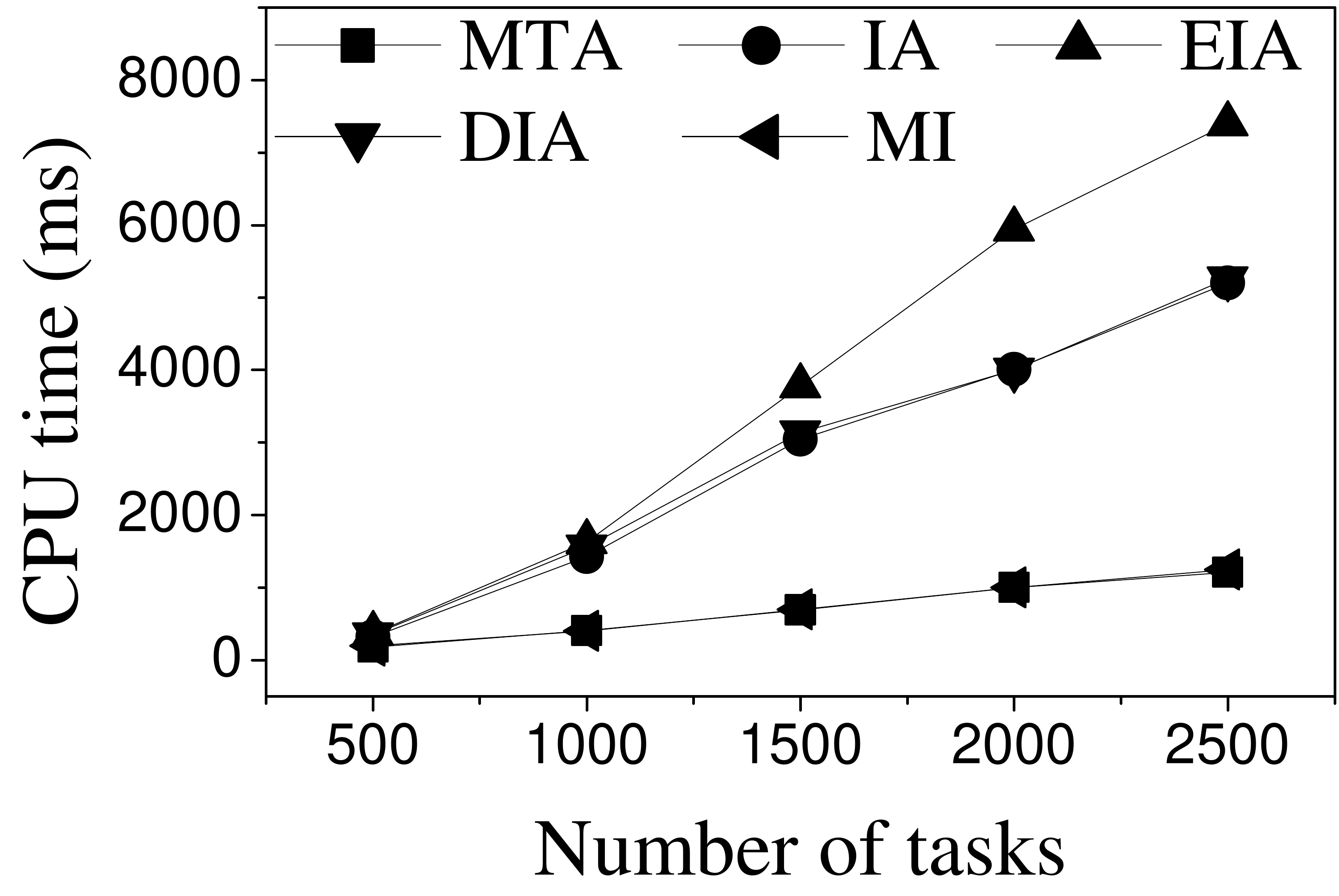}\label{fig:s-fs-cpu}}
\subfigure[Number of Assigned Tasks] {\includegraphics[width=0.19\textwidth]{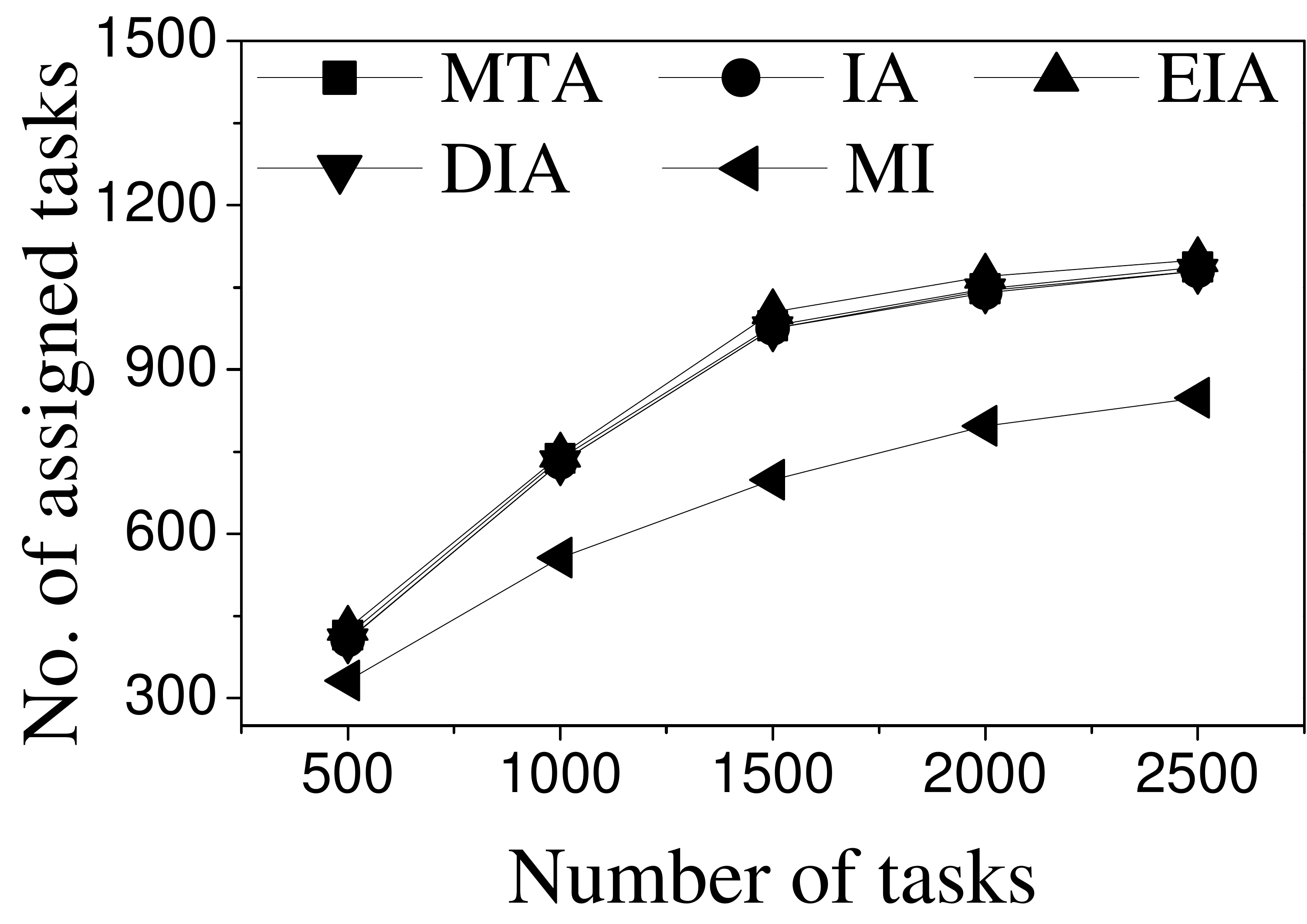}\label{fig:s-fs-s}}
\subfigure[Average Influence] {\includegraphics[width=0.188\textwidth]{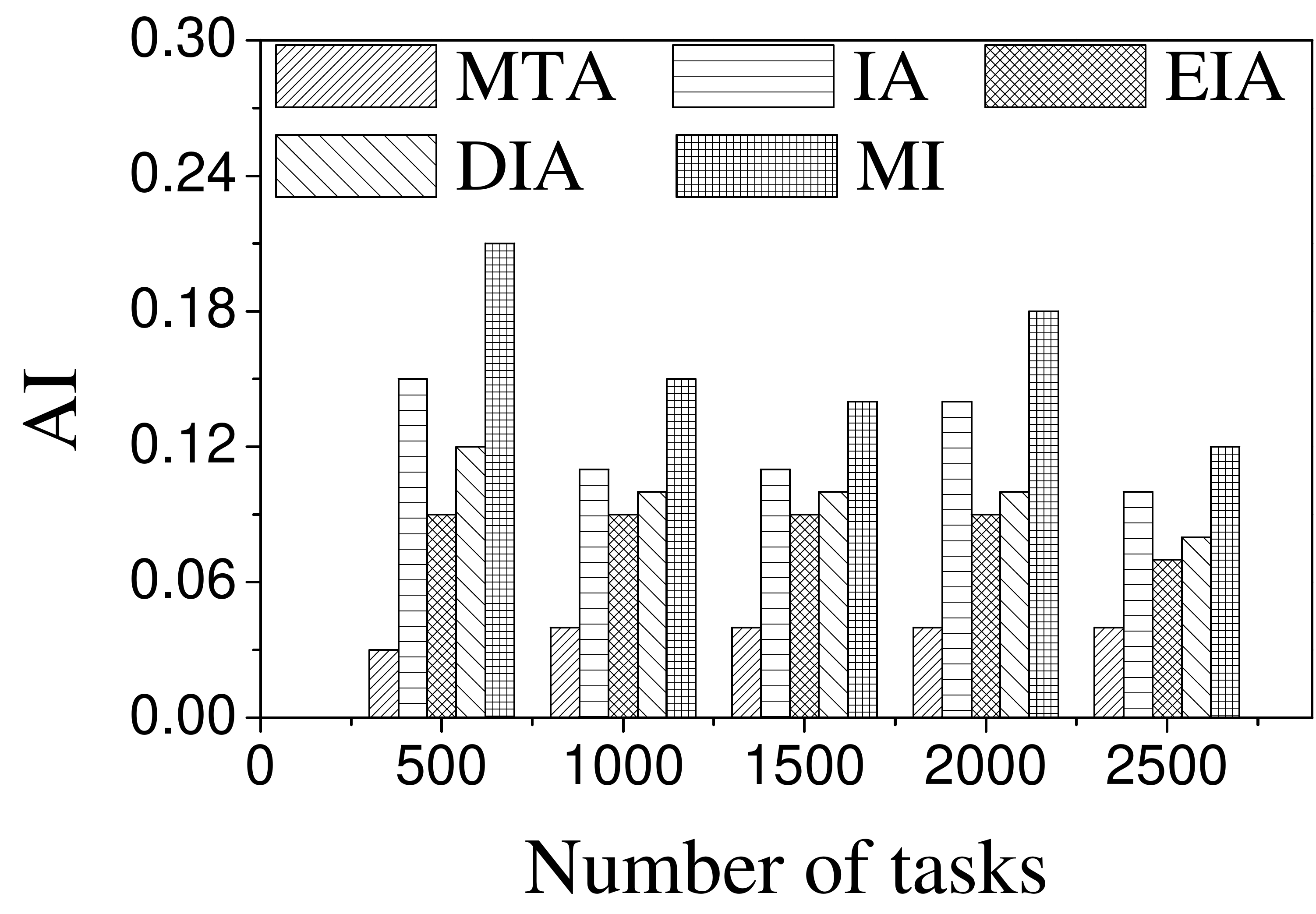}\label{fig:s-fs-ai}}
\subfigure[Average Propagation] {\includegraphics[width=0.183\textwidth]{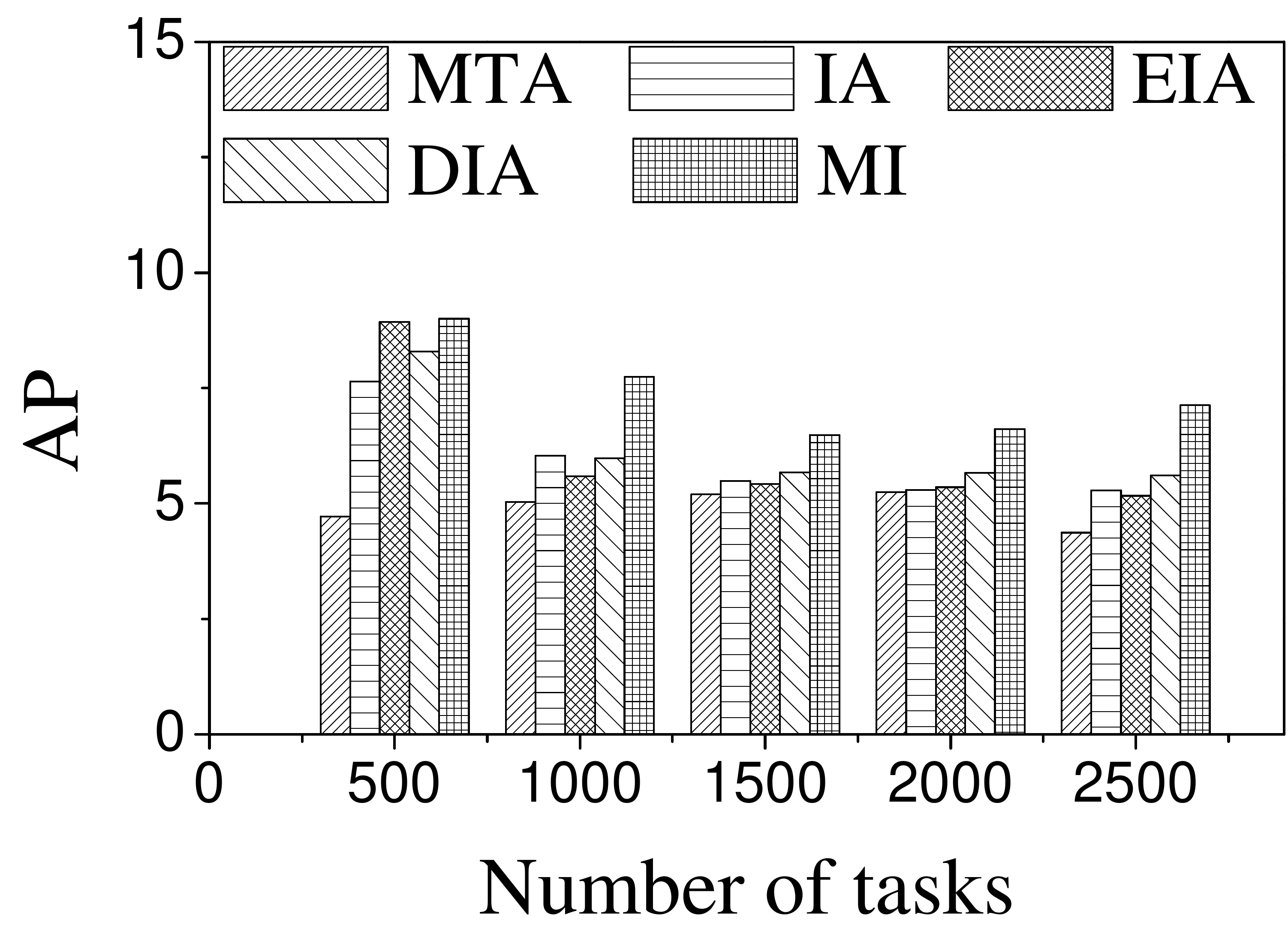}\label{fig:s-fs-ap}}
\subfigure[Travel Cost] {\includegraphics[width=0.183\textwidth]{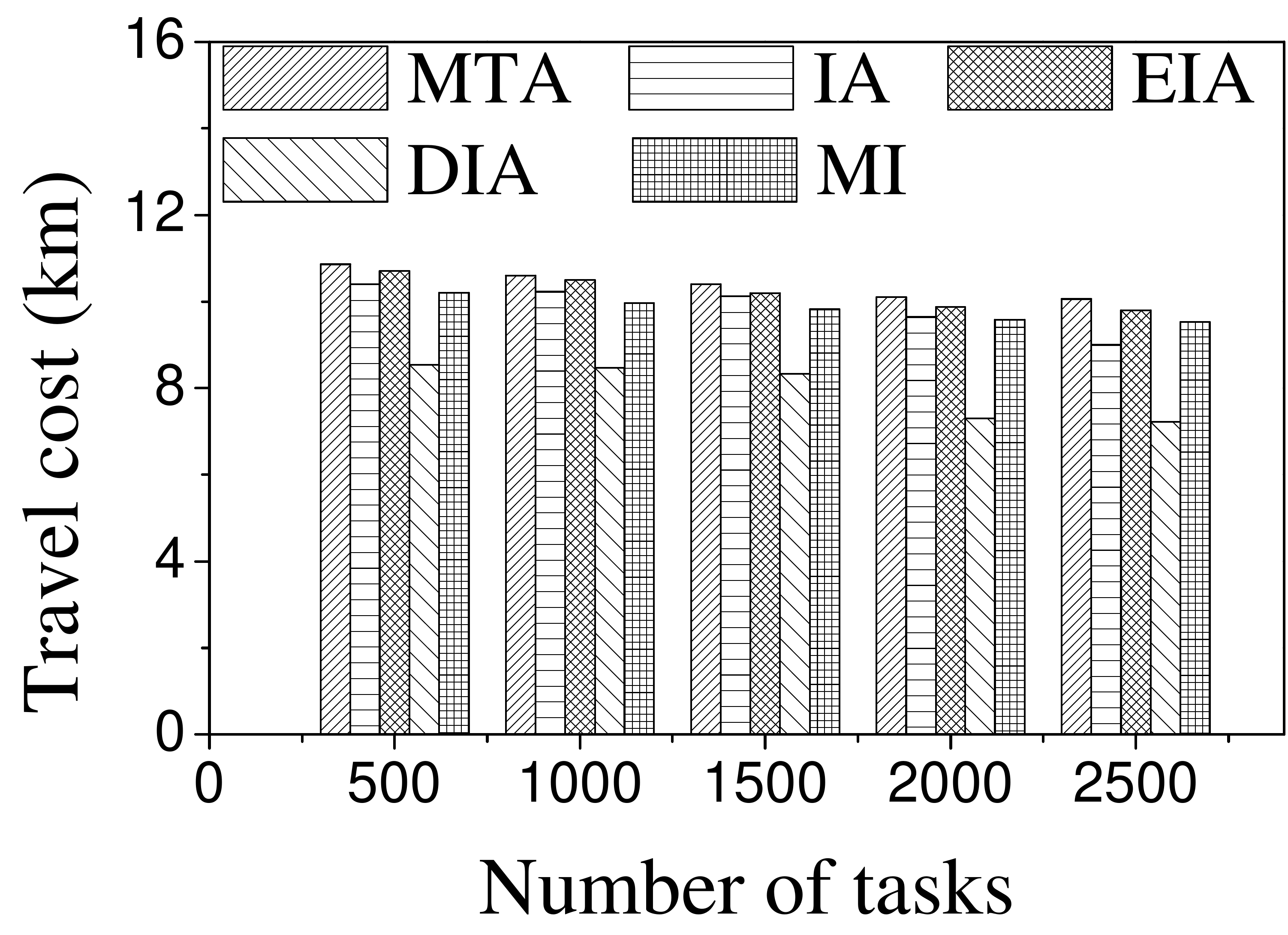}\label{fig:s-fs-d}}
\vskip -9pt
\caption{Effect of $|S|$ on FS}
\label{fig:s-fs}
\end{figure*}

i) MTA: The Maximum Task Assignment algorithm~\cite{kazemi2012geocrowd} that maximizes the number of assigned tasks by computing the maximum flow of the task assignment graph.

ii) IA: Our basic Influence-aware Assignment algorithm.

iii) EIA: Our Entropy-based IA algorithm. 

iv) DIA: Our Distance-based IA algorithm. 

v) MI: The Maximum Influence algorithm that is divided into two steps: 1) Select multiple workers for each task based on the spatio-temporal constraints; 2) assign a task to each worker to maximize the total worker-task influence.

When a worker knows a task, the worker will propagate the information of the task to other workers in the social network. More workers knowing the information of the task leads to larger worker-task influence. Thus we introduce a metric, Average Propagation, $\mathit{AP}$, to evaluate the performance of the task assignment algorithms.
\begin{equation}
\footnotesize
\label{equ:atp}
\mathit{AP}=\frac{\sum_{(s_i,w_i)\in A}\sum_{w_j\in W\setminus \{w_i\}}P_\mathit{pro}(w_i,w_j)}{\mathit{|A|}},
\end{equation}where $W$ is the set of all workers and $P_\mathit{pro}(w_i,w_j)$ is the probability that worker $w_j$ knows task $s_i$ from worker $w_i$.

Four additional metrics are also used to compare the algorithms: 1) CPU time: the CPU time costs for computing a task assignment during a time instance; 2) the total number of assigned tasks; 3) $\mathit{AI}$; and 4)
travel cost: the average travel costs for workers performing tasks.

\textbf{Effect of $|S|$.} We first study the effect of the number of tasks. We generate five datasets containing 500 to 2,500 tasks by random selection from the original dataset. 
As shown in Figures~\ref{fig:s-bk-cpu} and~\ref{fig:s-fs-cpu}, the CPU costs of all methods exhibit a similar increasing trend when $|S|$ grows. The reason is that a larger $|S|$ means that the task assignment graph becomes larger, which results in higher CPU time to compute task assignments. We can see that the CPU time is highest for EIA, followed by IA, DIA, MI and MTA. However, the number of tasks assigned by EIA is larger than those of the others (see Figures~\ref{fig:s-bk-s} and~\ref{fig:s-fs-s}), which demonstrates the superiority of the location entropy strategy. MI has the smallest number of assigned tasks while it has the largest Average Influence, $\mathit{AI}$, (see Figures~\ref{fig:s-bk-ai} and~\ref{fig:s-fs-ai}). This is due to the fact that MI aims to maximize the total worker-task influence and ignores to maximize the total number of assigned tasks, which increases the value of $\mathit{AI}$. The $\mathit{AI}$ of IA is larger than that of DIA, EIA, and MTA. The reason is that EIA and DIA adopt the location entropy and travel cost strategies, respectively, which reduces the effect of worker-task influence. DIA takes into account the travel cost of workers with the result that the worker willingness (see Equation~\ref{equ:offline}) of DIA is larger than that of EIA. Thus, the $\mathit{AI}$ of DIA is larger than that of EIA for all values of $|S|$. As expected, the Average Propagation $\mathit{AP}$ of MI, IA, EIA, and DIA is larger than that of MTA (see Figures~\ref{fig:s-bk-ap} and~\ref{fig:s-fs-ap}). The reason is that worker propagation is considered in MI, IA, EIA, and DIA, while being ignored in MTA. Since workers who can generate larger worker propagation have priority to perform tasks, we see that with the increase of $|S|$, workers with smaller worker propagation have more chances to perform tasks. 
Moreover, DIA yields the smallest average travel costs, as shown in Figures~\ref{fig:s-bk-d} and~\ref{fig:s-fs-d}. This is due to the fact that DIA takes into account the travel cost. Workers who are closer to tasks will be given higher priority to perform them. The average travel costs of all algorithms decrease with the increase of $|S|$, since the probability of assigned tasks located near workers increases.

\begin{figure*}
\centering
\subfigure[CPU Time] {\includegraphics[width=0.19\textwidth]{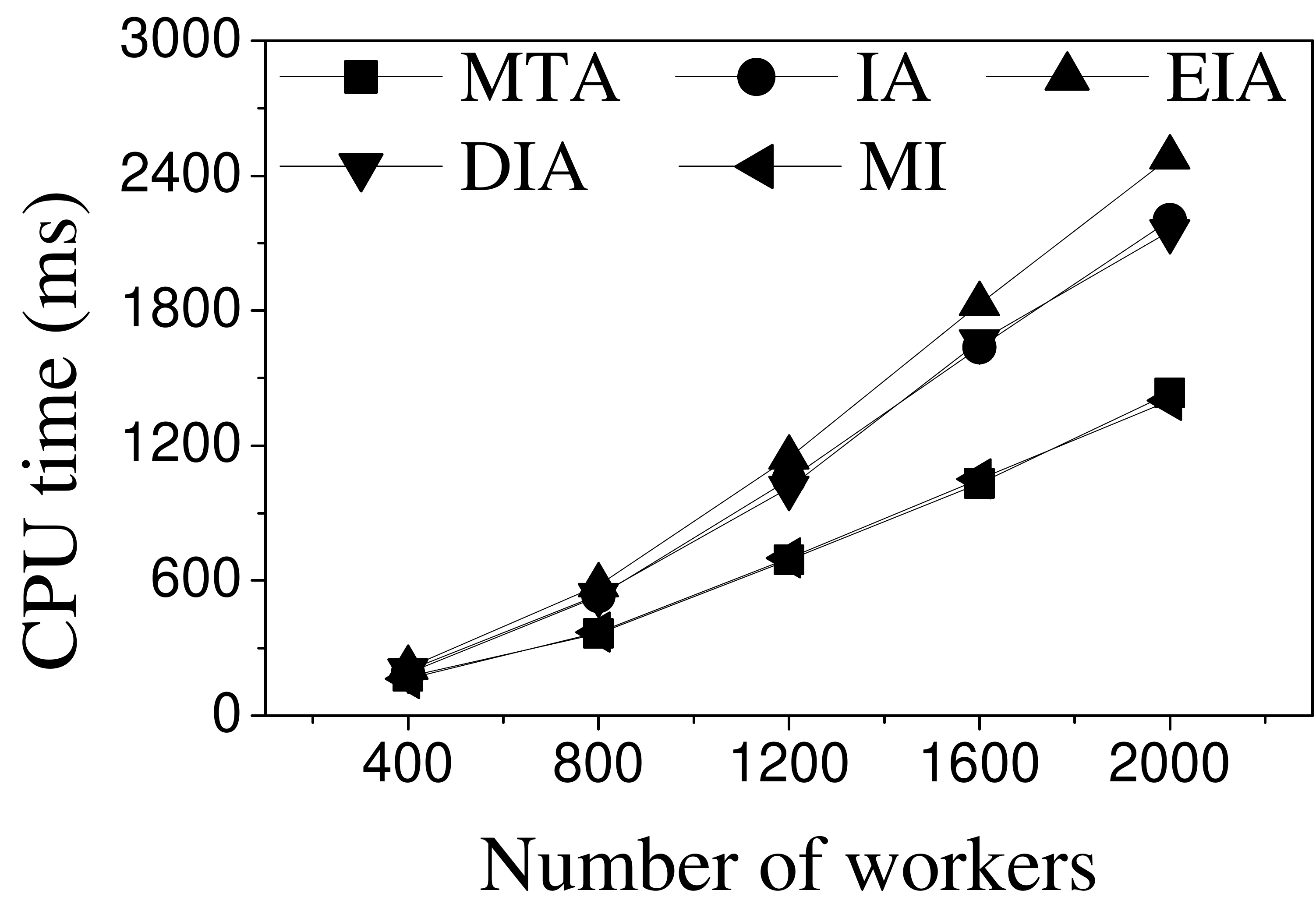}\label{fig:w-bk-cpu}}
\subfigure[Number of Assigned Tasks] {\includegraphics[width=0.19\textwidth]{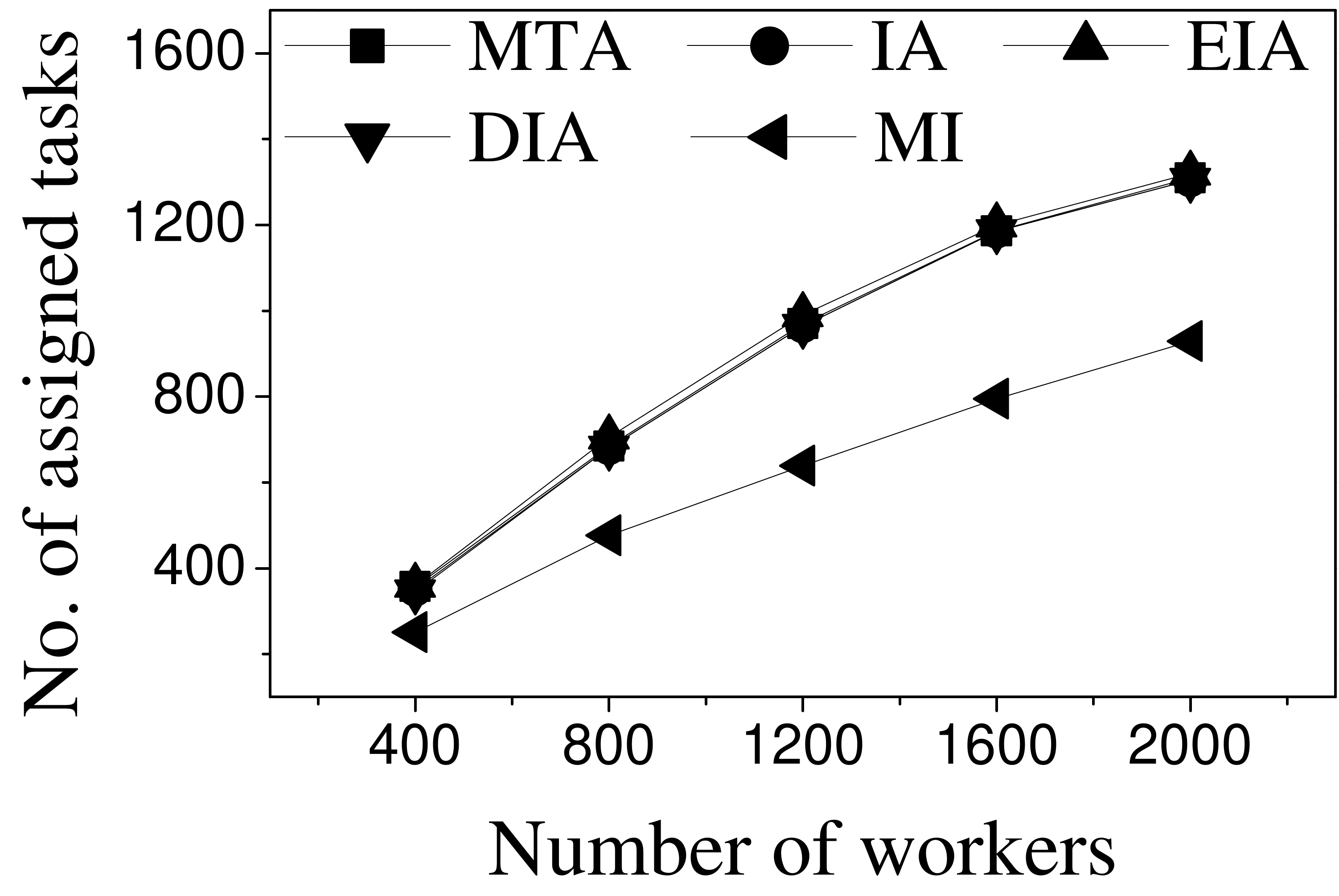}\label{fig:w-bk-s}}
\subfigure[Average Influence] {\includegraphics[width=0.185\textwidth]{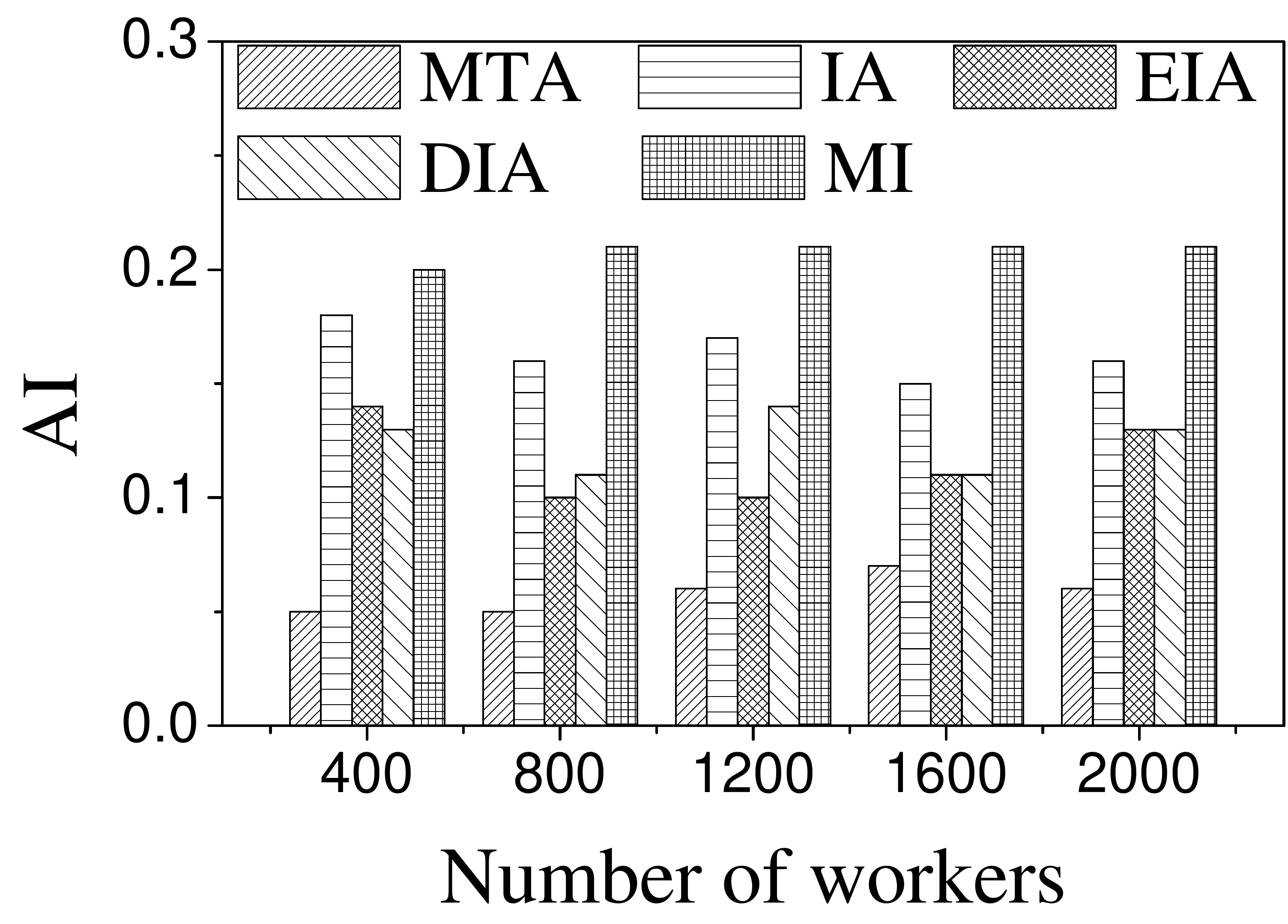}\label{fig:w-bk-ai}}
\subfigure[Average Propagation] {\includegraphics[width=0.185\textwidth]{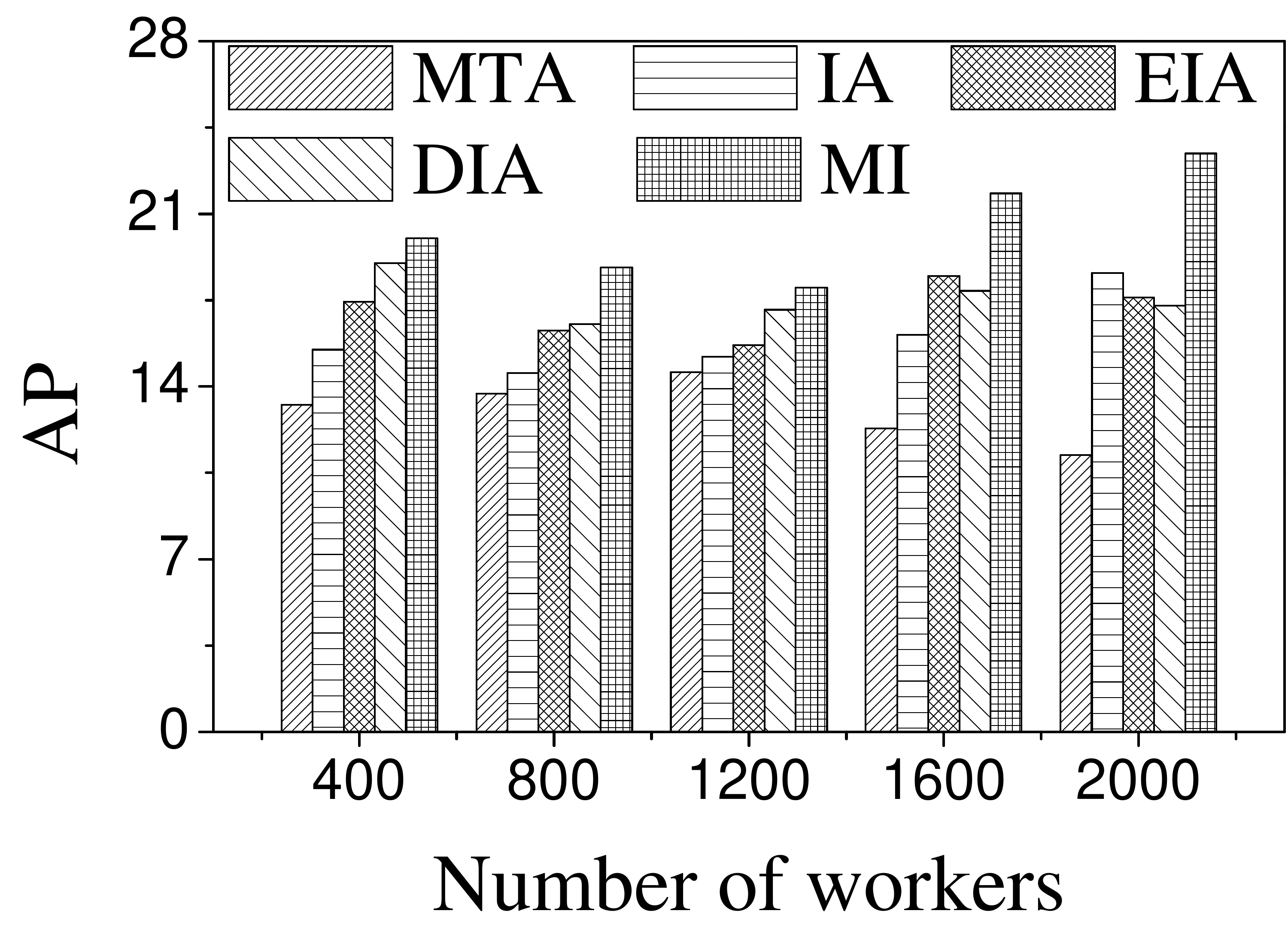}\label{fig:w-bk-ap}}
\subfigure[Travel Cost] {\includegraphics[width=0.185\textwidth]{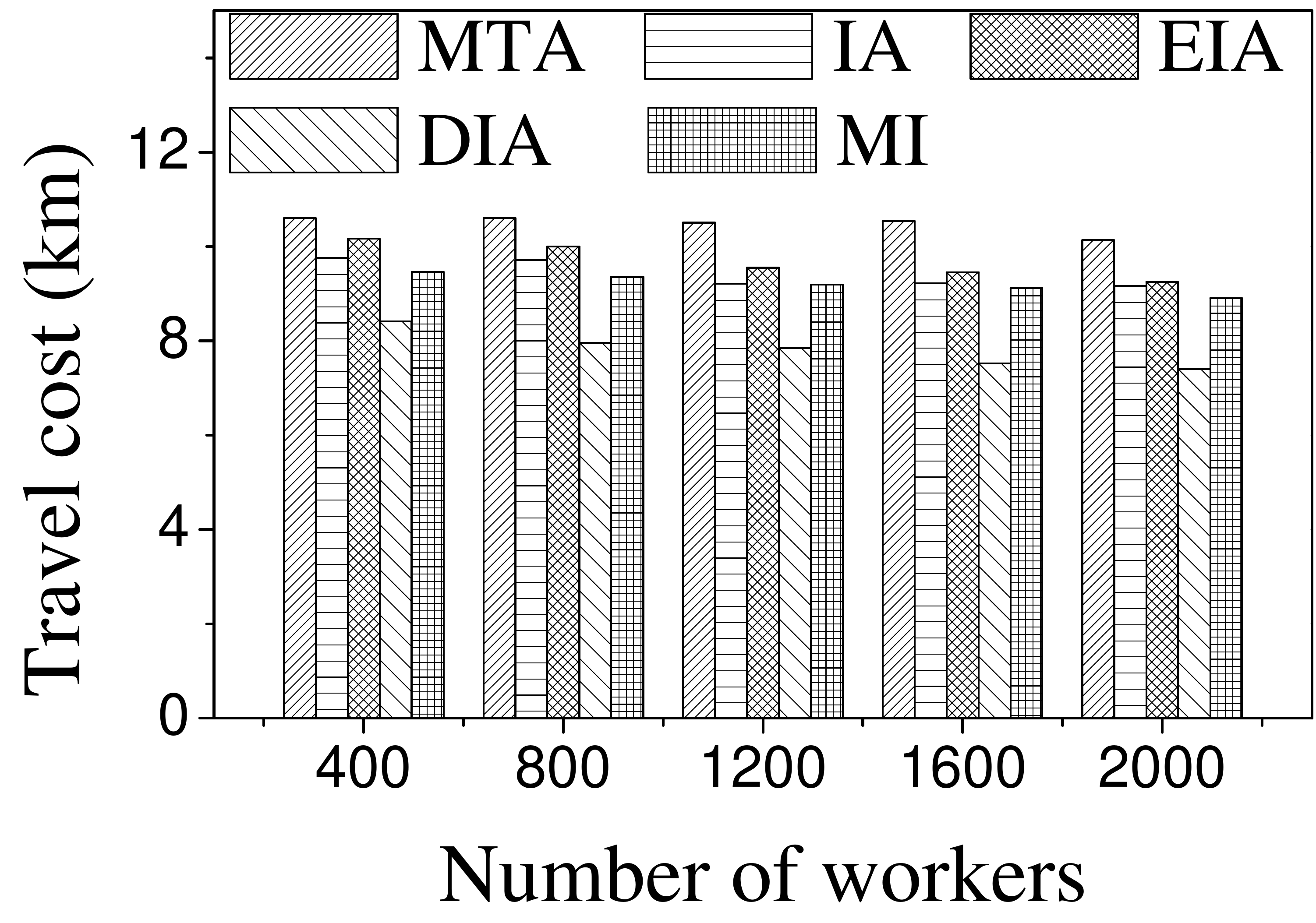}\label{fig:w-bk-d}}
\vskip -9pt
\caption{Effect of $|W|$ on BK}
\label{fig:w-bk}
\end{figure*}

\begin{figure*}
\centering
\subfigure[CPU Time] {\includegraphics[width=0.19\textwidth]{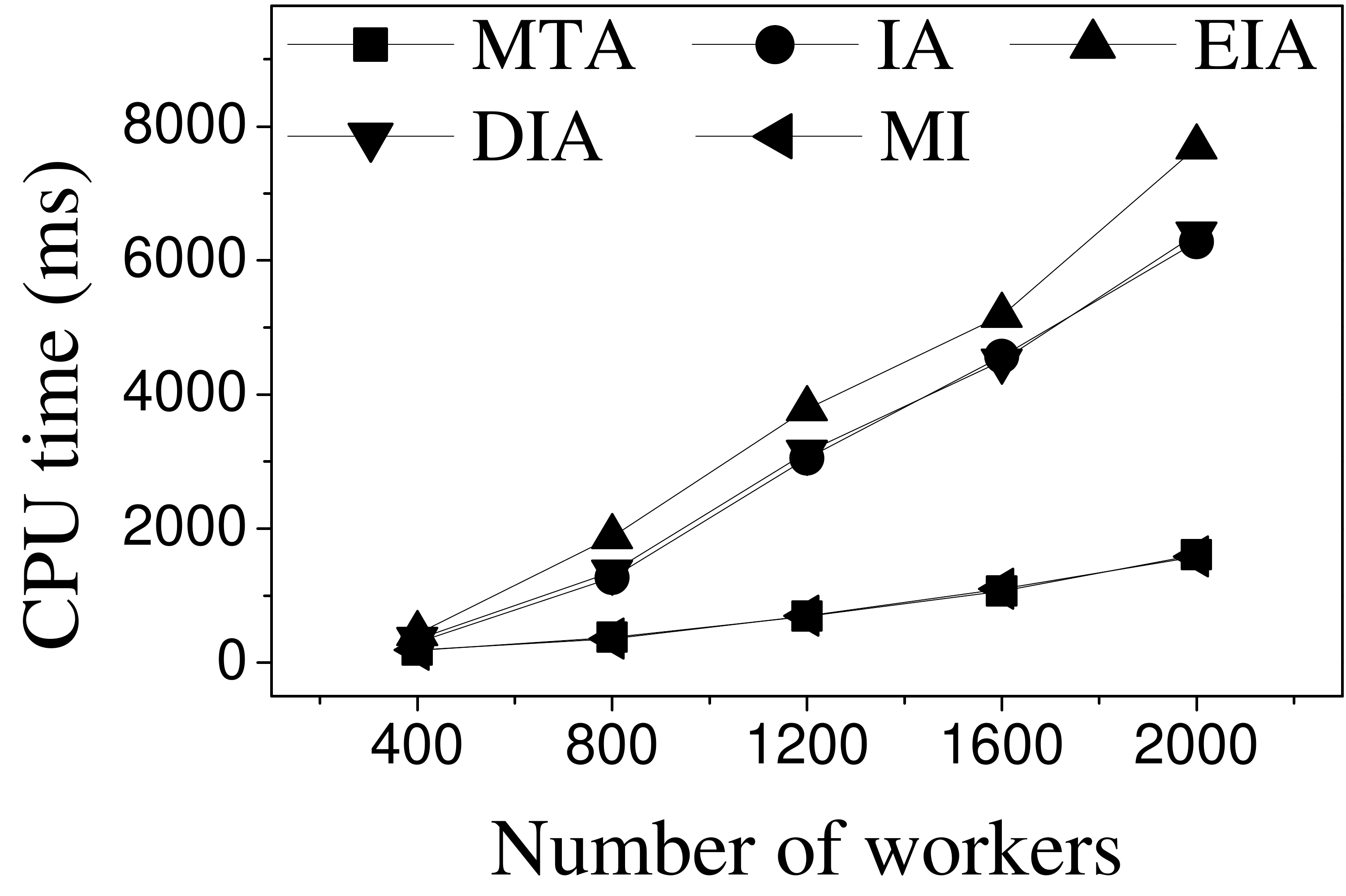}\label{fig:w-fs-cpu}}
\subfigure[Number of Assigned Tasks] {\includegraphics[width=0.19\textwidth]{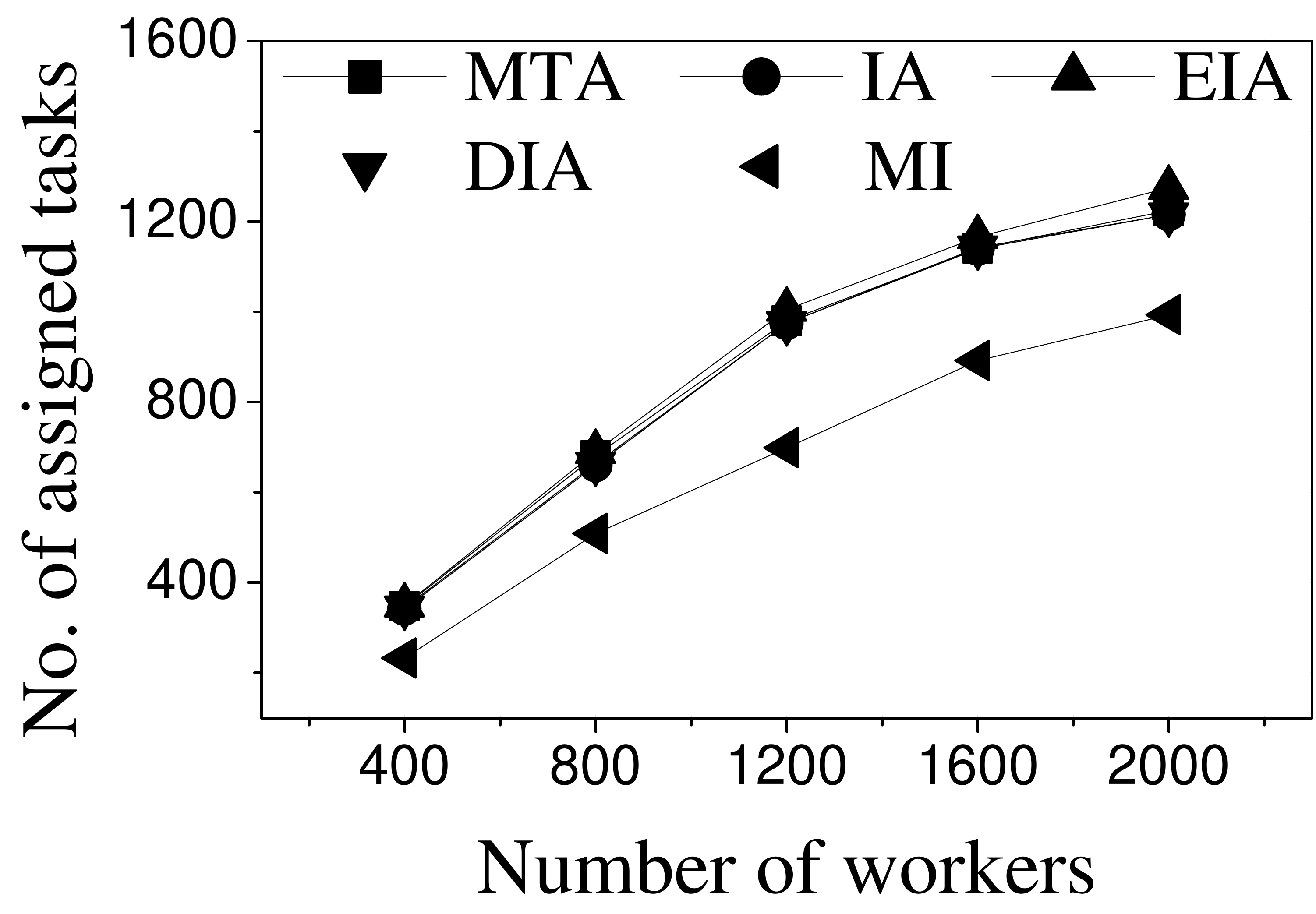}\label{fig:w-fs-s}}
\subfigure[Average Influence] {\includegraphics[width=0.19\textwidth]{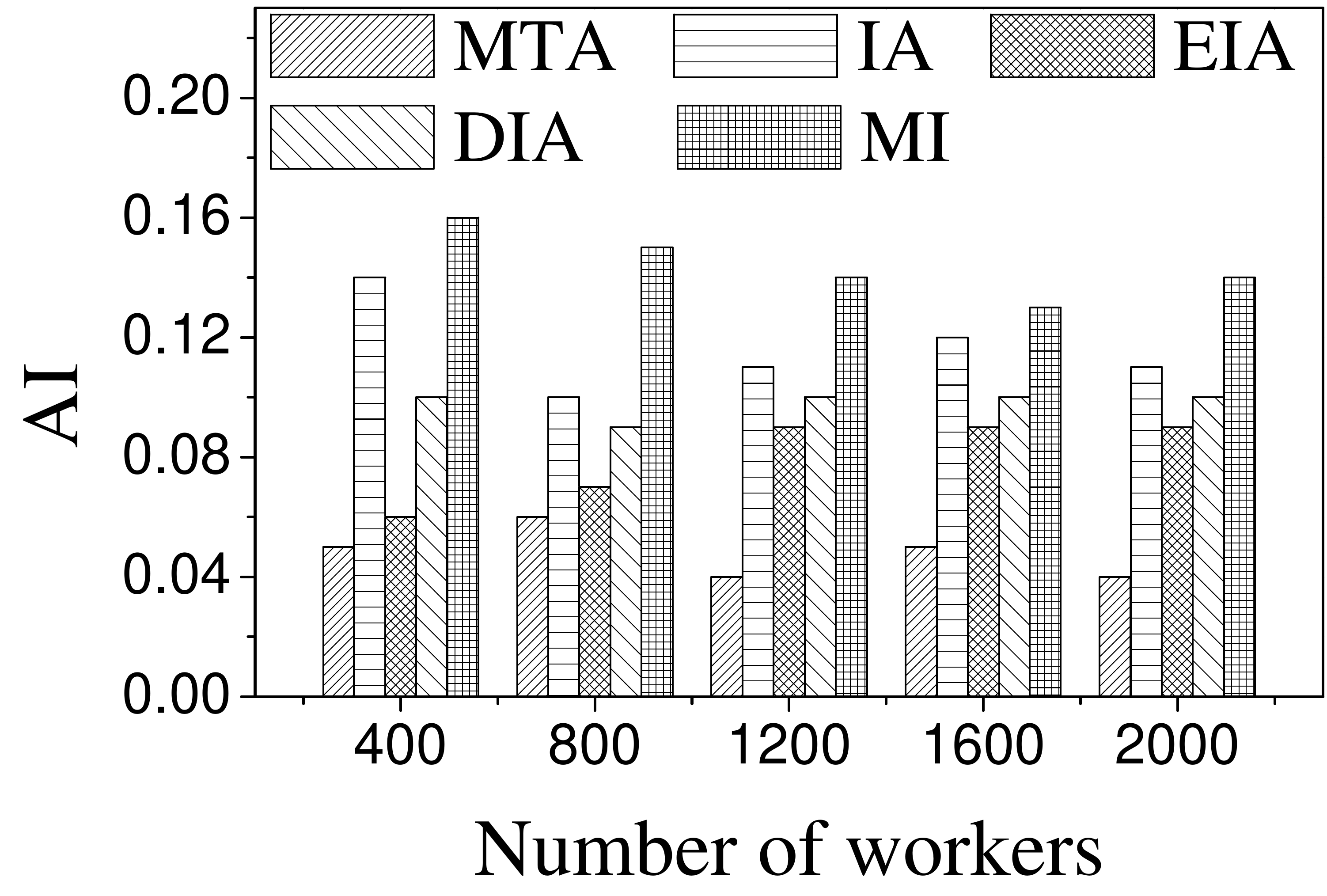}\label{fig:w-fs-ai}}
\subfigure[Average Propagation] {\includegraphics[width=0.18\textwidth]{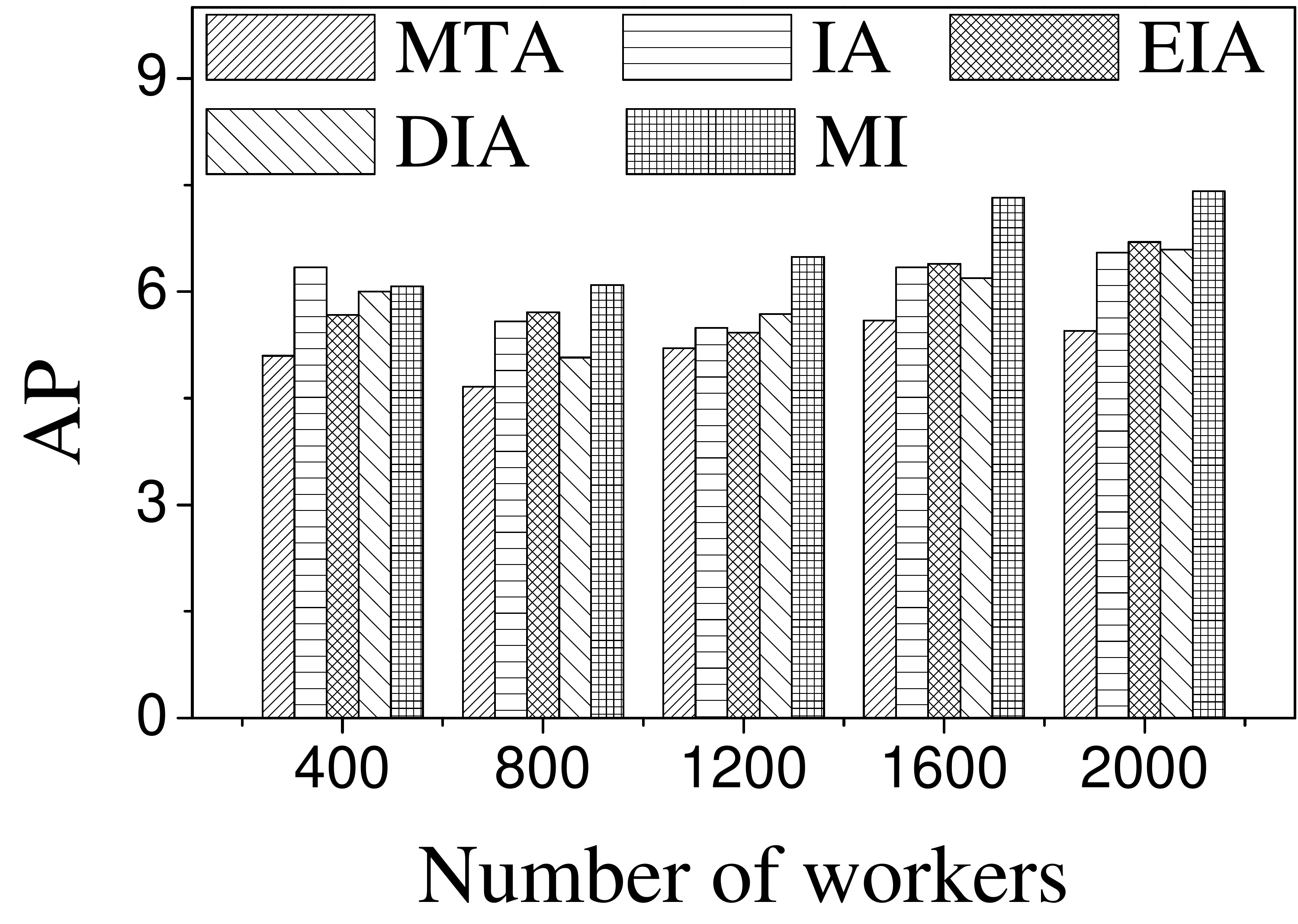}\label{fig:w-fs-ap}}
\subfigure[Travel Cost] {\includegraphics[width=0.183\textwidth]{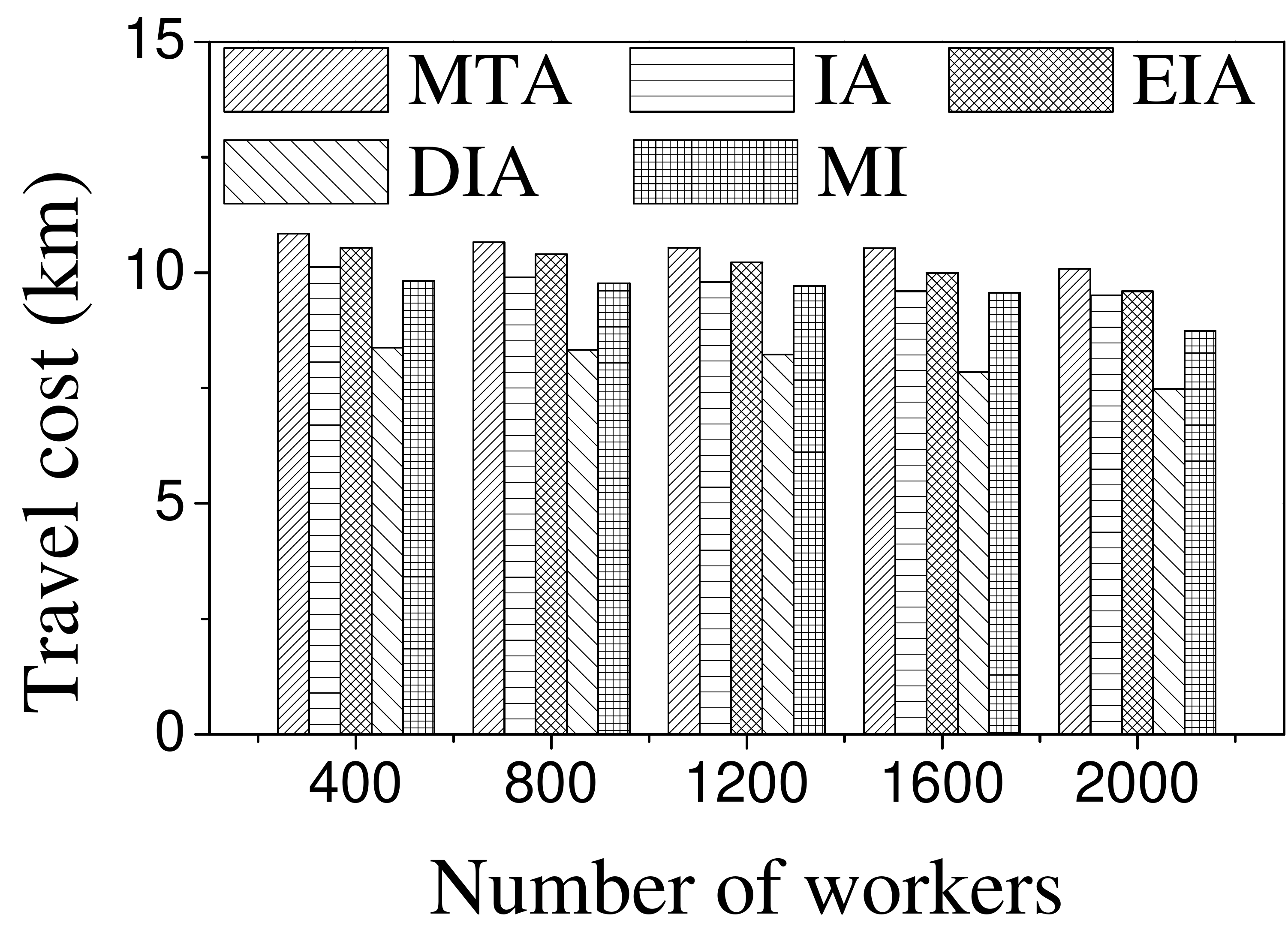}\label{fig:w-fs-d}}
\vskip -9pt
\caption{Effect of $|W|$ on FS}
\label{fig:w-fs}
\end{figure*}

\begin{figure*}
\centering
\subfigure[CPU Time] {\includegraphics[width=0.19\textwidth]{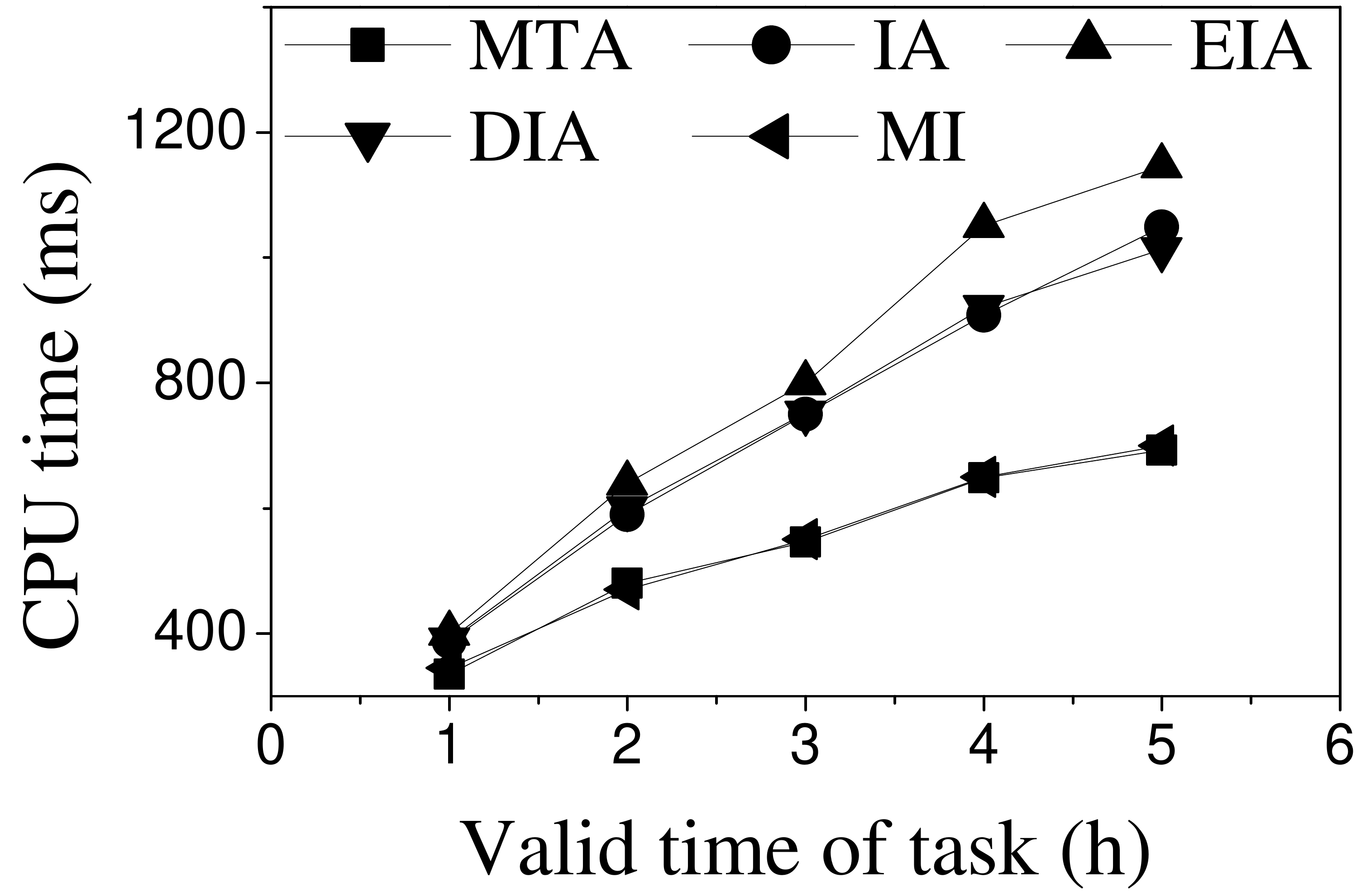}\label{fig:phi-bk-cpu}}
\subfigure[Number of Assigned Tasks] {\includegraphics[width=0.19\textwidth]{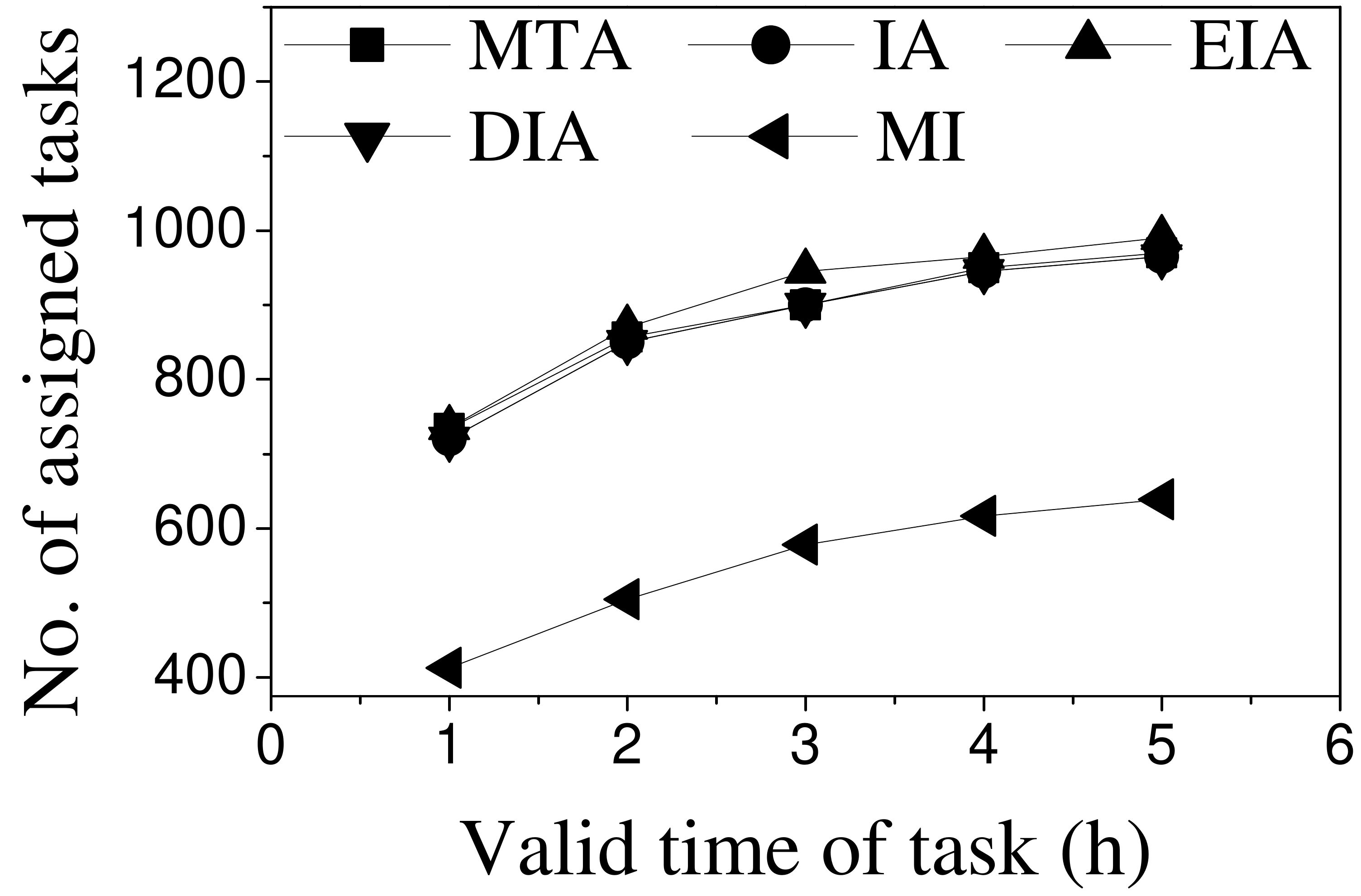}\label{fig:phi-bk-s}}
\subfigure[Average Influence] {\includegraphics[width=0.19\textwidth]{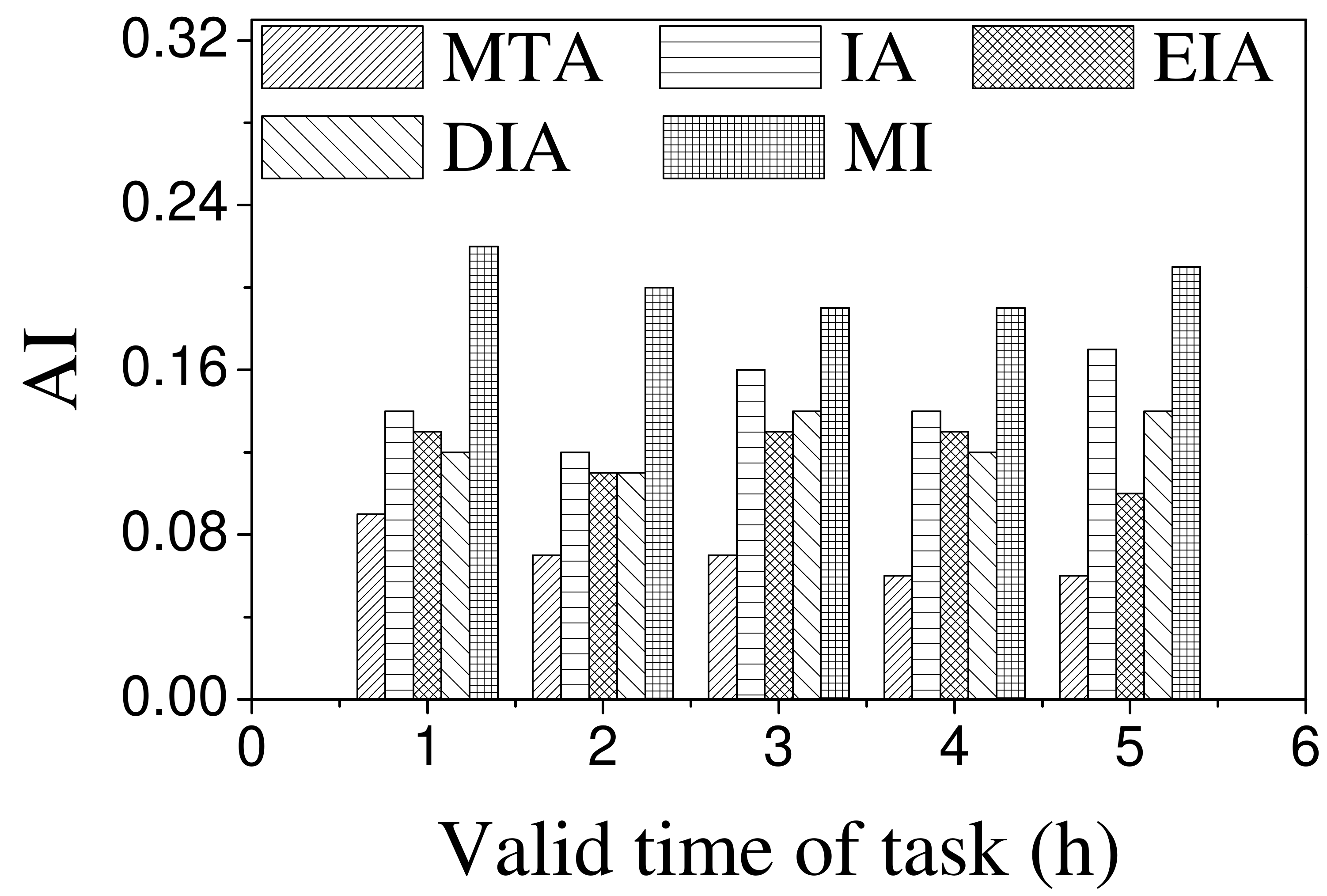}\label{fig:phi-bk-ai}}
\subfigure[Average Propagation] {\includegraphics[width=0.185\textwidth]{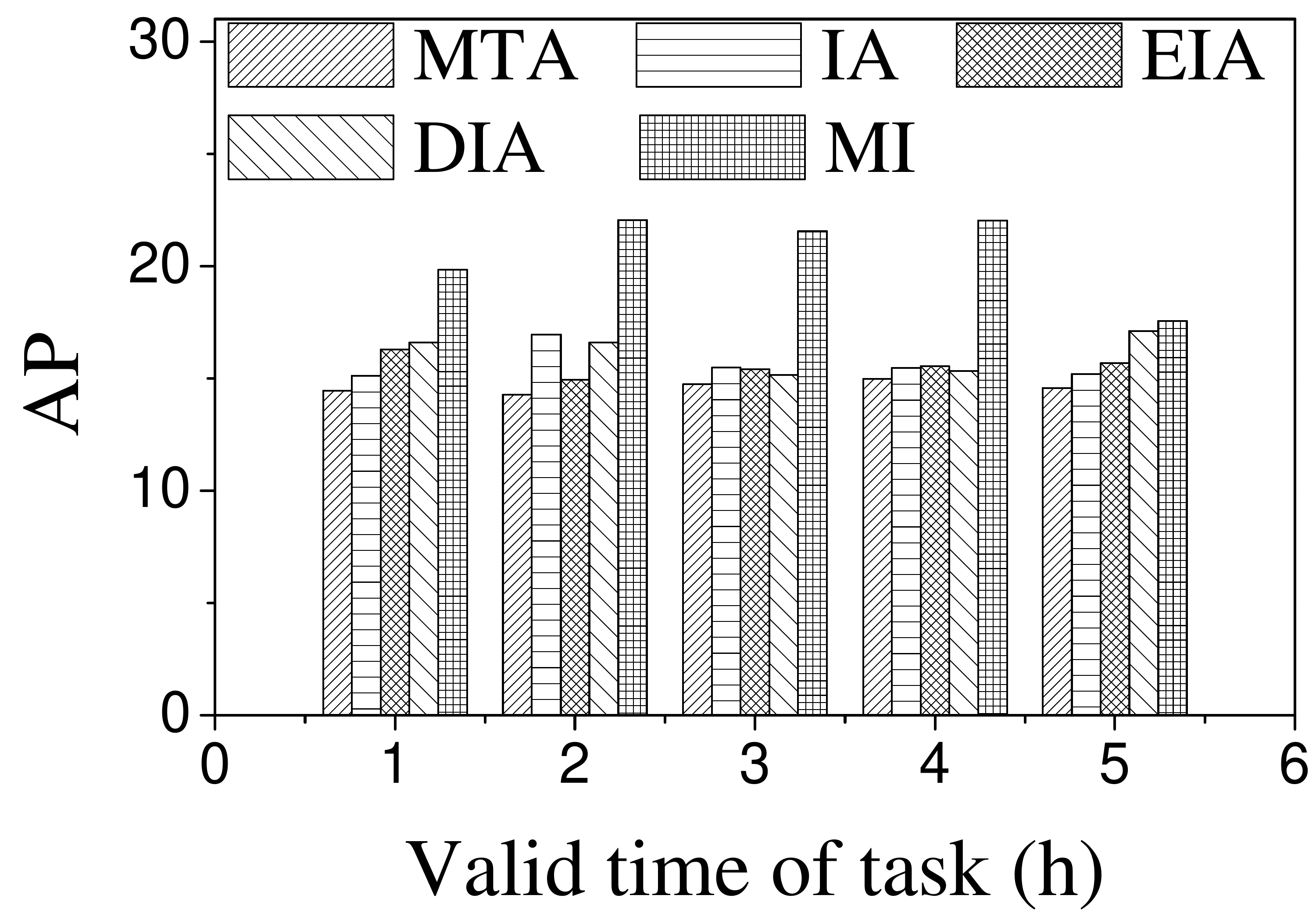}\label{fig:phi-bk-ap}}
\subfigure[Travel Cost] {\includegraphics[width=0.185\textwidth]{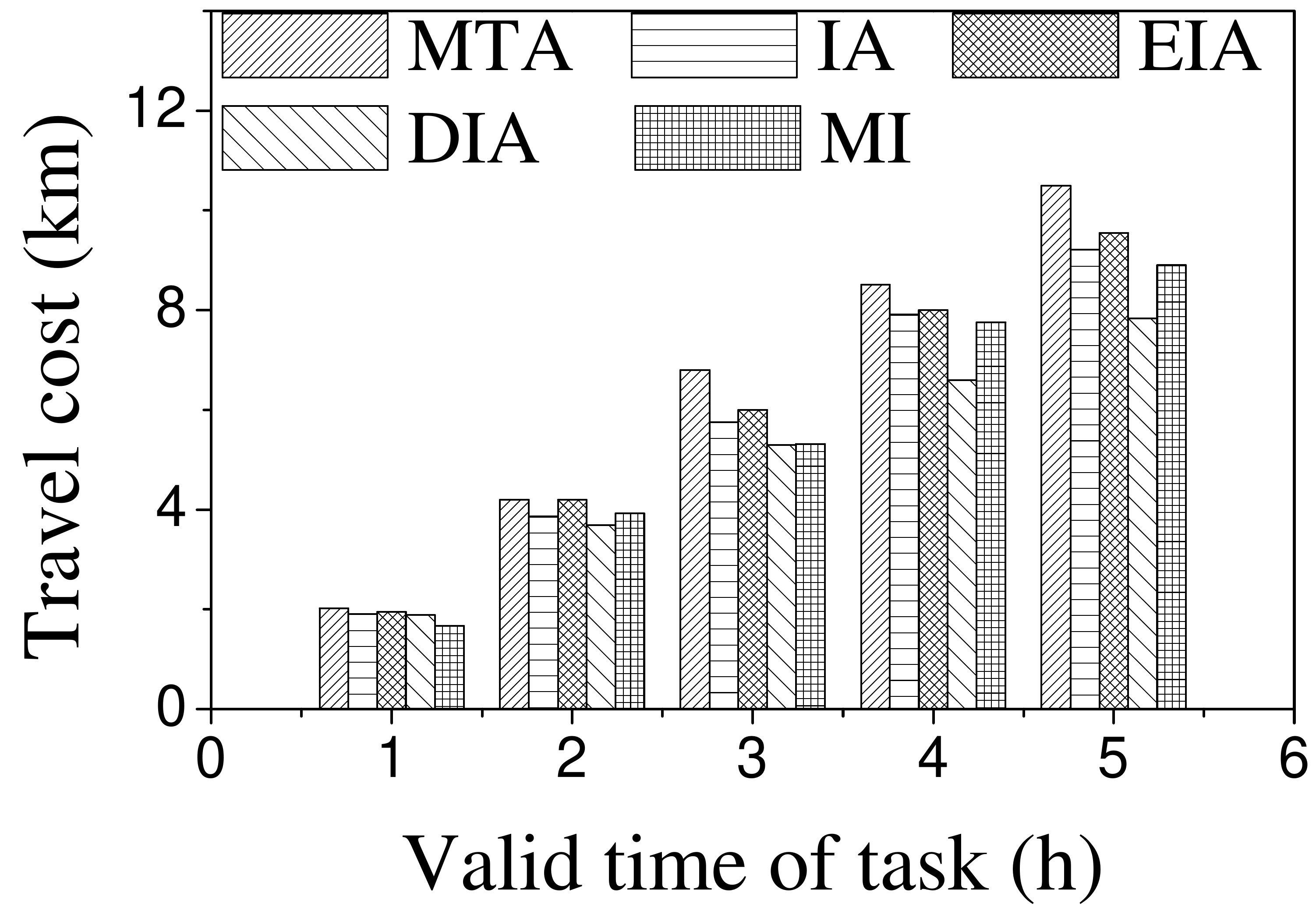}\label{fig:phi-bk-d}}
\vskip -9pt
\caption{Effect of $\varphi$ on BK}
\label{fig:phi-bk}
\end{figure*}

\begin{figure*}
\centering
\subfigure[CPU Time] {\includegraphics[width=0.19\textwidth]{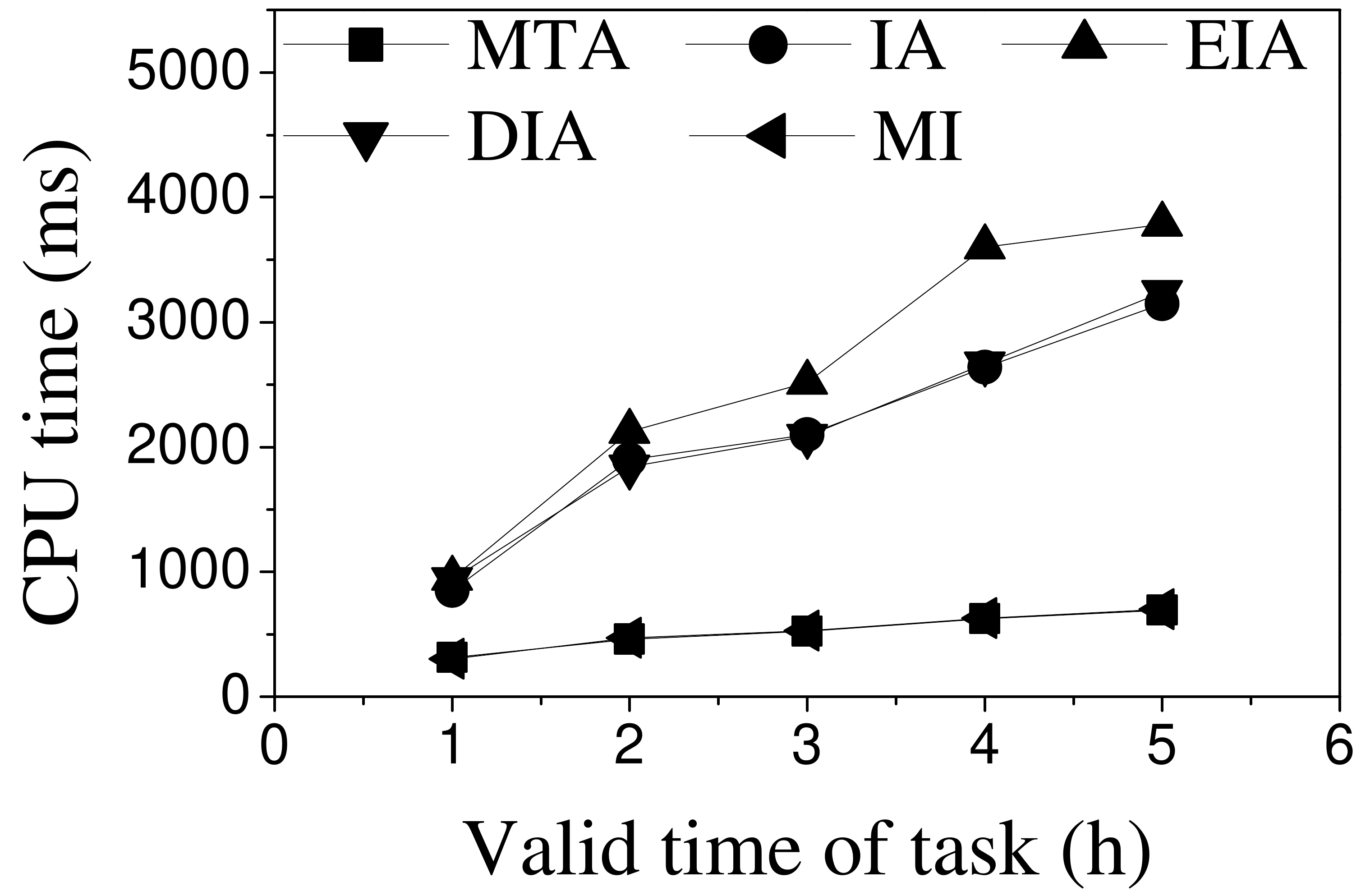}\label{fig:phi-fs-cpu}}
\subfigure[Number of Assigned Tasks] {\includegraphics[width=0.19\textwidth]{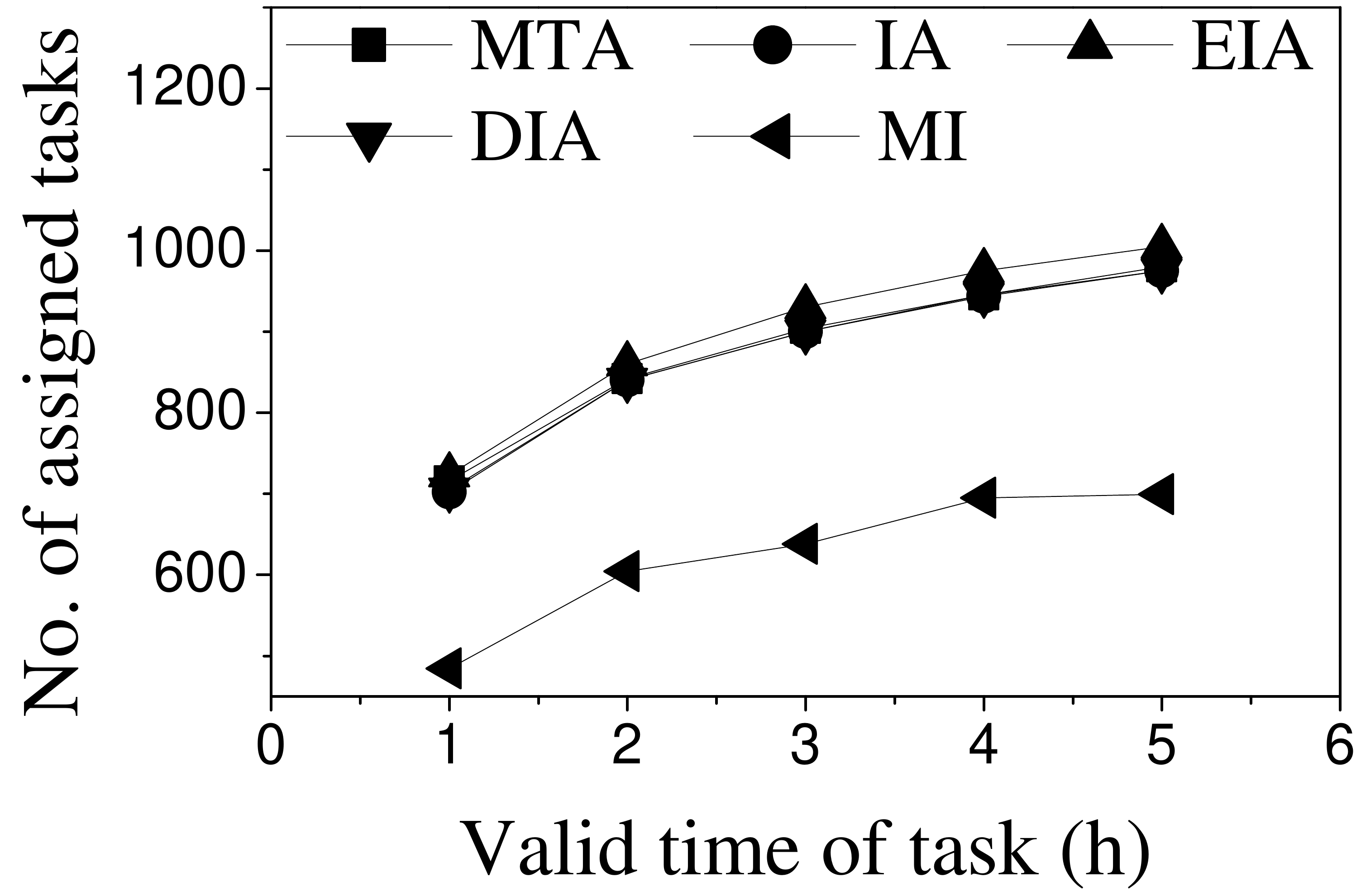}\label{fig:phi-fs-s}}
\subfigure[Average Influence] {\includegraphics[width=0.19\textwidth]{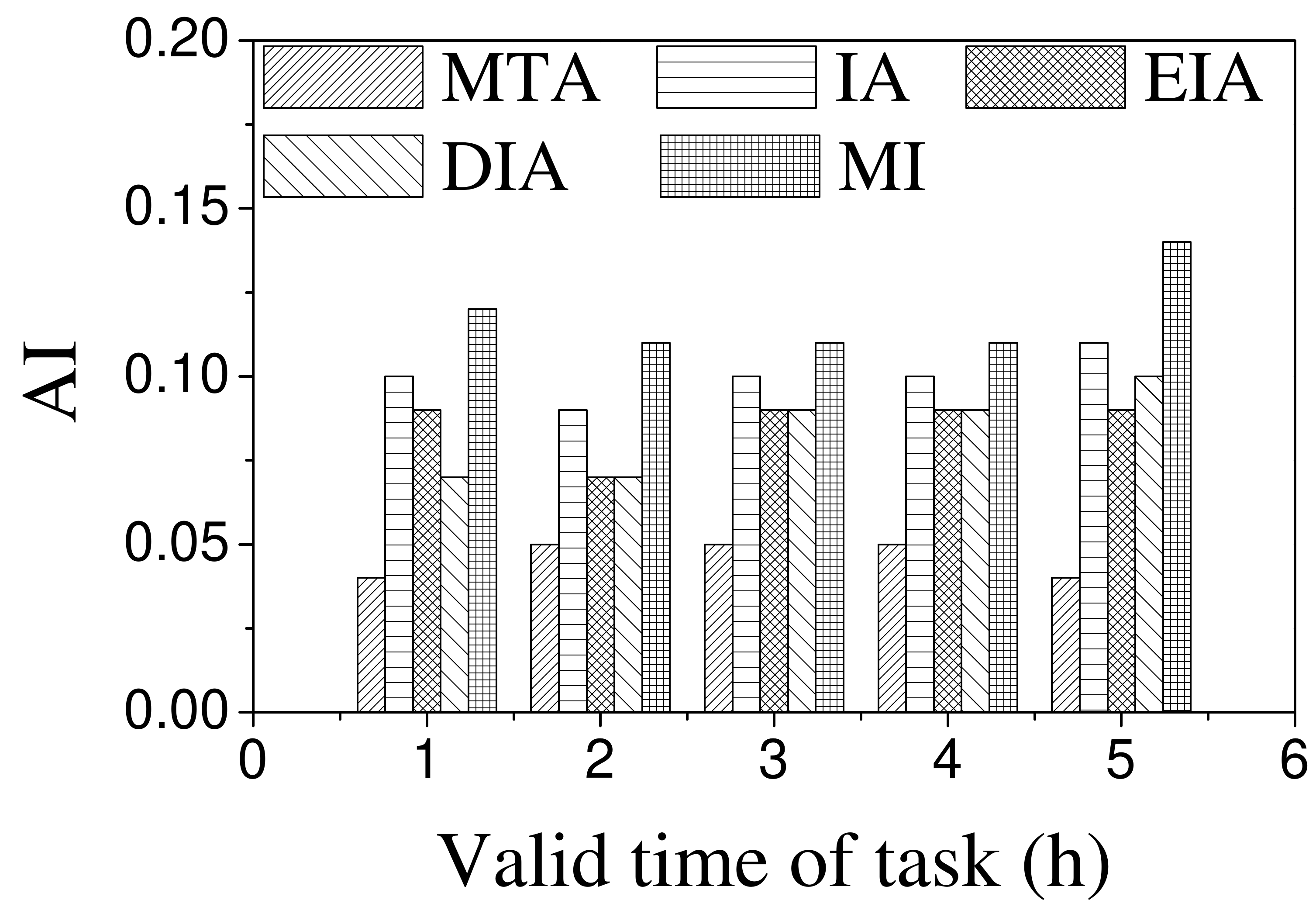}\label{fig:phi-fs-ai}}
\subfigure[Average Propagation] {\includegraphics[width=0.18\textwidth]{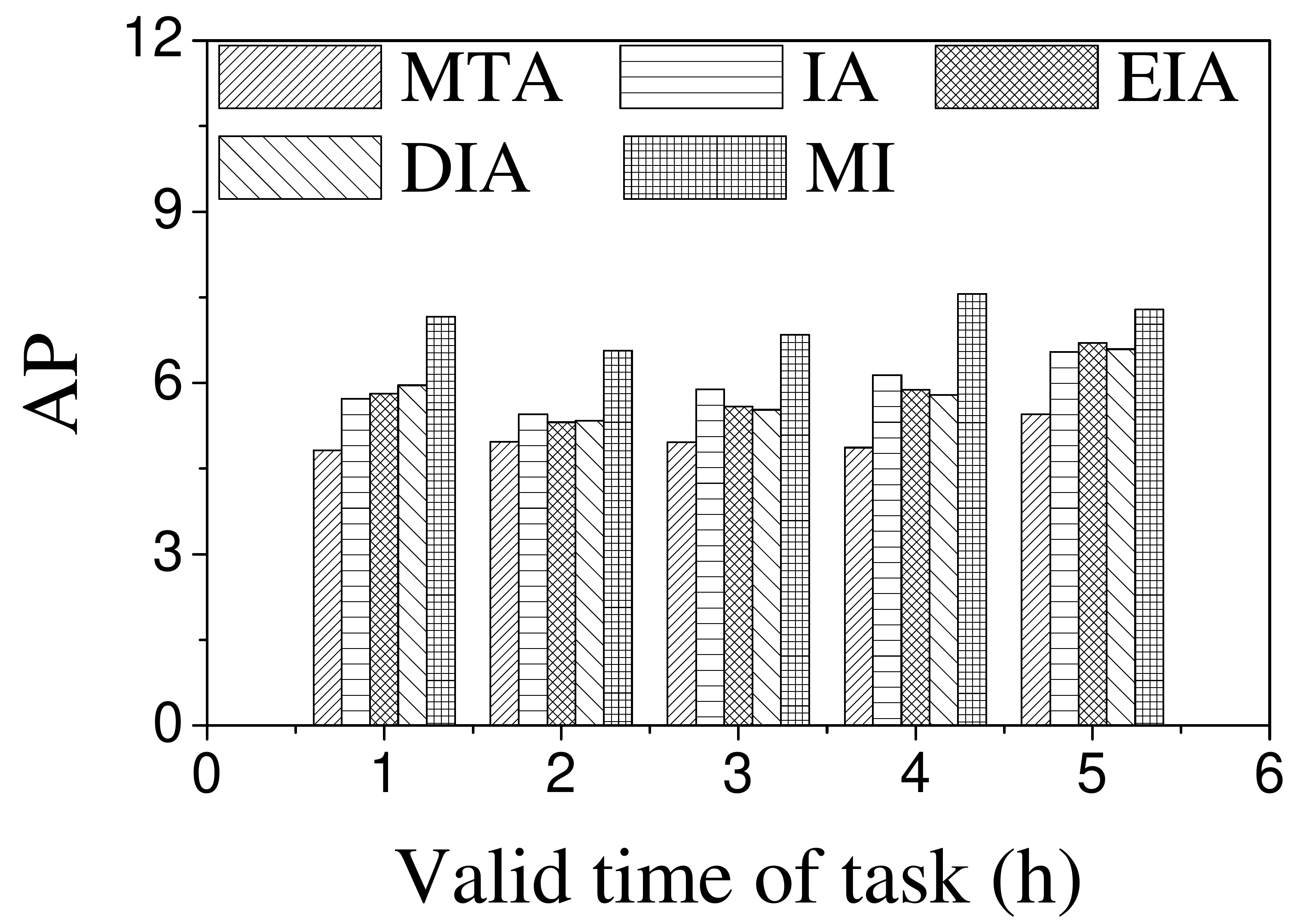}\label{fig:phi-fs-ap}}
\subfigure[Travel Cost] {\includegraphics[width=0.18\textwidth]{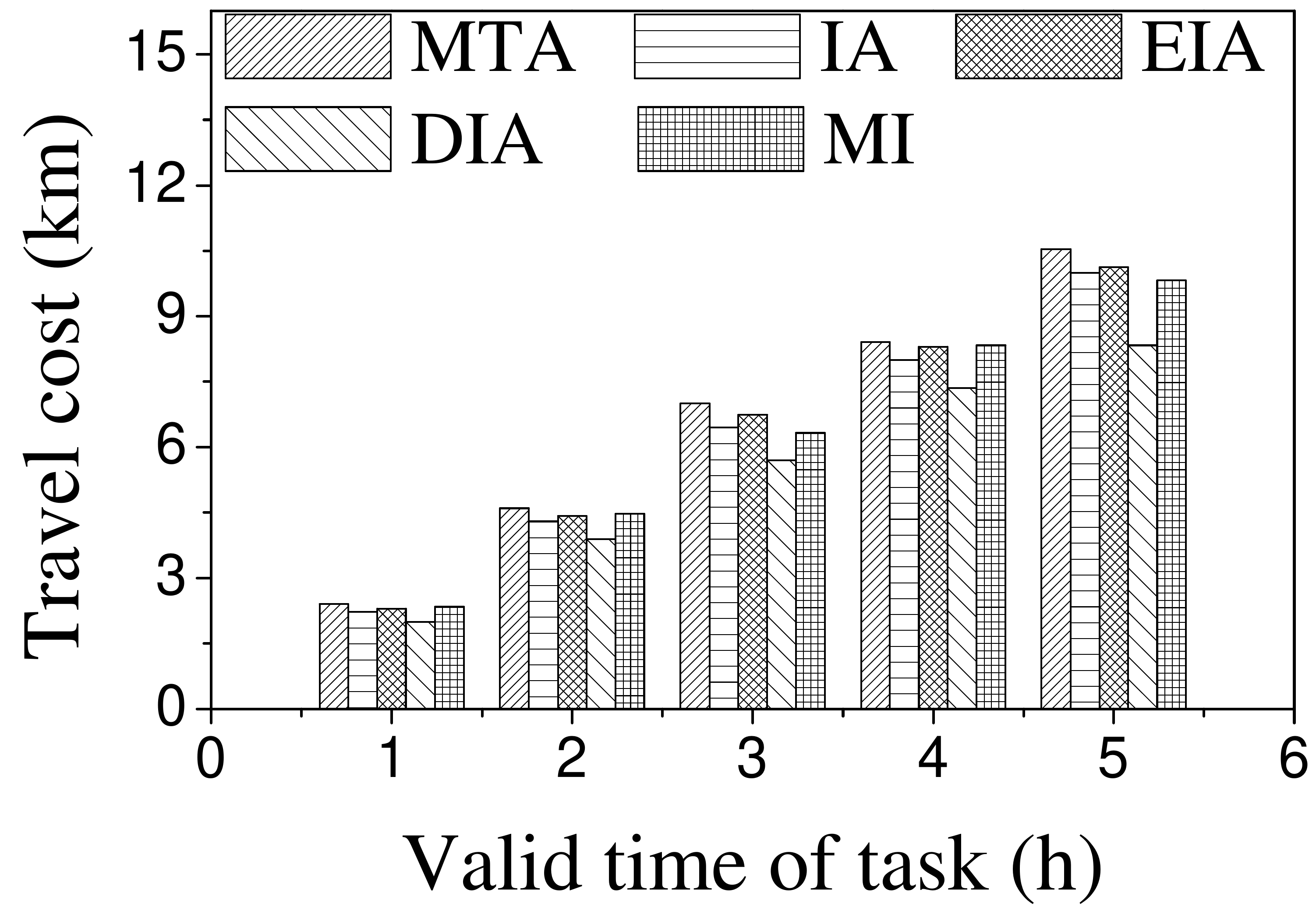}\label{fig:phi-fs-d}}
\vskip -9pt
\caption{Effect of $\varphi$ on FS}
\label{fig:phi-fs}
\end{figure*}

\emph{Effect of $|W|$:} Next, we study the effect of $|W|$ by varying it from 400 to 2,000. Figures~\ref{fig:w-bk-cpu} and~\ref{fig:w-fs-cpu} show that the CPU time increases when $|W|$ grows. The reason is that more workers tend to have more available task assignments, which leads to more edges in the task assignment graph. Since more workers can take part in task assignment, more tasks can be assigned, so the number of assigned tasks grows with the increase in the number of workers (see Figures~\ref{fig:w-bk-s} and~\ref{fig:w-fs-s}). As shown in Figures~\ref{fig:w-bk-ai} and~\ref{fig:w-fs-ai}, the $\mathit{AI}$ of MI, IA, EIA, and DIA are larger than that of MTA. Figures~\ref{fig:w-bk-ap} and~\ref{fig:w-fs-ap} show that the $\mathit{AP}$ of all methods changes randomly. The reason may be that workers are selected at random from the original datasets, which means that workers who can generate larger $\mathit{AP}$ have probability of being selected for $|W|$. Moreover, the average travel cost of DIA is the smallest, and that of MTA is the highest (see Figures~\ref{fig:w-bk-d} and~\ref{fig:w-fs-d}). The reason is that DIA takes workers' travel costs into account, while MTA disregards any location information.

\begin{figure*}
\centering
\subfigure[CPU Time] {\includegraphics[width=0.19\textwidth]{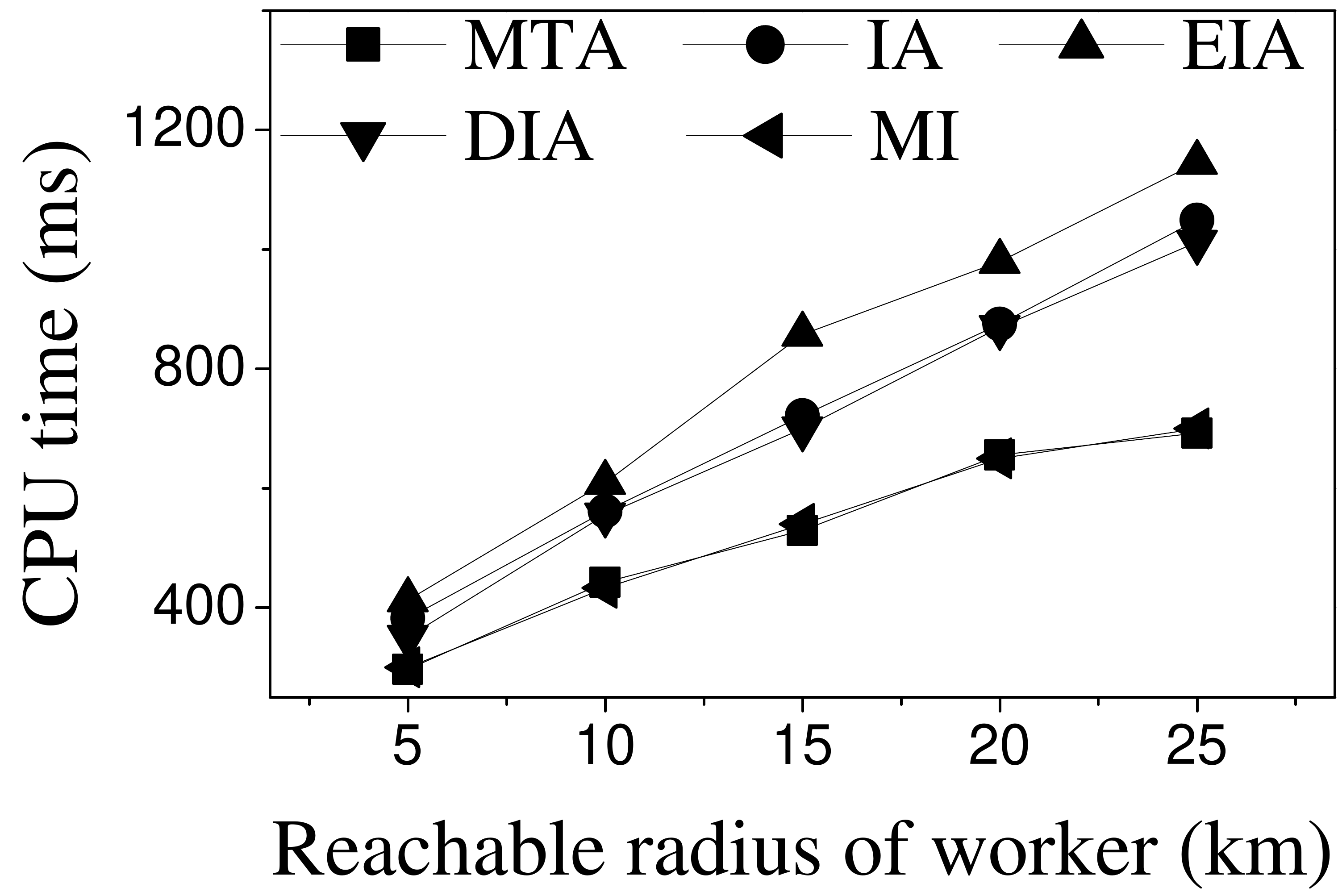}\label{fig:r-bk-cpu}}
\subfigure[Number of Assigned Tasks] {\includegraphics[width=0.19\textwidth]{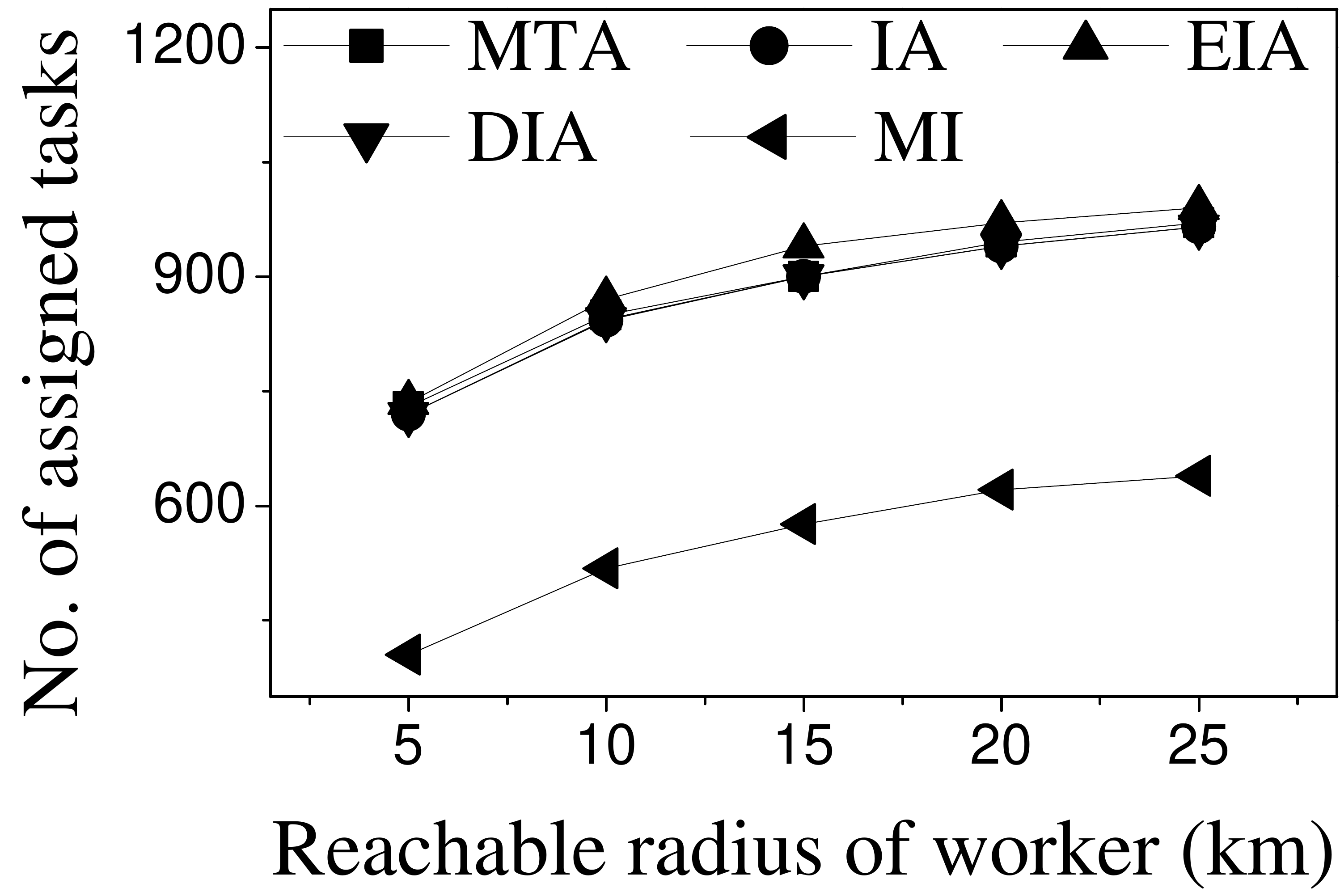}\label{fig:r-bk-s}}
\subfigure[Average Influence] {\includegraphics[width=0.19\textwidth]{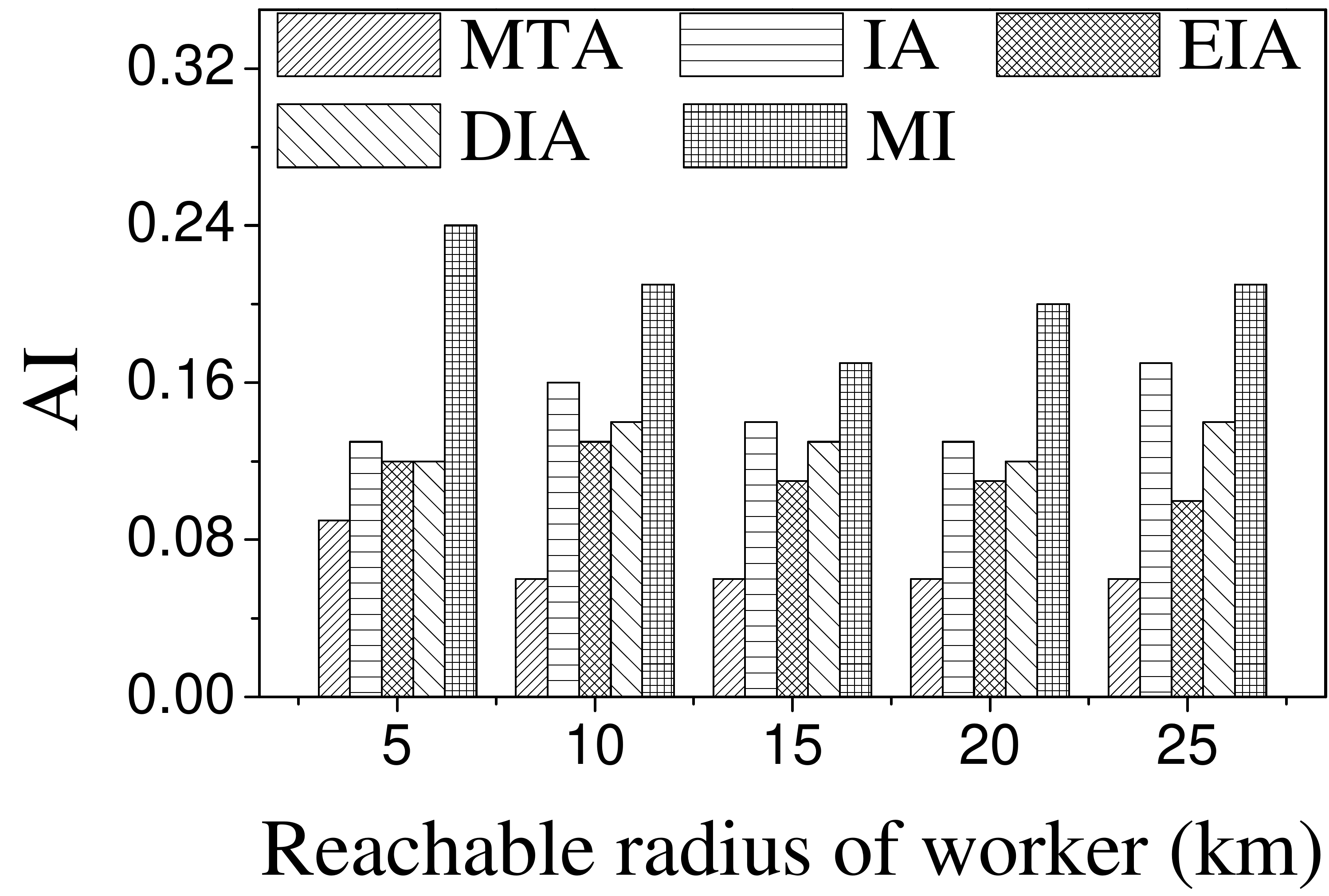}\label{fig:r-bk-ai}}
\subfigure[Average Propagation] {\includegraphics[width=0.184\textwidth]{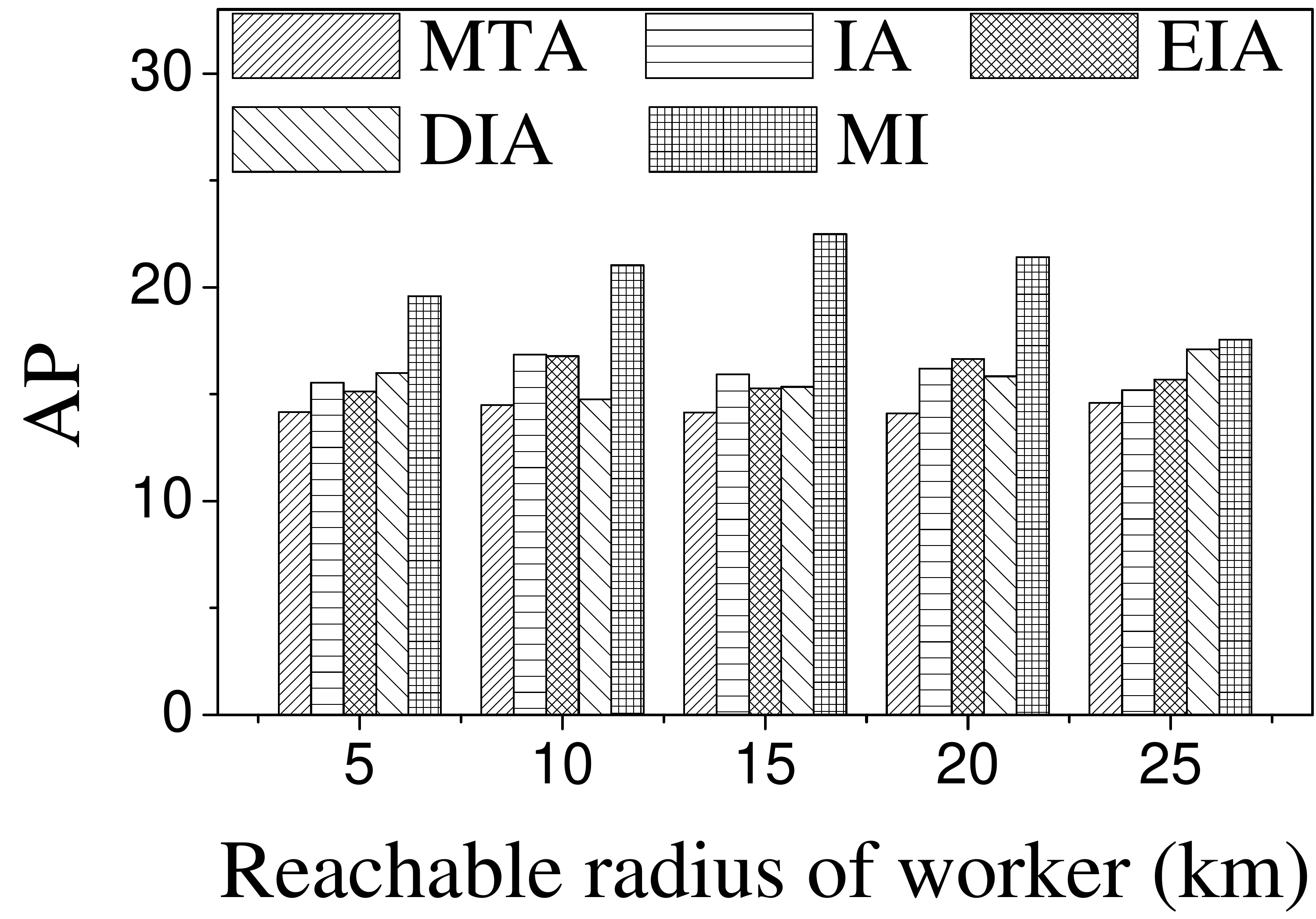}\label{fig:r-bk-ap}}
\subfigure[Travel Cost] {\includegraphics[width=0.18\textwidth]{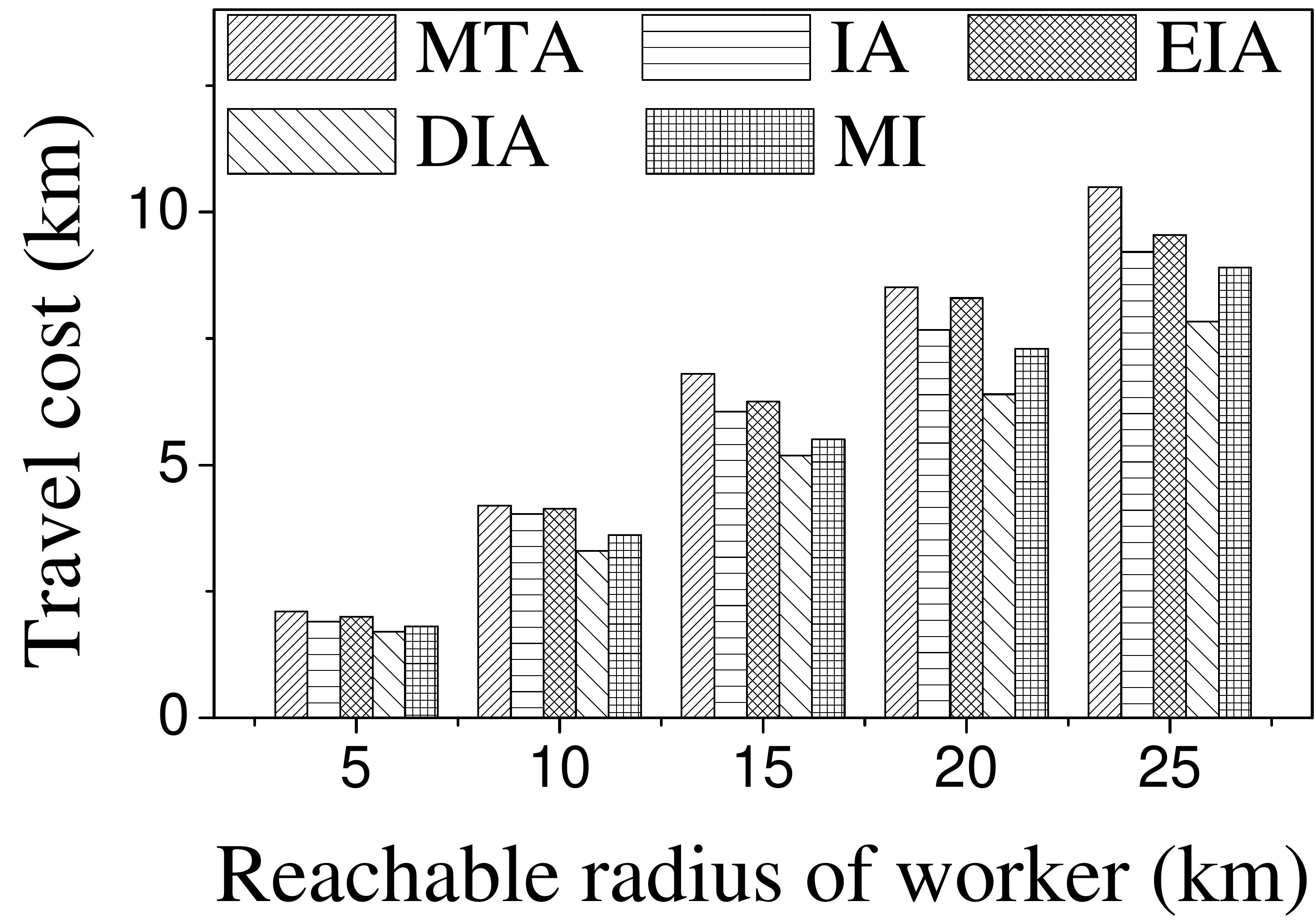}\label{fig:r-bk-d}}
\vskip -9pt
\caption{Effect of $r$ on BK}
\label{fig:r-bk}
\end{figure*}

\begin{figure*}
\centering
\subfigure[CPU Time] {\includegraphics[width=0.19\textwidth]{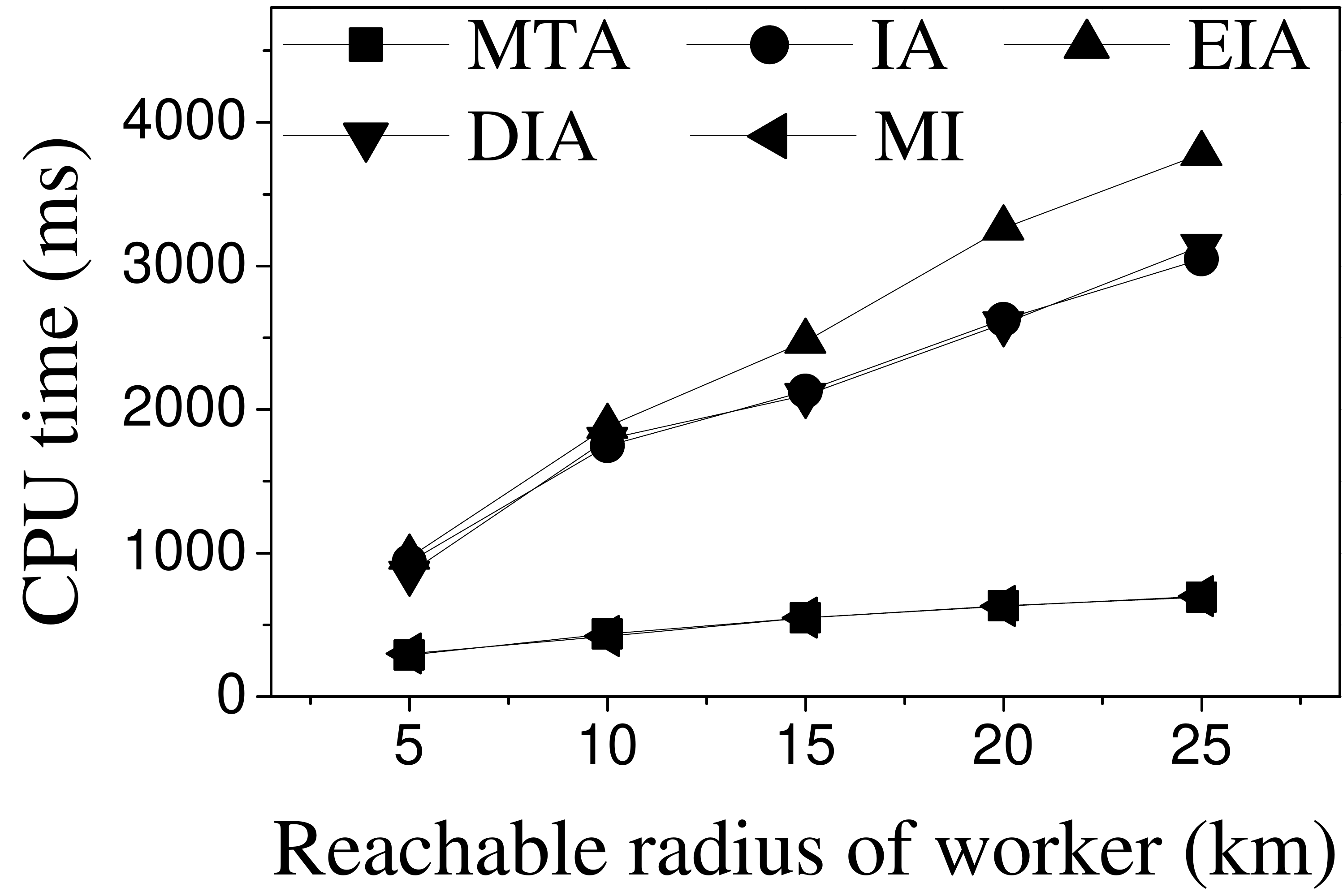}\label{fig:r-fs-cpu}}
\subfigure[Number of Assigned Tasks] {\includegraphics[width=0.19\textwidth]{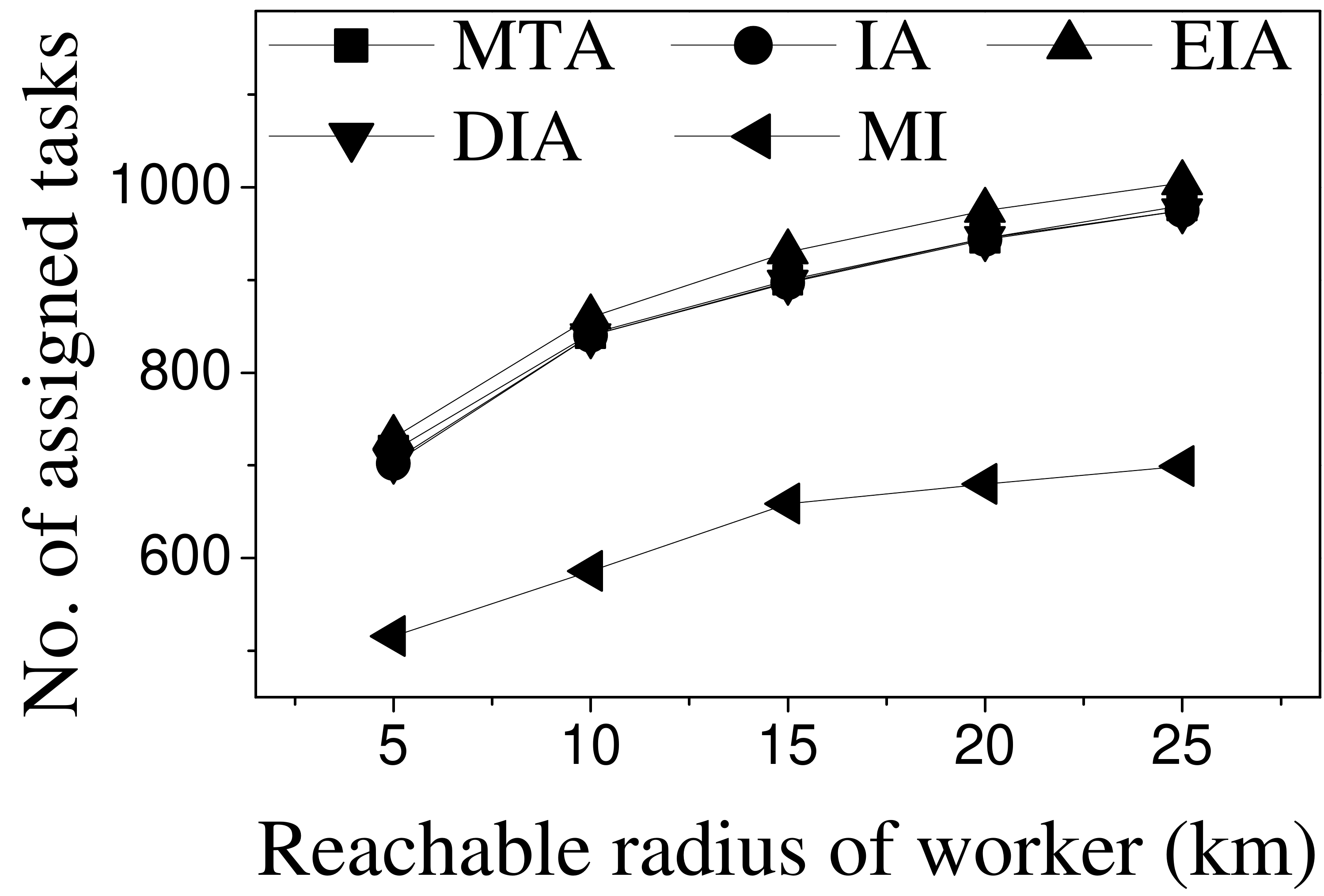}\label{fig:r-fs-s}}
\subfigure[Average Influence] {\includegraphics[width=0.19\textwidth]{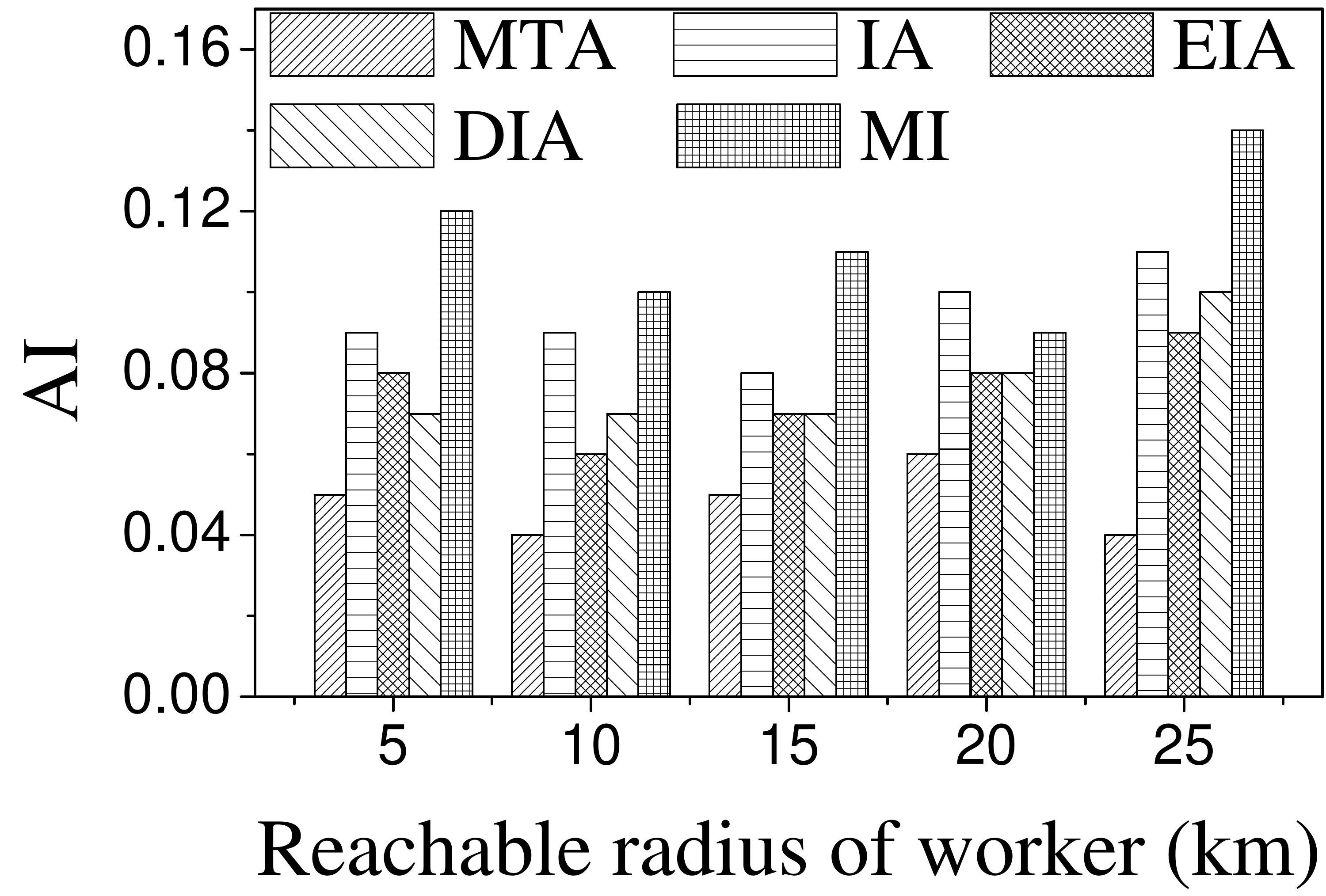}\label{fig:r-fs-ai}}
\subfigure[Average Propagation] {\includegraphics[width=0.18\textwidth]{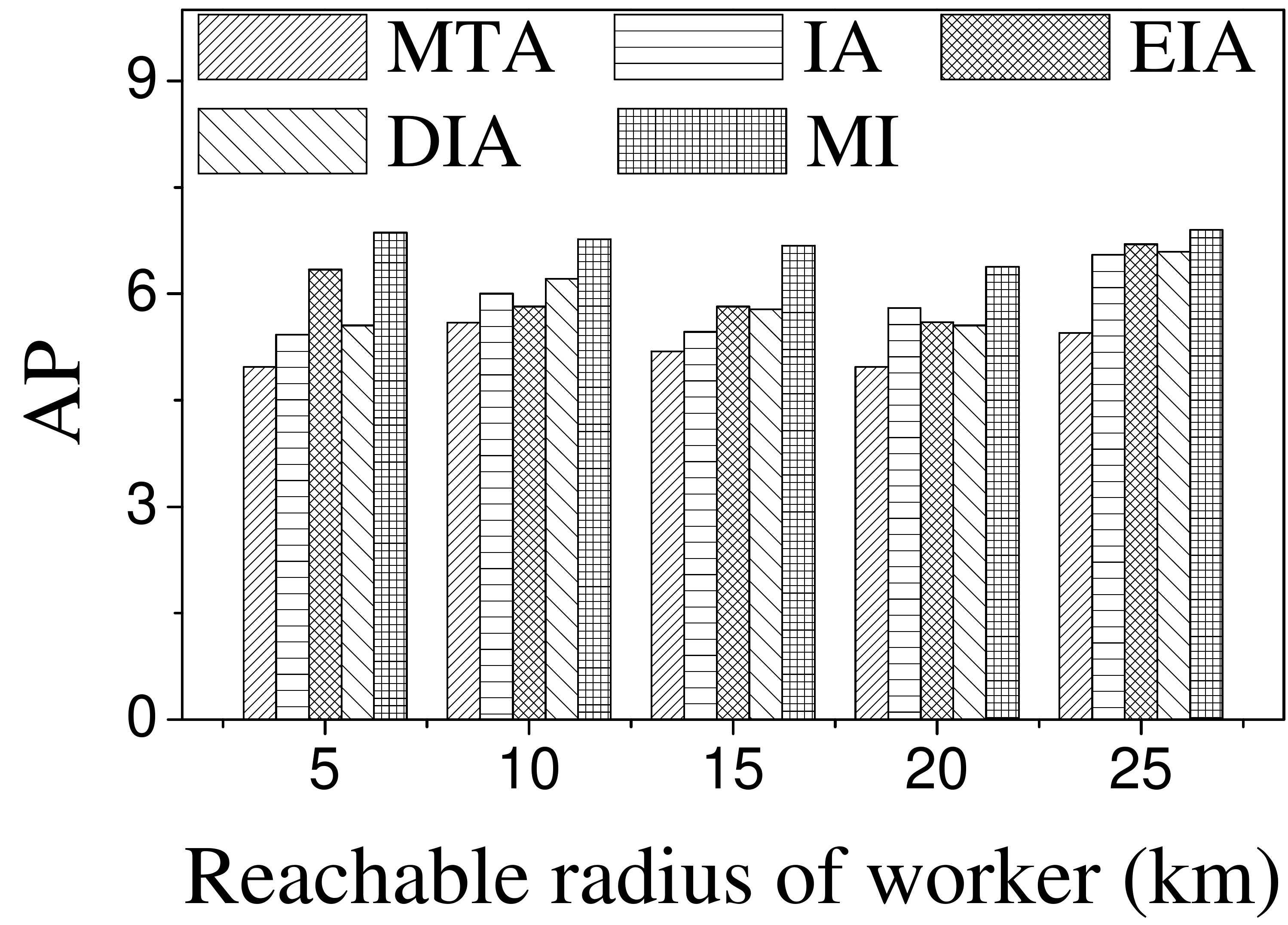}\label{fig:r-fs-ap}}
\subfigure[Travel Cost] {\includegraphics[width=0.181\textwidth]{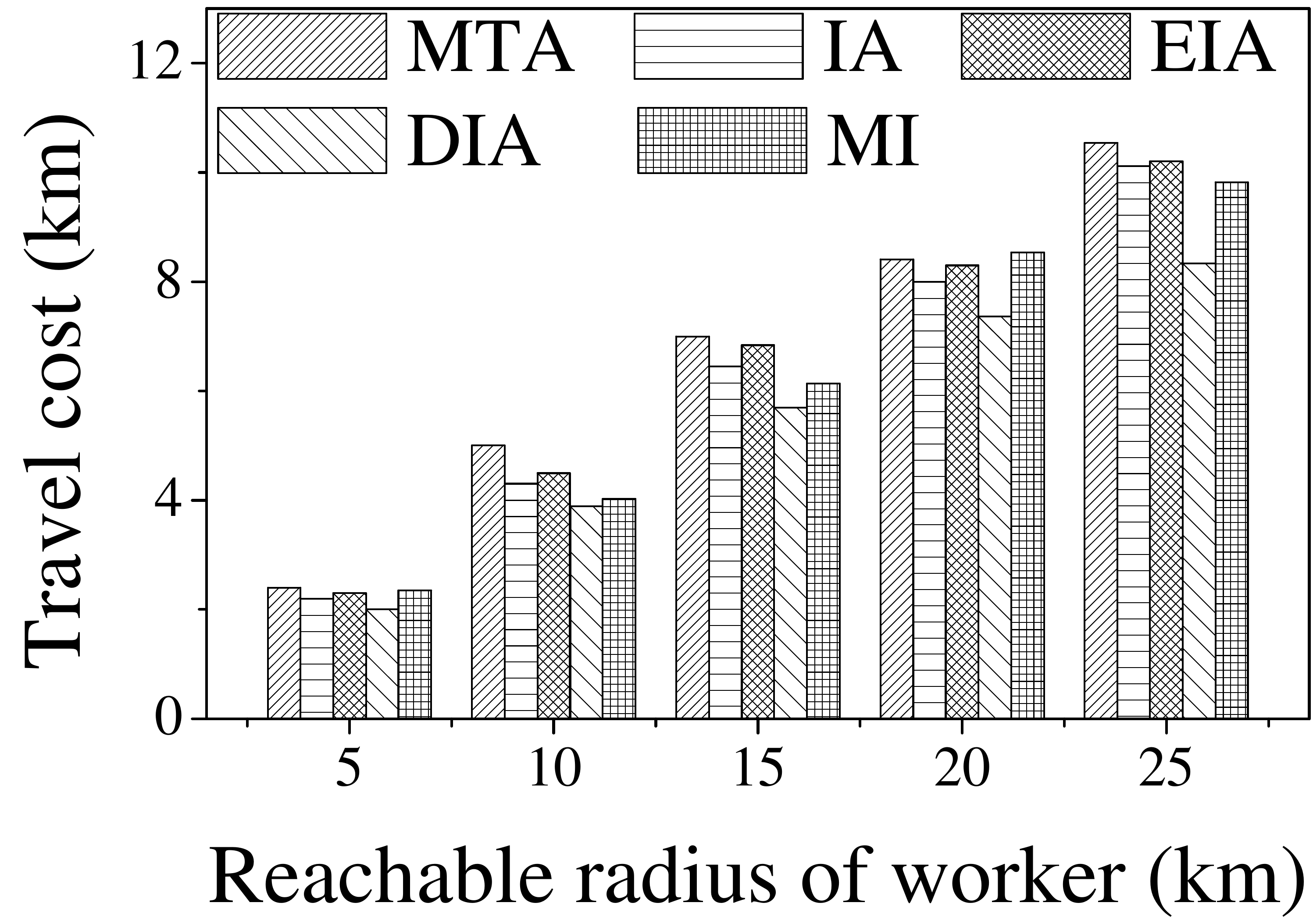}\label{fig:r-fs-d}}
\vskip -9pt
\caption{Effect of $r$ on FS}
\label{fig:r-fs}
\end{figure*}

\emph{Effect of $\varphi$:} As expected, the CPU costs of all methods increase when $\varphi$ grows (see Figures~\ref{fig:phi-bk-cpu} and~\ref{fig:phi-fs-cpu}). This occurs because workers can reach more tasks when $\varphi$ grows, which means that the number of available task assignments increases, i.e., more edges exist in the task assignment graph. As shown in Figures~\ref{fig:phi-bk-s} and~\ref{fig:phi-fs-s}, the number of assigned tasks of all methods grows with growing $\varphi$. The reason is that the task assignment graph becomes larger with larger $\varphi$, which means that the probability of workers being assigned a task increases. Figures~\ref{fig:phi-bk-ai},~\ref{fig:phi-bk-ap},~\ref{fig:phi-fs-ai}, and~\ref{fig:phi-fs-ap} show that the $\mathit{AI}$ and $\mathit{AP}$ of MI, IA, EIA, and DIA are larger than for MTA. The average travel costs of MTA are larger than those of other algorithms (see Figures~\ref{fig:phi-bk-d} and~\ref{fig:phi-fs-d}). Moreover, the average travel costs of all methods increase when $\varphi$ grows (see Figures~\ref{fig:phi-bk-d} and~\ref{fig:phi-fs-d}). The reason is that with the increase of $\varphi$, the probability of workers performing tasks with larger travel costs increases, which means that some workers are assigned tasks with larger travel costs. The average travel costs of EIA are larger than those of IA, DIA and MI since tasks with lower location entropy have higher priority to be assigned when applying EIA, which indicates workers travel longer to reach tasks.

\emph{Effect of $r$:} We proceed to consider the effect of $r$ by varying it from 5 to 25 km. Figures~\ref{fig:r-bk-cpu},~\ref{fig:r-bk-s},~\ref{fig:r-fs-cpu} and~\ref{fig:r-fs-s} show that the CPU time and the number of assigned tasks of all methods exhibit a similar increasing trend when $r$ grows. The reason is that with the increase of $r$, more tasks are available in each worker's reachable range, which means that each worker has higher probability to be assigned a task. It can also be seen that the gap in the number of assigned tasks between EIA and the other approaches increases. The reason is that when $r$ grows, the number of tasks that are far from workers increases, and the probability of workers accept tasks that are far from them is small. When applying EIA, the tasks with fewer workers nearby have higher priority of being assigned, increasing the probability that workers accept tasks that far from them, which leads to more assignments. As illustrated in Figures~\ref{fig:r-bk-ai},~\ref{fig:r-bk-ap},~\ref{fig:r-fs-ai}, and~\ref{fig:r-fs-ap}, the $\mathit{AI}$ and $\mathit{AP}$ of MTA are lower than for the other approaches, which demonstrates the superiority of the influence-aware assignment strategy. Since more tasks are assigned and workers can reach tasks with larger travel costs when $r$ grows, the average travel costs of all methods increase (cf. Figures~\ref{fig:r-bk-d} and~\ref{fig:r-fs-d}).

According to the above analysis, the time cost of MTA is the lowest, while the number of assigned tasks, Average Influence $(\mathit{AI})$ and Average Propagation $(\mathit{AP})$ of MTA are the smallest. The $\mathit{AI}$ of IA is larger than that of MTA, DIA and EIA because these algorithms adopt different strategies to improve the number of assigned tasks, which reduces the effect of worker-task influence. EIA is more time-consuming, but also achieved larger numbers of assigned tasks than the other algorithms. The travel cost of DIA is the smallest since it takes travel costs into account when assigning tasks. The $\mathit{AI}$ and $\mathit{AP}$ of MI are the largest because MI aims to maximize the total worker-task influence.

\section{Related Work}
\label{sec:related}
Spatial Crowdsourcing (SC) has been the subject of a range of studies~\cite{tong2020spatial,tong2018dynamic,zhao2021fairness,zhao2021coalition,xia2019profit,zhao2019destination,zhao2017destination,chen2020fair,li2021preference,yanwww2022}. One of the core problems in SC is task assignment. Kazemi et al.~\cite{kazemi2012geocrowd} consider two task publication modes, namely Worker Selected Tasks (WST) and Server Assigned Tasks (SAT). In WST mode, workers can choose nearby spatial tasks without the need to coordinate with the SC-server. In SAT mode, the server assigns tasks to workers with the aim of maximizing the number of assigned tasks~\cite{cheng2017prediction,tong2018slade,zhao2019preference,zhao2020preference} or maximizing the number of performed tasks for a worker with optimal schedule~\cite{deng2013maximizing}. 
Zeng et al.~\cite{zeng2018latency} study a latency-oriented task completion problem that addresses the trade-off between quality and latency for task assignment. Cheng et al.~\cite{cheng2019cooperation} focus on cooperation-aware spatial crowdsourcing, where more than one worker is required to complete a task. In contrast to these studies, we study a novel task assignment problem based on worker-task influence.

Next, quality assurance is a core challenge in spatial task assignment. Workers tend to complete tasks with good quality if a quality strategy exists. Zhao et al.~\cite{zhao2019preference} study preference-aware task assignment, which considers temporal preferences of workers.
Zhao et al.~\cite{zhao2019preference1} propose a preference-aware task assignment for on-demand taxi dispatching that aims to maximize the expected total profits. 
However, these studies simply infer workers' preferences from historical task-performing records, and they ignore workers' social impact.

Some recent studies try to improve task assignment based on social networks. Li et al.~\cite{li2020consensus} focus on group task assignment, which employs social features to learn social impact-based preferences of different worker groups. Wang et al.~\cite{wang2018social} propose two algorithms, Basic-Selector and Fast-Selector, to select a subset of workers to maximize the temporal-spatial coverage. However, these studies ignore the interactions among all workers in social networks and workers' long-term task performing patterns.


\section{Conclusion}
\label{sec:conclusion}
In this paper, we take an important step towards effective task assignment in spatial crowdsourcing that takes into account worker-task influence. Unlike most existing studies that only consider real-time worker and task locations, we further consider social networks to capture the interactions among workers, and we employ historical task-performing records to extract long-term task performing patterns of workers. We propose three task assignment algorithms that maximize the number of assigned tasks and worker-task influence. To the best of our knowledge, this is the first study in spatial crowdsourcing that considers worker-task influence in task assignment. An extensive empirical study based on real-world data demonstrates that the proposed methods can significantly improve the effectiveness of task assignment.

\section*{Acknowledgment}
This work is partially supported by NSFC (No. 61972069, 61836007 and 61832017), and Shenzhen Municipal Science and Technology R\&D Funding Basic Research Program (JCYJ20210324133607021).

\bibliography{ref}

\begin{thebibliography}{10}
\providecommand{\url}[1]{#1}
\csname url@samestyle\endcsname
\providecommand{\newblock}{\relax}
\providecommand{\bibinfo}[2]{#2}
\providecommand{\BIBentrySTDinterwordspacing}{\spaceskip=0pt\relax}
\providecommand{\BIBentryALTinterwordstretchfactor}{4}
\providecommand{\BIBentryALTinterwordspacing}{\spaceskip=\fontdimen2\font plus
\BIBentryALTinterwordstretchfactor\fontdimen3\font minus
  \fontdimen4\font\relax}
\providecommand{\BIBforeignlanguage}[2]{{%
\expandafter\ifx\csname l@#1\endcsname\relax
\typeout{** WARNING: IEEEtran.bst: No hyphenation pattern has been}%
\typeout{** loaded for the language `#1'. Using the pattern for}%
\typeout{** the default language instead.}%
\else
\language=\csname l@#1\endcsname
\fi
#2}}
\providecommand{\BIBdecl}{\relax}
\BIBdecl

\bibitem{zhao2020preference}
Y.~Zhao, K.~Zheng, H.~Yin, G.~Liu, J.~Fang, and X.~Zhou, ``Preference-aware
  task assignment in spatial crowdsourcing: from individuals to groups,''
  \emph{TKDE}, 2020.

\bibitem{cheng2016task}
P.~Cheng, X.~Lian, L.~Chen, J.~Han, and J.~Zhao, ``Task assignment on
  multi-skill oriented spatial crowdsourcing,'' \emph{TKDE}, vol.~28, no.~8,
  pp. 2201--2215, 2016.

\bibitem{cheng2017prediction}
P.~Cheng, X.~Lian, L.~Chen, and C.~Shahabi, ``Prediction-based task assignment
  in spatial crowdsourcing,'' in \emph{ICDE}, 2017, pp. 997--1008.

\bibitem{song2017trichromatic}
T.~Song, Y.~Tong, L.~Wang, J.~She, B.~Yao, L.~Chen, and K.~Xu, ``Trichromatic
  online matching in real-time spatial crowdsourcing,'' in \emph{ICDE}, 2017,
  pp. 1009--1020.

\bibitem{tong2018dynamic}
Y.~Tong, L.~Wang, Z.~Zhou, L.~Chen, B.~Du, and J.~Ye, ``Dynamic pricing in
  spatial crowdsourcing: A matching-based approach,'' in \emph{SIGMOD}, 2018,
  pp. 773--788.

\bibitem{tong2017flexible}
Y.~Tong, L.~Wang, Z.~Zimu, B.~Ding, L.~Chen, J.~Ye, and K.~Xu, ``Flexible
  online task assignment in real-time spatial data,'' \emph{PVLDB}, vol.~10,
  no.~11, pp. 1334--1345, 2017.

\bibitem{tong2018unified}
Y.~Tong, Y.~Zeng, Z.~Zhou, L.~Chen, J.~Ye, and K.~Xu, ``A unified approach to
  route planning for shared mobility,'' \emph{PVLDB}, vol.~11, no.~11, p. 1633,
  2018.

\bibitem{xia2019profit}
J.~Xia, Y.~Zhao, G.~Liu, J.~Xu, M.~Zhang, and K.~Zheng, ``Profit-driven task
  assignment in spatial crowdsourcing,'' in \emph{IJCAI}, 2019, pp. 1914--1920.

\bibitem{zhao2017destination}
Y.~Zhao, Y.~Li, Y.~Wang, H.~Su, and K.~Zheng, ``Destination-aware task
  assignment in spatial crowdsourcing,'' in \emph{CIKM}, 2017, pp. 297--306.

\bibitem{ye2021task}
G.~Ye, Y.~Zhao, X.~Chen, and K.~Zheng, ``Task allocation with geographic
  partition in spatial crowdsourcing,'' in \emph{CIKM}, 2021, pp. 2404--2413.

\bibitem{kazemi2012geocrowd}
L.~Kazemi and C.~Shahabi, ``Geocrowd: enabling query answering with spatial
  crowdsourcing,'' in \emph{SIGSPATIAL}, 2012, pp. 189--198.

\bibitem{cheng2014reliable}
P.~Cheng, X.~Lian, Z.~Chen, R.~Fu, L.~Chen, J.~Han, and J.~Zhao, ``Reliable
  diversity-based spatial crowdsourcing by moving workers,'' \emph{PVLDB},
  vol.~8, no.~10, pp. 1022--1033, 2015.

\bibitem{deng2013maximizing}
D.~Deng, C.~Shahabi, and U.~Demiryurek, ``Maximizing the number of worker's
  self-selected tasks in spatial crowdsourcing,'' in \emph{SIGSPATIAL}, 2013,
  pp. 324--333.

\bibitem{li2020consensus}
X.~Li, Y.~Zhao, X.~Zhou, and K.~Zheng, ``Consensus-based group task assignment
  with social impact in spatial crowdsourcing,'' \emph{Data Science and
  Engineering}, vol.~5, no.~4, pp. 375--390, 2020.

\bibitem{li2020group}
X.~Li, Y.~Zhao, J.~Guo, and K.~Zheng, ``Group task assignment with social
  impact-based preference in spatial crowdsourcing,'' in \emph{DASFAA}, 2020,
  pp. 677--693.

\bibitem{tang2018online}
J.~Tang, X.~Tang, X.~Xiao, and J.~Yuan, ``Online processing algorithms for
  influence maximization,'' in \emph{SIGMOD}, 2018, pp. 991--1005.

\bibitem{chen2020efficient}
X.~Chen, Y.~Zhao, G.~Liu, R.~Sun, X.~Zhou, and K.~Zheng, ``Efficient
  similarity-aware influence maximization in geo-social network,'' \emph{TKDE},
  2020.

\bibitem{yuen2012task}
M.-C. Yuen, I.~King, and K.-S. Leung, ``Task recommendation in crowdsourcing
  systems,'' in \emph{Proceedings of the first international workshop on
  crowdsourcing and data mining}, 2012, pp. 22--26.

\bibitem{to2014framework}
H.~To, G.~Ghinita, and C.~Shahabi, ``A framework for protecting worker location
  privacy in spatial crowdsourcing,'' \emph{PVLDB}, vol.~7, no.~10, pp.
  919--930, 2014.

\bibitem{vazirani2013approximation}
V.~V. Vazirani, \emph{Approximation algorithms}, 2013.

\bibitem{blei2003latent}
D.~M. Blei, A.~Y. Ng, and M.~I. Jordan, ``Latent dirichlet allocation,''
  \emph{the Journal of machine Learning research}, vol.~3, pp. 993--1022, 2003.

\bibitem{gong2018location}
W.~Gong, B.~Zhang, and C.~Li, ``Location-based online task assignment and path
  planning for mobile crowdsensing,'' \emph{IEEE Transactions on Vehicular
  Technology}, vol.~68, no.~2, pp. 1772--1783, 2018.

\bibitem{tao2020differentially}
Q.~Tao, Y.~Tong, Z.~Zhou, Y.~Shi, L.~Chen, and K.~Xu, ``Differentially private
  online task assignment in spatial crowdsourcing: A tree-based approach,'' in
  \emph{ICDE}, 2020, pp. 517--528.

\bibitem{han2011data}
J.~Han, J.~Pei, and M.~Kamber, \emph{Data mining: concepts and techniques},
  2011.

\bibitem{zhu2014exploiting}
W.-Y. Zhu, W.-C. Peng, and L.-J. Chen, ``Exploiting mobility for location
  promotion in location-based social networks,'' in \emph{DSAA}, 2014, pp.
  76--82.

\bibitem{zhu2015modeling}
W.-Y. Zhu, W.-C. Peng, L.-J. Chen, K.~Zheng, and X.~Zhou, ``Modeling user
  mobility for location promotion in location-based social networks,'' in
  \emph{SIGKDD}, 2015, pp. 1573--1582.

\bibitem{singhai2007novel}
R.~Singhai, S.~D. Joshi, and R.~K. Bhatt, ``A novel discrete distribution and
  process to model self-similar traffic,'' in \emph{ICT}, 2007, pp. 167--172.

\bibitem{kempe2003maximizing}
D.~Kempe, J.~Kleinberg, and {\'E}.~Tardos, ``Maximizing the spread of influence
  through a social network,'' in \emph{SIGKDD}, 2003, pp. 137--146.

\bibitem{chen2010scalable}
W.~Chen, C.~Wang, and Y.~Wang, ``Scalable influence maximization for prevalent
  viral marketing in large-scale social networks,'' in \emph{SIGKDD}, 2010, pp.
  1029--1038.

\bibitem{borgs2014maximizing}
C.~Borgs, M.~Brautbar, J.~Chayes, and B.~Lucier, ``Maximizing social influence
  in nearly optimal time,'' in \emph{SODA}, 2014, pp. 946--957.

\bibitem{tang2015influence}
Y.~Tang, Y.~Shi, and X.~Xiao, ``Influence maximization in near-linear time: A
  martingale approach,'' in \emph{SIGMOD}, 2015, pp. 1539--1554.

\bibitem{stoica2019fairness}
A.-A. Stoica and A.~Chaintreau, ``Fairness in social influence maximization,''
  in \emph{WWW}, 2019, pp. 569--574.

\bibitem{chen2021community}
X.~Chen, L.~Deng, Y.~Zhao, X.~Zhou, and K.~Zheng, ``Community-based influence
  maximization in location-based social network,'' \emph{WWWJ}, vol.~24, no.~6,
  pp. 1903--1928, 2021.

\bibitem{chung2006concentration}
F.~Chung and L.~Lu, ``Concentration inequalities and martingale inequalities: a
  survey,'' \emph{Internet Mathematics}, vol.~3, no.~1, pp. 79--127, 2006.

\bibitem{ford1956maximal}
L.~R. Ford and D.~R. Fulkerson, ``Maximal flow through a network,''
  \emph{Canadian journal of Mathematics}, vol.~8, pp. 399--404, 2009.

\bibitem{cranshaw2010bridging}
J.~Cranshaw, E.~Toch, J.~Hong, A.~Kittur, and N.~Sadeh, ``Bridging the gap
  between physical location and online social networks,'' in \emph{UbiComp},
  2010, pp. 119--128.

\bibitem{cho2011friendship}
E.~Cho, S.~A. Myers, and J.~Leskovec, ``Friendship and mobility: user movement
  in location-based social networks,'' in \emph{SIGKDD}, 2011, pp. 1082--1090.

\bibitem{likhyani2017locate}
A.~Likhyani, S.~Bedathur, and P.~Deepak, ``Locate: Influence quantification for
  location promotion in location-based social networks,'' in \emph{IJCAI},
  2017, pp. 2259--2265.

\bibitem{zhao2020predictive}
Y.~Zhao, K.~Zheng, Y.~Cui, H.~Su, F.~Zhu, and X.~Zhou, ``Predictive task
  assignment in spatial crowdsourcing: a data-driven approach,'' in
  \emph{ICDE}, 2020, pp. 13--24.

\bibitem{wang2021task}
Z.~Wang, Y.~Zhao, X.~Chen, and K.~Zheng, ``Task assignment with worker churn
  prediction in spatial crowdsourcing,'' in \emph{CIKM}, 2021, pp. 2070--2079.

\bibitem{jung2012irie}
K.~Jung, W.~Heo, and W.~Chen, ``Irie: Scalable and robust influence
  maximization in social networks,'' in \emph{ICDM}, 2012, pp. 918--923.

\bibitem{tong2020spatial}
Y.~Tong, Z.~Zhou, Y.~Zeng, L.~Chen, and C.~Shahabi, ``Spatial crowdsourcing: a
  survey,'' \emph{VLDBJ}, vol.~29, no.~1, pp. 217--250, 2020.

\bibitem{meng2013tracking}
D.~Meng, Y.~Jia, J.~Du, and F.~Yu, ``Tracking algorithms for multiagent
  systems,'' \emph{IEEE Transactions on neural networks and learning systems},
  vol.~24, no.~10, pp. 1660--1676, 2013.

\bibitem{zhao2021fairness}
Y.~Zhao, K.~Zheng, J.~Guo, B.~Yang, T.~B. Pedersen, and C.~S. Jensen,
  ``Fairness-aware task assignment in spatial crowdsourcing: Game-theoretic
  approaches,'' in \emph{ICDE}, 2021, pp. 265--276.

\bibitem{zhao2021coalition}
Y.~Zhao, J.~Guo, X.~Chen, J.~Hao, X.~Zhou, and K.~Zheng, ``Coalition-based task
  assignment in spatial crowdsourcing,'' in \emph{ICDE}, 2021, pp. 241--252.

\bibitem{zhao2019destination}
Y.~Zhao, K.~Zheng, Y.~Li, H.~Su, J.~Liu, and X.~Zhou, ``Destination-aware task
  assignment in spatial crowdsourcing: A worker decomposition approach,''
  \emph{TKDE}, vol.~32, no.~12, pp. 2336--2350, 2019.

\bibitem{chen2020fair}
Z.~Chen, P.~Cheng, L.~Chen, X.~Lin, and C.~Shahabi, ``Fair task assignment in
  spatial crowdsourcing,'' \emph{PVLDB}, vol.~13, no.~12, pp. 2479--2492, 2020.

\bibitem{li2021preference}
Y.~Li, Y.~Zhao, and K.~Zheng, ``Preference-aware group task assignment in
  spatial crowdsourcing: A mutual information-based approach,'' in \emph{ICDM},
  2021, pp. 350--359.

\bibitem{yanwww2022}
Y.~Zhao, X.~Chen, L.~Deng, T.~Kieu, C.~Guo, B.~Yang, K.~Zheng, and C.~S.
  Jensen, ``Outlier detection for streaming task assignment in crowdsourcing,''
  in \emph{WWW}, 2022.

\bibitem{tong2018slade}
Y.~Tong, L.~Chen, Z.~Zhou, H.~V. Jagadish, L.~Shou, and W.~Lv, ``Slade: A smart
  large-scale task decomposer in crowdsourcing,'' \emph{TKDE}, vol.~30, no.~8,
  pp. 1588--1601, 2018.

\bibitem{zhao2019preference}
Y.~Zhao, J.~Xia, G.~Liu, H.~Su, D.~Lian, S.~Shang, and K.~Zheng,
  ``Preference-aware task assignment in spatial crowdsourcing,'' in
  \emph{AAAI}, vol.~33, no.~01, 2019, pp. 2629--2636.

\bibitem{zeng2018latency}
Y.~Zeng, Y.~Tong, L.~Chen, and Z.~Zhou, ``Latency-oriented task completion via
  spatial crowdsourcing,'' in \emph{ICDE}, 2018, pp. 317--328.

\bibitem{cheng2019cooperation}
P.~Cheng, L.~Chen, and J.~Ye, ``Cooperation-aware task assignment in spatial
  crowdsourcing,'' in \emph{ICDE}, 2019, pp. 1442--1453.

\bibitem{zhao2019preference1}
B.~Zhao, P.~Xu, Y.~Shi, Y.~Tong, Z.~Zhou, and Y.~Zeng, ``Preference-aware task
  assignment in on-demand taxi dispatching: An online stable matching
  approach,'' in \emph{AAAI}, vol.~33, no.~01, 2019, pp. 2245--2252.

\bibitem{wang2018social}
J.~Wang, F.~Wang, Y.~Wang, D.~Zhang, L.~Wang, and Z.~Qiu,
  ``Social-network-assisted worker recruitment in mobile crowd sensing,''
  \emph{TMC}, vol.~18, no.~7, pp. 1661--1673, 2018.

\end{thebibliography}
\bibliographystyle{IEEEtran}
\end{document}